\documentclass[11pt,a4paper]{scrartcl}

\usepackage[bf]{caption}
\usepackage[utf8]{inputenc} 
\usepackage[T1]{fontenc} 
\usepackage[UKenglish]{babel} 
\usepackage[section]{placeins} 
\usepackage{a4, 
            amsfonts, 
            amsmath, 
            amssymb, 
            amsthm, 
            appendix, 
            arydshln, 
            booktabs, 
            calligra, 
            empheq, 
            etex, 
            etoolbox, 
            fancyvrb, 
            geometry, 
            hyperref, 
            lipsum, 
            listings, 
            longtable, 
            lscape, 
            mathrsfs, 
            pbox, 
            scrlayer-scrpage, 
            subfig, 
            tikz, 
            cite,
            todonotes, 
            xy 
}

\xyoption{all}

\usetikzlibrary{matrix,arrows,automata,calc,chains,intersections,decorations.markings,decorations.pathmorphing,shapes.geometric,positioning,plotmarks,scopes,
                patterns,fit}

\tikzstyle{vecArrow} = [thick, decoration={markings,mark=at position
   1 with {\arrow[semithick]{open triangle 60}}},
   double distance=1.4pt, shorten >= 5.5pt,
   preaction = {decorate},
   postaction = {draw,line width=1.4pt, white,shorten >= 4.5pt}]

\hypersetup{
    pdftitle={Algebraic Cycles and Local Anomalies in F-Theory},
    pdfauthor={Martin Bies, Christoph Mayrhofer, Timo Weigand},
    pdfsubject={F-Theory, 4d/6d SQFTs, Algebraic Cycles}
} 

\definecolor{FireBrick}{rgb}{0.5812,0.0074,0.0083}
\definecolor{RoyalBlue}{rgb}{0.0236,0.0894,0.6179}
\definecolor{RoyalGreen}{rgb}{0.0236,0.6179,0.0894}
\definecolor{RoyalRed}{rgb}{0.6179,0.0236,0.0894}
\definecolor{LightBlue}{rgb}{0.8544,0.9511,1.0000}
\definecolor{Black}{rgb}{0.0,0.0,0.0}
\definecolor{linkColor}{rgb}{0.0,0.0,0.554}
\definecolor{citeColor}{rgb}{0.0,0.0,0.554}
\definecolor{fileColor}{rgb}{0.0,0.0,0.554}
\definecolor{urlColor}{rgb}{0.0,0.0,0.554}
\definecolor{promptColor}{rgb}{0.0,0.0,0.589}
\definecolor{brkpromptColor}{rgb}{0.589,0.0,0.0}
\definecolor{gapinputColor}{rgb}{0.589,0.0,0.0}
\definecolor{gapoutputColor}{rgb}{0.0,0.0,0.0}
\definecolor{FuncColor}{rgb}{0.0,0.0,0.0}
\definecolor{Chapter}{rgb}{0.0,0.0,0.0}
\definecolor{DarkOlive}{rgb}{0.1047,0.2412,0.0064}
\definecolor{darkgreen}{rgb}{0.05,0.6,0.1}
\definecolor{LightCyan}{rgb}{0.88,1,1}

\newcommand*\widefbox[1]{\fbox{\hspace{2em}#1\hspace{2em}}}

\newcommand{\cE}{\mathcal{E}}

\newcommand{\vect}[1]{\boldsymbol{#1}}
\newcommand{\Mint}{\int \limits}

\renewcommand{\hat}{\widehat}

\newcommand{\bfR}{\mathbf{R}}

\newcommand{\ie}{i.e.\ }
\newcommand{\eg}{e.g.\ }

\renewcommand{\bf}[1]{\textbf{#1}}

\renewcommand{\rm}[1]{\textrm{#1}}

\renewcommand{\[}{\begin{equation}}
\renewcommand{\]}{\end{equation}}

\newcommand{\be}{\begin{equation}}
\newcommand{\ee}{\end{equation}}

\newcommand{\beq}{\begin{equation}}
\newcommand{\eeq}{\end{equation}}

\newcommand{\bal}{\begin{aligned}}
\newcommand{\eal}{\end{aligned}}

\newcommand{\bea}{\begin{eqnarray}}
\newcommand{\eea}{\end{eqnarray}}

\DeclareMathOperator{\Hom}{\mathscr{H}\text{\kern -3pt {\calligra\large om}}\,}

\makeatletter
\def\enumfix{%
\if@inlabel
 \noindent \par\nobreak\vskip-\topsep\hrule\@height\z@
\fi}
\let\olditemize\itemize
\def\itemize{\enumfix\olditemize}
\makeatother

\makeatletter
\def\enumfix{%
\if@inlabel
 \noindent \par\nobreak\vskip-\topsep\hrule\@height\z@
\fi}
\let\oldenumerate\enumerate
\def\enumerate{\enumfix\oldenumerate}
\makeatother

\newtheoremstyle{break}  
  {\topsep}   
  {\topsep}   
  {}  
  {0pt}       
  {\bfseries} 
  {:}         
  {\newline}  
  {}          

\theoremstyle{break}

\makeatletter
\let\@addpunct\@gobble
\makeatother

\lstnewenvironment{mat}
{\lstset{language=mathematica,mathescape,columns=flexible}}
{}

\sbox0{\small\ttfamily A}
\edef\mybasewidth{\the\wd0 }
\lstnewenvironment{gap}
{\lstset{language=GAP,
	frame=single,
	basicstyle=\ttfamily\small,
	numbers=left,
	numberstyle=\small,
	keywordstyle=\color{red},
	stringstyle=\color{blue},
	commentstyle=\color{green!70!black},
	columns=fullflexible,
	basewidth=\mybasewidth,
	keepspaces=true,
	stepnumber=1,
	showstringspaces=false,
	breaklines=true}}
{}

\def\hybrid{\topmargin -20pt    \oddsidemargin 0pt
        \headheight 15.2pt \headsep 0pt
        \textwidth 6.25in       
        \textheight 9 in       
        \marginparwidth .875in
        \parskip 5pt plus 1pt 
        \jot = 1.5ex
   }
\hybrid

\numberwithin{equation}{section}
\numberwithin{table}{section}

\def\sectionautorefname{Section}
\def\subsectionautorefname{Section}

\addto\extrasUKenglish{%
  \renewcommand{\sectionautorefname}{Section}%
}

\makeatletter
\patchcmd{\hyper@makecurrent}{%
    \ifx\Hy@param\Hy@chapterstring
        \let\Hy@param\Hy@chapapp
    \fi
}{%
    \iftoggle{inappendix}{
        \@checkappendixparam{chapter}%
        \@checkappendixparam{section}%
        \@checkappendixparam{subsection}%
        \@checkappendixparam{subsubsection}%
        \@checkappendixparam{paragraph}%
        \@checkappendixparam{subparagraph}%
    }{}%
}{}{\errmessage{failed to patch}}

\newcommand*{\@checkappendixparam}[1]{%
    \def\@checkappendixparamtmp{#1}%
    \ifx\Hy@param\@checkappendixparamtmp
        \let\Hy@param\Hy@appendixstring
    \fi
}
\makeatletter

\newtoggle{inappendix}
\togglefalse{inappendix}

\apptocmd{\appendix}{\toggletrue{inappendix}}{}{\errmessage{failed to patch}}
\apptocmd{\subappendices}{\toggletrue{inappendix}}{}{\errmessage{failed to patch}}

\begin{document}

\let\subsectionautorefname\sectionautorefname
\let\subsubsectionautorefname\sectionautorefname

\baselineskip=14pt
\parskip 5pt plus 1pt

\vspace*{-1.5cm}
\begin{flushright}    
  {\small

  }
\end{flushright}

\vspace{2cm}
\begin{center}        
  {\LARGE Algebraic Cycles and Local Anomalies in F-Theory}
\end{center}

\vspace{0.75cm}
\begin{center}        
Martin Bies\textsuperscript{1}, Christoph Mayrhofer\textsuperscript{2}, and Timo Weigand\textsuperscript{1,3}
\end{center}

\vspace{0.15cm}
\begin{center}        
  \emph{\textsuperscript{1} Institut f\"ur Theoretische Physik, Ruprecht-Karls-Universit\"at, \\
             Philosophenweg 19, 69120 
             Heidelberg, Germany}
             \\[0.15cm]
 \emph{\textsuperscript{2} Arnold Sommerfeld Center for Theoretical Physics, \\
             Theresienstra{\ss}e 37, 80333 M\"unchen, Germany}
             \\[0.15cm]
 \emph{\textsuperscript{3} CERN, Theory Division, \\
             CH-1211 Geneva 23, Switzerland}
             \\[0.15cm]
   
 \end{center}

\vspace{2cm}

\begin{abstract}

We introduce a set of  identities in the cohomology ring of elliptic fibrations which are equivalent to the cancellation of gauge and mixed gauge-gravitational anomalies in F-theory compactifications to four and six dimensions. The identities consist in (co)homological relations between complex codimension-two cycles. The same set of relations, once evaluated on elliptic Calabi-Yau three-folds and four-folds, is shown to universally govern the structure of anomalies and their Green-Schwarz cancellation in six- and four-dimensional F-theory vacua, respectively. We furthermore conjecture that these relations hold not only within the cohomology ring, but even at the level of the Chow ring, \ie as relations among codimension-two cycles modulo rational equivalence. We verify this conjecture in non-trivial examples with Abelian and non-Abelian gauge groups factors. Apart from governing the structure of local anomalies, the identities in the Chow ring relate different types of gauge backgrounds on elliptically fibred Calabi-Yau four-folds.

\end{abstract}

\thispagestyle{empty}
\clearpage


\newpage
\setcounter{tocdepth}{2}
\tableofcontents


\section{Introduction} \label{sec_Intro}

Anomalies provide a beautiful window of opportunity to study the consistency of quantum field theories and their coupling to gravity in various dimensions. As a celebrated example, the absence of gauge and gravitational anomalies in perturbative string theory \cite{Green:1984sg} implies the cancellation of one-loop anomalies in the effective 10-dimensional supergravity by the classical Green-Schwarz counterterms. Given the absence of anomalies in ten dimensions, compactifications of string theory to lower dimensions are automatically free of gauge and gravitational anomalies provided all string consistency conditions are correctly obeyed. For instance, in perturbative orientifold compactifications, the cancellation of gauge or gravitational anomalies in the low-energy effective action is ensured by the tadpole conditions \cite{Green:1984ed,Green:1984qs,Polchinski:1987tu,Sagnotti:1995ga}, as reviewed e.g. in \cite{Blumenhagen:2006ci,Ibanez:2012zz} and references therein. 

In F-theory, tadpole cancellation is automatically built into the topological and geometric properties of the torus-fibration underlying the definition of the F-theory vacuum. Hence the absence of anomalies in the F-theory effective action can serve as a unique tool to gain valuable insights into the mathematical structure of such fibrations. In this work we find a number of non-trivial relations in the cohomology ring of elliptic fibrations which are equivalent to the cancellation of all gauge and gauge-gravitational anomalies in F-theory compactified to four and six dimensions.

Assuming the absence of anomalies in the effective theory, as suggested by the above reasoning, therefore provides a physics proof for these novel and non-trivial mathematical relations among codimension-2 cycles; these identities must then be valid  on every flat, smooth, elliptically fibred Calabi-Yau 3- and 4-fold irrespective of its interpretation as the compactification space of an F-theory. 

The constraints imposed by absence of anomalies in an effective supergravity are particularly strong in 6-dimensional $N=(1,0)$ supersymmetric vacua. These chiral theories are obtained, in the context of  F-theory, via compactifications on elliptic Calabi-Yau 3-folds.\footnote{The constraints imposed by anomalies on globally consistent 6-dimensional F-theory vacua have been studied in detail in the literature, including \cite{Kumar:2009us,Kumar:2009ac,Kumar:2010ru,Seiberg:2011dr,Park:2011wv} and references therein. Some extensions of this reasoning to 4-dimensional compactifications have appeared in \cite{Grimm:2012yq}.}
The 6-dimensional generalisation of the Green-Schwarz mechanism \cite{Green:1984bx,Sagnotti:1992qw} acts on the self-dual tensor fields in the effective action such as to cancel all factorisable gauge and gravitational anomalies. Its realisation in F-theory was first described for ADE type gauge algebras in \cite{Sadov:1996zm} and extended further to all simple non-Abelian gauge algebras in \cite{Grassi:2000we,Grassi:2011hq}. In particular, \cite{Grassi:2011hq} has identified a set of relations in the Chow group $\mathrm{CH}_0(B_2)$ of points on the base $B_2$ of an elliptically fibred 3-fold which are sufficient to guarantee the cancellation of all gauge-gravitational one-loop anomalies by the Green-Schwarz terms. The cancellation of 6-dimensional anomalies in F-theory has been extended in \cite{Park:2011ji} to include also Abelian gauge group factors. The same reference furthermore shows that a certain set of intersection theoretic identities ensures cancellation of all gauge and gravitational anomalies. These identities relate the intersection numbers between divisors and curves on the fibration which, respectively, represent the coroot/coweight and the root/weight lattice of the effective gauge theory. Similarly to the set of relations investigated in \cite{Grassi:2011hq} they hold for any elliptically fibred Calabi-Yau 3-fold even though the proofs presented in \cite{Grassi:2011hq,Park:2011ji}  proceed in purely physics terms via the connection to gauge anomalies. 

By comparing the 6-dimensional F-theory effective action to its dual M-theory in five dimensions\footnote{Similar relations hold for dual F/M-theory pairs in lower dimensions.}, the cancellation of anomalies in F-theory is deeply related to the structure of Chern-Simons terms in M-theory, as studied in detail in \cite{Grimm:2011fx,Bonetti:2011mw,Cvetic:2012xn,Grimm:2015zea,Esole:2015xfa}. According to \cite{Grimm:2015zea}, such reasoning implies that the cancellation for instance of all Abelian gauge anomalies in the F-theory effective action is equivalent to a symmetry among all sections as to which serves as the zero-section of the F-theory fibration defining the Kaluza-Klein $U(1)$ in the dual M-theory. 

Our analysis of anomalies begins in the 4-dimensional context. Local anomalies in 4-dimen-sional $\mathcal{N}=1$ supergravity theories occur only if the massless spectrum is chiral. In F-theory compactifications on elliptically fibred 4-folds $\hat Y_4$ this requires the presence of gauge flux. Duality with M-theory translates  the latter into a flux for the field strength $G_4$ of the M-theory 3-form potential $C_3$, as will be briefly reviewed in \autoref{sec_RevF}. As in six dimensions the anomalies comprise two parts, the 1-loop and the Green-Schwarz terms: The one-loop induced gauge-gravitational anomalies involve chiral fermions running in triangle diagrams. The chiral index of the relevant fermions is determined by the intersection theoretic pairing of the flux $G_4$ with certain matter surfaces inside $\hat Y_4$. In any consistent vacuum, these 1-loop anomalies are cancelled by the 4-dimensional Green-Schwarz terms. For F-theory compactification on $\hat Y_4$ the latter have been derived in \cite{Cvetic:2012xn}. As will be reviewed in \autoref{sec_revGS}, the Green-Schwarz terms can likewise be formulated as the intersection theoretic paring between $G_4$ and a set of 4-cycle classes on $\hat Y_4$. Cancellation of all anomalies implies a relation between these intersection numbers which must be valid for all types of consistent gauge fluxes $G_4$ \cite{Cvetic:2012xn}. 

Starting from this formulation we will derive, in \autoref{sec:genrelChow}, a set of relations between the cohomology classes of the matter surfaces and the divisors spanning the coroot/coweight lattice of an elliptically fibred Calabi-Yau 4-fold. These relations go beyond the intersection theoretic identities of \cite{Cvetic:2012xn} in that they hold independently of any projection onto a gauge flux $G_4$. To arrive at such relations it is important to note that the structure of the matter surfaces must not only reflect consistency of the anomalies of all geometrically realised gauge symmetries, but also include consistency of anomalies if the gauge algebra on a 7-brane is broken by a gauge background to some subalgebra. This point will be discussed in \autoref{sec_Extension}. Taking into account the contribution of charged matter states in the bulk of the 7-branes allows us to isolate a set of identities which is summarized, for a general fibration, in equation \eqref{summary4dCoho1} and \eqref{summary4dCoho2}. These identities extend and generalize previous observations in \cite{Lin:2016vus}, which attributed the vanishing of anomalies in a class of fibrations with the MSSM gauge group to homological relations among 4-cycles. 

In fact, in \autoref{sec:ChowRelationsExamples} we will verify, for prototypical example fibrations containing Abelian and non-Abelian gauge algebras, that these relations hold not only as relations within the cohomology ring, but as relations among algebraic complex 2-cycles with rational coefficients modulo rational equivalence. We conjecture this to be the case more generally as made precise in equation (\ref{summary4dChow1}) and (\ref{summary4dChow2}). The Chow ring of algebraic cycles modulo rational equivalence is a refinement of the cohomology ring and general properties of this ring in complicated geometries are hard to come by. The specific set of relations in the Chow ring described in this article implies useful relations between the possible gauge backgrounds on an elliptic fibration $\hat Y_4$. Indeed, the gauge backgrounds can be represented by complex 2-cycles modulo rational equivalence \cite{Bies:2014sra}. Relation (\ref{summary4dChow2}) already entered the computation of the exact massless charged matter spectrum on the background analyzed in \autoref{subsec:SpecialFTheoryGUTModel} in \cite{Bies:2017fam}. In \autoref{sec_SU4example} we present another verification and application of  (\ref{summary4dChow2}): It has been known that in sufficiently simple fibrations no vertical gauge fluxes can exist, an example being the family of  generic Tate models with gauge group $SU(n)$ for $n < 5$ \cite{Krause:2012yh}. As we will exemplify, this is a direct consequence of the general constraint (\ref{summary4dChow2}) which consistent flux backgrounds have to obey.

The proof of the cohomological relations in the examples of \autoref{sec:ChowRelationsExamples} makes use of the structure of the 4-fold fibration, but not of the base. This raises the question of their validity also for elliptic fibrations of different complex dimension. Somewhat surprisingly, in \autoref{sec_Implications6d}  we will see that the same relations (\ref{summary4dCoho1}) and (\ref{summary4dCoho2}), applied now to elliptically fibred Calabi-Yau 3-folds, imply the cancellation of all gauge and mixed gauge-gravitational anomalies in 6-dimensional F-theory vacua, and in turn follows from these. More precisely, intersecting the cohomological identities (\ref{summary4dCoho1}) with generators of the coroot/coweight lattice reproduces the intersection theoretic relations of \cite{Park:2011ji} governing the structure of 6-dimensional anomaly cancellation. Intuitively, this connection may be understood as follows: (\ref{summary4dCoho1}) directly incorporates the cubic nature of the gauge-gravitational anomalies in four dimensions. Intersection with another divisor adds an external leg to the anomaly it represents as required to make the connection to the quartic anomalies in six dimensions. Nonetheless, given the rather different structure of Green-Schwarz counterterms in dimensions four and six, we find it intriguing that the same set of cohomological identities governs the cancellation of the 1-loop induced anomalies. Again we conjecture that these relations hold more generally within the Chow ring, as suggested by the analysis in \autoref{sec:ChowRelationsExamples}. In this sense, (\ref{summary4dCoho1}) can be viewed as a fundamental set of identities which universally governs the structure of gauge and gauge-gravitational anomalies in 4 and 6-dimensional F-theory compactifications. In \autoref{sec_Conclusions} we conclude and list some open questions. 

\section{Anomalies in 4-Dimensional F-Theory Vacua}

This preparatory section reviews  the intersection theoretic identities implied by anomaly cancellation in 4-dimensional F-theory vacua. We define our notation and conventions in \autoref{sec_RevF}, which collects the most relevant properties of F-theory compactifications on elliptic 4-folds used in the sequel. In \autoref{sec_revGS} we review the geometric formulation of the Green-Schwarz mechanism in four-dimensional F-theory provided in \cite{Cvetic:2012xn}.

\subsection{F-Theory on Calabi-Yau 4-Folds} \label{sec_RevF}

Our primary interest in this paper is in F-theory compactifications to four dimensions with gauge group 
\bea \label{Gtot}
G_{\mathrm{tot}} = \prod_{I=1}^{n_G} G_I \times \prod_{A=1}^{n_{U \left( 1 \right) }}{U \left( 1 \right)_A} \,.
\eea
Here $G_I$ denotes a non-Abelian Lie group and $U(1)_A$ an extra Abelian gauge group factor. Such a vacuum is described in terms of an elliptically fibred Calabi-Yau 4-fold $Y_4$ with projection
\bea
\pi \colon Y_4 \twoheadrightarrow B_3 \,.
\eea
The codimension-1 singularities of the fibration lie over divisors $W_I$ in the base $B_3$. They are associated with 7-branes carrying gauge group $G_I$. Resolving the singularities of $Y_4$ leads to a smooth fibration $\hat Y_4$, which  we assume to be Calabi-Yau. 

Each of the  non-abelian gauge group factors $G_I$ is related to the resolution of the codimension-1 singularities over the respective divisor $W_I$. This process introduces the resolution divisors $E_{i_I}$, $i = 1, \ldots, \mathrm{rk}(G_I)$, which take the role of the generators of the Cartan subgroup of $G_I$. In short, the gauge potential $\mathbf{A}_{i_I}$ associated with $E_{i_I}$ arises by expanding the M-theory 3-form as
\bea \label{C3expansion}
C_3 = \mathbf{A}_{i_I} \wedge E_{i_I} + \ldots \,.
\eea
More precisely, each resolution divisor $E_{i_I}$ is rationally fibred over $W_I$. Its fibre is given by one of the rational curves $\mathbb P^1_{i_I}$ inserted at the singular point in the fibre of $Y_4$ in the process of the resolution.\footnote{This simple picture is true for the simply-laced, \ie ADE, gauge groups. For non-simply laced gauge groups there exist exceptional divisors $E_i$, such that the fibre of the $E_i$'s split into several $\mathbb P^1$s, which are exchanged by monodromies along $W_I$ \cite{Bershadsky:1996nh}. This is nicely explained in \cite{Park:2011ji} and references therein. In this case we denote by $\mathbb P^1_{i_I}$ the basis of linearly independent rational curves in the fibre associated with the negative of the simple roots.} 
The group theoretic interpretation of the $E_{i_I}$ and the resolution curves $\mathbb P^1_{j_J}$ follows from their intersection numbers 
\bea
\left[E_{i_I}\right] \cdot \left[\mathbb P^1_{j_J}\right] = - \delta_{IJ} \, C_{i_ I j_I} \,,
\eea
where $C_{i_ I j_I}$ denotes the Cartan matrix of $G_I$. Here $[E_{i_I}]$ denotes the (dual) cohomology class associated with the divisor $E_{i_I} \in \mathrm{Cl}(\hat Y_4 )$.
Up to a minus sign, this matches the action 
\bea
\mathcal{T}_{i_I} |\alpha_{j_J} \rangle = C_{i_I j_J} |\alpha_{j_J} \rangle
\eea
 of a $U(1)_{i_I}$ generator $\mathcal{T}_{i_I}$ in the coroot basis on a simple root 
$\alpha_{j_J}$. We therefore identify  $\mathbf{A}_{i_I}$ with the $U(1)_{i_I}$ Cartan in this coroot basis, and an M2-brane
 wrapping one of the linearly independent fibral curves $\mathbb P^1_{j_J}$ with a state of weight vector $-\alpha_{j_J}$.

Of some relevance for us will be the fact that the projection $\hat{\pi}$ of the fibration $\hat Y_4$ induces the pushforward $\hat{\pi}_*: H_k(\hat Y_4) \rightarrow H_{k}(B_3)$. It is defined such that for instance
\bea
[ E_{i,I} ] \cdot_{\hat Y_4} [ E_{j,J} ] \cdot_{\hat Y_4} [ D_\alpha^\mathrm{b}] \cdot_{\hat Y_4} [ D_\beta^\mathrm{b} ] = \hat{\pi}_\ast \left( \left[ E_{i,I} \right] \cdot_{\hat Y_4} \left[ E_{j,J} \right] \right) \cdot_{B_3} [ D_\alpha^\mathrm{b} ] \cdot_{B_3} [ D_\beta^\mathrm{b} ]
\eea
for all $[ D^{\mathrm{b}}_\alpha ]$ representing basis elements of $H^{1,1}(B_3)$. Here $\cdot_{\hat Y_4}$ denotes the intersection product on $\hat Y_4$ and $\cdot_{B_3}$ the intersection product on $B_3$, but we will mostly drop the subscripts when the context allows it.\footnote{For future reference, we use the same symbol $\cdot$ to denote the intersection product within the cohomology and in the Chow ring. The context will make clear which is being used.} 
With this notation, we have the important relation
\[ \hat{\pi}_* \left( \left[ E_{i,I} \right] \cdot \left[ E_{j,J} \right] \right) = - \delta_{IJ} \, \mathfrak{C}_{i_I j_I} \, \left[ W_I \right] \,, \label{intersectionEiEj} \]
where
\bea \label{DefcalC}
\mathfrak{C}_{ij} = \frac{\langle \alpha, \alpha\rangle_\mathrm{max}}{\langle\alpha_j,\alpha_j\rangle} \frac{2 \langle \alpha_i, \alpha_j\rangle}{\langle \alpha_j, \alpha_j\rangle} = \frac{2}{\lambda} \frac{1}{\langle\alpha_j,\alpha_j\rangle} C_{ij} \,.
\eea
The object $\langle \alpha, \alpha\rangle_\mathrm{max}$ denotes the length of the longest root and is related to the so-called Dynkin index of the fundamental representation, $\lambda$, via 
\bea \label{lambdadef}
\lambda = \frac{2}{\langle \alpha, \alpha\rangle_\mathrm{max}} \,.
\eea
For the simply laced Lie algebras, \ie the Lie algebras of ADE type, 
\bea
\mathfrak{C}_{ij} \left( \mathrm{ADE} \right) = C_{ij} \left( \mathrm{ADE} \right)
\eea
but both quantities differ for the non-simply laced Lie algebras. 
As a last technicality that will appear again later we note that the same object $\mathfrak{C}_{ij}$ governs the normalization of the coroot basis via
\bea \label{trace-coroot}
\mathrm{tr}_{\mathbf{R}}  \mathcal{T}_i \mathcal{T}_j = \lambda \, \mathfrak{C}_{ij} \, c^{(2)}_{\mathbf{R}} \,.
\eea
In (\ref{trace-coroot}) we introduced the group theoretic constants $c_{\mathbf{R}}^{(n)}$ which interpolate between the trace in representation $\mathbf{R}$ and the trace in the fundamental representation via
\bea \label{cRn-def}
\rm{tr}_{\bf{R}} F^n = c^{ \left( n \right)}_{\mathbf{R}} {\rm{tr}}_{\rm{fund}} F^n \,,\qquad n=2,3 \, .
\eea
For more details we refer to the group theoretic discussion in \cite{Park:2011ji}.

Let us now turn to the Abelian gauge group factors $U(1)_A$, which are related to the existence of extra rational sections $S_A$ in addition to the zero-section $S_0$ of the fibration.
To each rational section $S_A$ we assign an element $U_A \in {\mathrm{CH}}^1(\hat Y_4)$ via the Shioda map, 
\bea \label{UAdefintion}
U_A = S_A - S_0 - D^{\mathrm{b}} + \sum_{i_I} k_{i_I} E_{i_I} \in {\mathrm{CH}}^1 ( \hat Y_4 ) \,.
\eea
Here the divisor class $D^{\mathrm{b}}$ on $B_3$ and the coefficients $k_{i_I}$ are to be chosen such that $U_A$ satisfies the following transversality conditions for all $\alpha$, $\beta$, $\gamma$,
\begin{align}
& [ U_A ] \cdot [D_\alpha^{\mathrm{b}}] \cdot [D_\beta^{\mathrm{b}}] \cdot [D_\gamma^{\mathrm{b}}] = 0 \,, \qquad 
  [ U_A ] \cdot [D_\alpha^{\mathrm{b}}] \cdot [D_\beta^{\mathrm{b}}] \cdot  \left[S_0\right] = 0 \,, \cr
& [ U_A ] \cdot [D_\alpha^{\mathrm{b}}] \cdot [D_\beta^{\mathrm{b}}] \cdot \left[E_{i_I}\right] = 0 \, .
\end{align}
In these equations $[D^{\mathrm{b}}_\alpha]$ represents basis elements of $H^{1,1}(B_3)$. Such $U_A$ serves as the generator of the Abelian gauge group $U(1)_A$. The $U(1)_A$ gauge potential $\mathbf{A}_A$ arises by replacing $E_{i_I}$ in the expansion (\ref{C3expansion}) with $U_A$.

For future purposes it is useful to introduce the combined notation
\bea \label{FSigma}
F_\Sigma \in \{ E_{i_I}, U_A \}
\eea
for these two types of divisors. By the Shioda-Tate-Wazir theorem \cite{Shioda, Tate1, Tate2, 2001math.....12259W}, $H^{1,1}(\hat Y_4)$ is given by
\bea \label{STW}
H^{1,1} ( \hat Y_4 ) = \text{span}_{\mathbb{C}} \left\{ [ S_0 ], \, [ F_\Sigma ], \, [D^{\mathrm{b}}_{\alpha} ] \right\} \,.
\eea

There are two types of massless matter in such F-theory compactifications: The so-called bulk matter arises from states which are free to propagate over the divisors $W_I$ and which transform, in absence of gauge flux, in the adjoint representation of $G_I$. We will come back to this type of matter in section \ref{sec_Extension}. For now our prime interest is in the matter localised in codimension two in the base and which transforms in some representation $\mathbf{R}$ of the full gauge group $G$.
We will denote the matter curve on $B_3$ on which these massless states localise by $C_\bfR$. The origin of such matter states are M2-branes wrapping a suitable combination of rational curves in the fibre of $\hat Y_4$ over $C_\bfR$. Let us denote by $\beta^a_{i_I}(\mathbf{R})$, $a=1, \ldots, \mathrm{dim}(\mathbf{R})$, the weight vector associated with representation $\mathbf{R}$. Its entries are the charges of the $a$-th state in the representation under the Cartan $U(1)$ with generator $E_{i_I}$. To each state in representation $\mathbf{R}$ we have an associated matter surface $S^a_\bfR$, given by a linear combination of fibral curves over $C_\bfR$. M2-branes wrapping this linear combination of fibral curves $\mathcal{S}^a_\bfR$ give rise to the $a$-th state. The weight vector is identified with the intersection number\footnote{In order to ensure for (\ref{weightdefinition}) we allow $\mathcal{S}^a_\bfR$ to be either a linear combinations of holomorphic fibral curves with non-negative coefficients or a linear combination of anti-holomorphic fibral curves with non-negative coefficients. Alternatively, we could stick to linear combinations of holomorphic fibral curves with non-negative coefficients only, but then we would have to include suitable minus signs in (\ref{weightdefinition}), depending on the phase of the resolution \cite{Intriligator:1997pq,Grimm:2011fx,Hayashi:2014kca,Esole:2014bka}.}
\bea \label{weightdefinition}
\beta^a_{i_I} \left( \mathbf{R} \right) \, \left[ C_{\mathbf{R}} \right] = \hat{\pi}_{\ast} \left( \left[ E_{i_I} \right] \cdot \left[ S^a_{\mathbf{R}} \right] \right) \,.
\eea 
Note that in some cases it may be necessary to allow for fractional coefficients in the definition of $S^a(\bfR)$ in order to achieve (\ref{weightdefinition}), \ie in general a matter surface defines an element in $\mathrm{CH^2}(\hat Y_4) \otimes \mathbb Q$. An example of this phenomenon is presented in \autoref{fibre_structure_SU4}.
Each of these states carries furthermore charge 
\bea \label{qAdefinition}
q_A \left( \mathbf{R} \right) \, \left[ C_{\mathbf{R}} \right] = \hat{\pi}_{\ast} \left( \left[ U_A \right] \cdot \left[ S^a_{\mathbf{R}} \right] \right)
\eea
under the non-Cartan $U(1)_A$. In the spirit of (\ref{FSigma}), we will sometimes find it useful to collectively denote the Cartan and non-Cartan charges by $\beta^a_\Sigma(\mathbf{R})$.

A gauge flux $G_4$ is described, at the level of cohomology, by an element of $H^{2,2}_{\mathbb{Z}/2}(\hat Y_4)$ which satisfies the so-called transversality conditions
\bea \label{verticality1}
G_4 \cdot [ S_0 ] \cdot [D^{\mathrm{b}}_\alpha] = 0, \qquad G_4 \cdot [D^{\mathrm{b}}_\alpha] \cdot [D^{\mathrm{b}}_\beta] = 0 \, ,
\eea
for any two $D_\alpha^{\mathrm{b}}, D_\beta^{\mathrm{b}} \in \text{Div} \left( B_3 \right)$.\footnote{Note that the superindex `b' will be used throughout this article to indicate divisors in the base space $B_3$ or more generally $B_n$.} In order for $G_4$ not to break the non-Abelian gauge group factors $G_I$, we require in addition
\bea \label{gaugeinvariantflux}
G_4 \cdot [ E_{i_I} ] \cdot [D^{\mathrm{b}}_\alpha] = 0
\eea
for all exceptional divisors $E_{i_I}$ and all base divisors $D_\alpha^{\mathrm{b}} \in \text{Div} \left( B_3 \right)$. For later purposes we note that the full cohomology group $H^{2,2}(\hat Y_4)$ in which the fluxes are valued enjoys a decomposition (over $\mathbb Q$ or $\mathbb{R}$ or $\mathbb{C}$) \cite{Greene:1993vm,Braun:2014xka}
\bea \label{H22decompo}
H^{2,2} ( \hat Y_4 ) = H^{2,2}_{\mathrm{vert}} ( \hat Y_4 ) \oplus H^{2,2}_{\mathrm{hor}} ( \hat Y_4 ) \oplus H^{2,2}_{\mathrm{rem}}( \hat Y_4 ) \,,
\eea
where elements of the three subspaces are mutually orthogonal with respect to the homological intersection pairing on $\hat Y_4$.

An important property of gauge flux is that it induces a chiral index of the form
\bea \label{chidef}
\chi \left( \mathbf{R} \right) = G_4 \cdot \left[S^a_\bfR\right] \,.
\eea
For gauge invariant flux, this result is independent of the choice of $a$ as a consequence of (\ref{gaugeinvariantflux}), and with this understanding we have suppressed this index in writing $\chi(\mathbf{R})$. 

A more refined characterization of the gauge background is possible by specifying an equivalence class  of complex 2-cycles modulo rational equivalence, \ie an element of the Chow group $A \in \mathrm{CH}^2(\hat Y_4)$. Its associated cohomology class describes the gauge flux $[A] = G_4$, but the Chow class $A$ contains much more information about the gauge background including the information about the $C_3$ Wilson line degrees of freedom not encapsulated in the field strength $G_4$. More details can be found in \cite{Bies:2014sra,Bies:2017fam}. Note furthermore that the Chow group $\mathrm{CH}^1(\hat Y_4)$, \ie codimension-1 cycles modulo rational equivalence, coincides with the divisor class group $\mathrm{Cl}(\hat Y_4)$.

\subsection{The Green-Schwarz Mechanism with Gauge Invariant Fluxes} \label{sec_revGS}

Anomaly cancellation implies a number of relations among the chiral indices of the matter states in the presence of a gauge flux which satisfies both (\ref{verticality1}) and (\ref{gaugeinvariantflux}). To gain some intuition let us assume for a moment that the gauge group is of the form $G \times U(1)_A$, before treating the case (\ref{Gtot}) in full generality.

First, the absence of cubic non-Abelian anomalies requires that 
\bea \label{cubic-anom1}
\sum_{\bf{R}} c^{(3)}_{\bf{R}} \, \chi \left( {\bf{R}} \right) = 0  \,.
\eea
Here we are using the group theoretic constants defined in (\ref{cRn-def}). Consider now a matter surface $S^a_\bfR$, $a = 1, \ldots, {\rm{dim}}({\bf{R}})$, associated to one of the weights $\beta^a(\mathbf{R})$ of representation $\mathbf{R}$. In view of (\ref{chidef}), relation (\ref{cubic-anom1}) is equivalent to the statement that
\bea
\sum_{\bf{R}} c^{(3)}_{\bf{R}} \, \left[S^a_\bfR \right] \cdot G_4 = 0    
\eea
for all vertical fluxes $G_4$ which leave the non-Abelian gauge group factor $G$ unbroken, \ie for all $G_4 \in H^{2,2}(\hat Y_4)$ satisfying both (\ref{verticality1}) and (\ref{gaugeinvariantflux}).

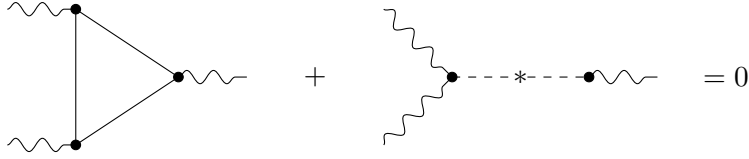
\begin{figure}[tb]
\begin{center}
\begin{tikzpicture}[scale = 0.9]
  
  \def\shift{1};
  
  
  \draw[decorate, decoration=snake] (0,1) -- (1,1);
  \draw[decorate, decoration=snake] (0,-1) -- (1,-1);
  \draw[decorate, decoration=snake] (2.5,0) -- (3.5,0);
  
  \draw (1,1) -- (2.5,0) -- (1,-1) -- (1,1);
  
  \draw [fill] (1,1) circle (2pt);
  \draw [fill] (2.5,0) circle (2pt);
  \draw [fill] (1,-1) circle (2pt); 

  \draw (3.5 + \shift, 0) node {$+$};


  \draw[decorate, decoration=snake] (3.5 + 2 * \shift,1) -- (4.5 + 2 * \shift,0);
  \draw[decorate, decoration=snake] (3.5 + 2 * \shift,-1) -- (4.5 + 2 * \shift,0);
  \draw[dashed] (4.5 + 2 * \shift, 0) -- (5.4 + 2 * \shift, 0 );
  \draw[dashed] (5.6 + 2 * \shift, 0) -- (6.5 + 2 * \shift, 0 );
  \draw (5.5 + 2 * \shift, 0 ) node {$\ast$};
  \draw[decorate, decoration=snake] (6.5 + 2 * \shift,0) -- (7.5 + 2 * \shift,0);
  \draw [fill] (4.5 + 2 * \shift, 0) circle (2pt);
  \draw [fill] (6.5 + 2 * \shift, 0,0) circle (2pt);
  
  \draw (7.5 + 3 * \shift,0) node {$= 0$};
  
\end{tikzpicture}
\end{center}
\caption{4d cubic anomalies and their cancellation via Green-Schwarz counterterms.}
\label{figure:4DAnomaly}
\end{figure}

Extra relations of similar type ensure the consistent cancellation of the mixed Abelian-non-Abelian, the pure Abelian, and the mixed Abelian-gravitational anomalies. 
The one-loop generated anomalies induced by the localised matter representations $\mathbf{R}$ of $U(1)_A$ charge $q_A(\mathbf{R})$ need not vanish by themselves, but must be cancelled by the classical generalised Green-Schwarz counterterms \cite{Sagnotti:1992qw} (see \autoref{figure:4DAnomaly}). First, the mixed $U(1)_A - G^2$ anomalies cancel if
\bea \label{anom-G11-a}
\sum_{\rm{\bf{R}}} q_A \left( {\mathrm{\bf{R}}} \right) \, c^{(2)}_{\mathrm{\bf{R}}} \, \chi \left( {\rm{\bf{R}}} \right) \equiv \sum_{\rm{\bf{R}}} q_A \left( {\rm{\bf{R}}} \right) \, c^{(2)}_{\rm{\bf{R}}} \,  G_4 \cdot \left[S^a_\bfR \right] = G_4 \cdot [ U_A ] \cdot [ D^\mathbf{b}_{AGG} ] \,. 
\eea
The Green-Schwarz counterterm on the RHS is expressed in terms of a certain base divisor class $D^\mathbf{b}_{AGG} \in H^{1,1}(B_3)$ \cite{Cvetic:2012xn} and the generator $U_A \in H^{1,1}(\hat Y_4)$ associated with gauge group $U(1)_A$ as defined in (\ref{UAdefintion}). We will discuss $D^\mathbf{b}_{AGG}$ momentarily. In particular, for fluxes orthogonal to the subspace spanned by $\text{span}_{\mathbb{C}} \{ U_A \wedge D^{\mathbf{b}}_\alpha \; | \; D^{\mathbf{b}}_\alpha \in H^{1,1} ( B_3 ) \}$ these Green-Schwarz counterterms vanish. This is in agreement with the fact that such fluxes do not render $U(1)_A$ St\"uckelberg massive and hence the field theoretic anomalies in (\ref{anom-G11-a}) vanish identically.

Similarly, cancellation of the $U(1)_A^3$ anomalies and the mixed gravitational $U(1)_A - R^2$ anomalies imply relations of the form
\begin{align}
\sum_{\rm{\bf{R}}} q^3_A \left( \mathbf{R} \right) \rm{dim} \left( \bf{R} \right) \chi \left( \bf{R} \right) \equiv \sum_{\bf{R}}{q^3_A \left( \mathbf{R} \right) \rm{dim} \left( \bf{R} \right) \, G_4 \cdot \left[ S^a_\bfR  \right]} &= G_4 \cdot [ U_A ] \cdot [ D^\mathbf{b}_{AAA} ] \label{anom-111-a} \\
\sum_{\rm{\bf{R}}} q_A \left( {\mathbf{R}} \right) \rm{dim} \left( \rm{\bf{R}} \right) \chi \left( \rm{\bf{R}} \right) \equiv \sum_{\rm{\bf{R}}} q_A \left( \mathbf{R} \right) \rm{dim} \left( \rm{\bf{R}} \right) \, G_4 \cdot \left[ S^a_\bfR \right] & = G_4 \cdot [ U_A ] \cdot [ D^\mathbf{b}_{ARR} ]
\label{anom-1RR-a}
\end{align}
for suitable divisor classes  $D^\mathbf{b}_{AAA}, D^\mathbf{b}_{ARR} \in \mathrm{CH}^1(B_3)$.

The form of the Green-Schwarz counterterms in F/M-theory appearing on the RHS has been derived in \cite{Cvetic:2012xn}, for a very general F-theory vacuum with a gauge group of the form (\ref{Gtot}). As shown in \cite{Cvetic:2012xn} the relations (\ref{cubic-anom1}), (\ref{anom-G11-a}), (\ref{anom-111-a}), generalised to multiple gauge group factors, can be elegantly summarized in the single compact expression\footnote{Note that in our normalisation the symmetrisation of $\Gamma$, $\Lambda$, $\Sigma$ on the RHS comes with a factor of $1/3!$.}
\bea \label{anomalies-all}
\frac{1}{3} \sum_{\mathbf{R}} \sum_{a} n^a_{\Lambda \Sigma \Gamma} \left( \bf{R} \right) \, \left( G_4 \cdot \left[ S^a_{\bf{R}} \right] \right) &=& G_4 \cdot \left[ F_{ (\Gamma} \right] \cdot \left[ \hat{\pi}_* \left( F_{\Lambda} \cdot F_{\Sigma)} \right) \right]\,.
\eea
The numerical coefficients
\bea \label{naRdef}
n^a_{\Lambda \Sigma \Gamma} \left( \bfR \right) = \beta^a_{\Lambda} \left( \mathbf{R} \right) \, \beta^a_{\Sigma} \left( \mathbf{R} \right) \, \beta^a_{\Gamma}\left( \mathbf{R} \right) \,
\eea
are computed geometrically via (\ref{weightdefinition}) or (\ref{qAdefinition}). 

Expression (\ref{anomalies-all}) is valid for all fluxes $G_4$ which leave the non-Abelian gauge group unbroken. The RHS of (\ref{anomalies-all}) represents the Green-Schwarz counterterms derived in \cite{Cvetic:2012xn}, with the indices $\Gamma$, $\Lambda$ and $\Sigma$ totally symmetrised. Recall that the operation $\hat{\pi}_*$ denotes projection of a complex 2-cycle class in $\hat{Y}_4$ onto the base $B_3$, where it yields a divisor class. For instance if all $F_\Lambda$ are chosen to represent some of the non-Abelian resolution divisors, then the RHS of (\ref{anomalies-all}) vanishes identically since, by assumption, $G_4$ does not break the non-Abelian gauge group. Thereby we recover (\ref{cubic-anom1}). 

Similarly, the mixed Abelian-gravitational anomaly cancellation condition takes the form \cite{Cvetic:2012xn}
\[ \frac{1}{3} \sum_{\mathbf{R}} \sum_{a} \beta^a_\Lambda \left( \mathbf{R} \right) \, \left( G_4 \cdot \left[ S^a_{\mathbf{R}} \right] \right) \, = - 2 \,G_4 \cdot [ \overline{K}_{B_3} ] \cdot \left[F_{\Lambda}\right] \label{grav-anom-gen} \]
with $\overline{K}_{B_3} = c_1 (B_3)$ the anti-canonical class of the base $B_3$.

It was stressed already in \cite{Cvetic:2012xn} that (\ref{anomalies-all}) and (\ref{grav-anom-gen}) can be viewed as a set of relations satisfied by the matter surface classes $[S^a_\bfR]$ on any smooth elliptically fibred Calabi-Yau 4-fold $\hat Y_4$. More precisely it is a relation satisfied by the intersection products of matter surfaces with any element $G_4 \in H^{2,2}(\hat Y_4)$ subject to the conditions (\ref{verticality1}) and (\ref{gaugeinvariantflux}). 

\section{General Relations in the Chow Ring and Anomalies} \label{sec:genrelChow-A}

In this section we promote the identities (\ref{anomalies-all}) and (\ref{grav-anom-gen}) to a considerably stronger set of relations between the involved matter surfaces $S^a_\bfR$. 
To this end we first extend, in \autoref{sec_Extension}, the Green-Schwarz terms in such a way as to allow also for fluxes breaking the geometrically realized gauge group. This will then lead, in \autoref{sec:genrelChow}, to a set of identities within $H^{2,2}(\hat Y_4)$ and even $\mathrm{CH}^2(\hat Y_4)$.

\subsection{Non-Gauge Invariant Fluxes and Their Anomaly Relations} \label{sec_Extension}

It would be tempting to declare  (\ref{anomalies-all}) and (\ref{grav-anom-gen}) to be valid as a relation among complex 2-cycles even without the need to project onto the flux $G_4$. This, however, cannot be correct because (\ref{anomalies-all}) and (\ref{grav-anom-gen}) do not hold if $G_4$ breaks the non-Abelian gauge group factors $G_I$ by violating (\ref{gaugeinvariantflux}). The physical reason is that in such a situation it is not the anomalies of the original non-Abelian gauge group $G_I$ which need to be considered, but rather the anomalies of its subgroups to which it gets broken by the flux. Decomposing the massless spectrum into irreducible representations of these subgroups we find extra contributions to the anomalies which must be taken into account. These extra contributions are due to  matter states localised in the bulk of the 7-brane divisors $W_I$ associated with the non-Abelian gauge group $G_I$. While for gauge invariant flux, the bulk matter transforms in the adjoint representation of $G$ and hence does not contribute to the anomalies, more generally one finds extra chiral massless states which must not be neglected. 

To determine these extra contributions suppose that the flux $G_4$ breaks $G_I$ as
\bea \label{G-breaking}
G_I &\rightarrow& H_I \times U \left( 1 \right)_{H_I} \\
\textbf{adj} \left( G_I \right)& \rightarrow& \bigoplus_{m_I} {\mathbf r}_{m_I} \,,
\eea
where the irreducible representations ${\mathbf r}_{m_I}$ of the non-Abelian subgroup $H_I$ carry $U(1)_{H_I}$ charges $q_I({\mathbf r}_{m_I})$.
To each such ${\mathbf r}_{m_I}$ we associate the weight vector $\beta^a({\mathbf r}_{m_I})$.\footnote{For notational simplicity we do not introduce a new parameter to label the different weights of each subgroup $H_I$. It should always be clear from the context which values the index $a$ takes.} The state with weight $\beta^a({\mathbf r}_{m_I})$ arises from an M2-brane wrapping a linear combination of rational curves in the fibre over the 7-brane divisor $W_I$. Since the representation ${\mathbf r}_{m_I}$ descends from the adjoint representation of $G_I$, the rational curves in question are just suitable combinations of rational curves $\mathbb P^1_{i_I}$ associated with minus one times the simple roots $\alpha_{i_I}$. 
Let us denote the linear combination of $\mathbb P^1_{i_I}$ with charge $\beta^a({\mathbf r}_{m_I})$ as
\bea \label{rmfibre}
C \left( \beta^a \left( {\mathbf{r}}_{m_I} \right) \right) = \sum_{i_I} a_{i_I} \left( {\mathbf{r}}_{m_I}, a \right) \mathbb{P}^1_{i_I} \, . 
\eea

An important result for our analysis is that the chiral index of these states induced by the flux $G_4$ is
\bea \label{chi-bulkmatter}
\chi \left( {\mathbf r}_{m} \right) = G_4 \cdot \left( - c_1 \left( W_I \right) \right) \cdot \sum_{i_I} \hat a_{i_I}({\mathbf r}_{m_I},a) \, \left[ E_{i_I} \right]
\eea
for any choice of weight $a$ associated with representation ${\mathbf r}_{m_I}$. Here $c_1(W_I)$ denotes the first Chern class of the divisor $W_I$.
For simply laced Lie algebras, $\hat a_{i_I}({\mathbf r}_{m_I},a) = a_{i_I} \left( {\mathbf{r}}_{m_I}, a \right)$ because each $\mathbb P^1_{i_I}$ is the fibre of the resolution divisor $E_{i_I}$, but for non-simply laced Lie algebras, the fibre of $E_{i_I}$ splits locally into various rational curves, all homologous to $\mathbb P^1_{i_I}$, which are exchanged by a monodromy along $W_I$. In this case $\hat a_{i_I}({\mathbf r}_{m_I},a)$ includes a fractional correction factor to account for this monodromy.

To see this one first identifies the cohomology groups on $W_I$ counting massless bulk matter in representation  ${\mathbf r}_{m_I}$. 
If we denote by $L_{m_I}$ the line bundle induced by the flux background to which the bulk states in representation ${\mathbf r}_{m_I}$ couple, the results of \cite{Beasley:2008dc} imply for the associated chiral index 
\bea \label{indexrmlocal}
\chi \left( \mathbf{r}_{m_I} \right) = - \Mint_{W_I}{c_1 \left( W_I \right)} \cdot c_1 \left( L_{m_I} \right) \,.
\eea

In order to connect this to the formula (\ref{chi-bulkmatter})  we need to extract the line bundle $L_{m_I}$ on $W_I$ from the flux data incorporated in the $G_4$ flux on $\hat Y_4$. This step has been spelled out in \cite{Bies:2017fam}, to which we refer for more details. The prescription is to intersect the algebraic complex 2-cycle class, \ie the element in $\mathrm{CH}^2(\hat Y_4)$, underlying the definition of $G_4$ with the linear combination $\sum_{i_I} \hat a_{i_I}({\mathbf r}_{m_I},a) \, E_{i_I}$ of resolution divisors. Projecting this onto $W_I$ defines an element in $\mathrm{CH}_1(W_I)$, \ie a line bundle on $W_I$, which we identify with $L_{m_I}$. Due to the projection onto $W_I$, the result is the same for each choice of weight vector $\beta^a({\mathbf r}_{m_I})$ in (\ref{rmfibre}). The chiral index (\ref{indexrmlocal}) can then be rewritten as in (\ref{chi-bulkmatter}).

For compactness of notation we define a complex 2-cycle associated with each representation ${\mathbf r}_{m_I}$ of the form
\bea \label{SarmI}
S^{a}_{\mathbf{r}_{m_I}} = \sum_{i_I} \hat a_{i_I} \left( \mathbf{r}_{m_I},a \right) \, \left. E_{i_I} \right|_{K_{W_I}} \, .
\eea
Here $\left. E_{i_I} \right|_{K_{W_I}}$ denotes the restriction of the resolution divisor $E_{i_I}$, which is a fibration over $W_I$, to the canonical divisor $K_{W_I}$ on $W_I$. 
Let us point out that in general $S^{a}_{\mathbf{r}_{m_I}} $ defines a class in $\mathrm{CH}^2(\hat Y_4) \otimes \mathbb Q$, as do the matter surfaces $S^a(\bfR)$ of the localised matter states.
Note furthermore that $[K_{W_I}] = - c_1(W_I)$. Then the expression (\ref{chi-bulkmatter}) for the chiral index of the massless bulk matter can be rewritten as
\bea
\chi \left( \mathbf{r}_{m}, \alpha_m \right) = G_4 \cdot [ S^{a}_{\mathbf{r}_{m_I}}  ] \,.
\eea

We are now in a position to generalise (\ref{anomalies-all}) such as to be valid also for fluxes $G_4$ not respecting the non-Abelian gauge algebra. If $G_4$ is responsible for a breaking of the form (\ref{G-breaking}), the correct modification of (\ref{anomalies-all}) is to add on the LHS the contribution to the anomalies from the chiral bulk matter states in representation ${\mathbf r}_{m_I}$. Group theoretically, the set of all weights $\beta^a({\mathbf r}_{m_I})$ equals the set of all weights of the adjoint representation of $G_I$, \ie the complete set of roots of $G_I$. Labelling the roots of $G_I$ by $\rho_I = 1, \ldots, \mathrm{dim}(G_I)$, we define the complex 2-cycle
\bea \label{def-SrhoI}
S^{\rho_I} = \sum_{i_I} \hat a_{i_I} \left( \rho_I \right) \, \left. E_{i_I} \right|_{K_{W_I}} \, ,
\eea
which is to be viewed as the analogue of (\ref{SarmI}), but directly for the adjoint representation of $G_I$. With this notation in place the generalisation of (\ref{anomalies-all}) is
\begin{align} \label{G4dotanomaly1}
& G_4 \cdot \left( \sum_{\mathbf{R} \neq \mathbf{adj}} \sum_{a} n^a_{\Lambda \Sigma \Gamma}(\bfR) \, \left[ S^a_{\mathbf{R}} \right] \, + \frac{1}{2} \sum_{\rho_I} n^{\rho_I}_{\Lambda \Sigma \Gamma} \, \left[ S^{\rho_I} \right] - 3 \, \cdot \left[ F_{(\Gamma} \right] \cdot \left[ \hat{\pi}_* \left( F_{\Lambda} \cdot F_{\Sigma)} \right) \right] \, \right) = 0  \,.
\end{align}
Here we are introducing, in analogy to (\ref{naRdef}), the abbreviation
\bea
n^{\rho_I}_{\Lambda \Sigma \Gamma} = \beta^{\rho_I}_{\Lambda} \, \beta^{\rho_I}_{\Sigma} \, \beta^{\rho_I}_{\Gamma}\,.
\eea
The extra factor of $1/2$ in the second line appears because the adjoint is a real representation and we are summing over all roots, positive and negative. 
Similarly, relation (\ref{grav-anom-gen}) generalises to
\bea \label{G4dotanomaly2}
G_4 \cdot \left( \sum_{\mathbf{R} \neq \mathbf{adj}} \sum_{a} \, \beta^a_\Lambda \left( \mathbf{R} \right) \, \left[ S^a_{\mathbf{R}} \right] + \frac{1}{2} \sum_{\rho_I} \, \beta^{\rho_I}_\Lambda \, \left[ S^{\rho_I} \right] + 6 \, [ \overline{K}_{B_3} ] \cdot [ F_{\Lambda} ] \right) = 0 \,.
\eea

\subsection{Anomaly Relations in the Chow Ring} \label{sec:genrelChow}
 
According to (\ref{G4dotanomaly1}) and (\ref{G4dotanomaly2}), the sum of the expressions in the brackets is orthogonal to each element of $H^{2,2}(\hat Y_4,\mathbb Q)$ which satisfies the transversality condition (\ref{verticality1}).\footnote{We are writing here $H^{2,2}(\hat Y_4,\mathbb Q)$ because in general $G_4$ is only half-integer quantised upon demanding that $G_4 + \frac{1}{2} c_2(\hat Y_4) \in H^{2,2}(\hat Y_4) \cap H^{4}(\hat Y_4, \mathbb Q)$. Unless stated otherwise, in the sequel we will always mean $H^{2,2}(\hat Y_4,\mathbb Q)$ when writing $H^{2,2}(\hat Y_4)$.} This is equivalent to a consistent cancellation of all anomalies in F-theory including those of the $G_I$ subgroups in the presence of non-gauge-invariant flux. We will now promote (\ref{G4dotanomaly1}) and (\ref{G4dotanomaly2}) to relations among cohomology classes within the cohomology ring $H^{2,2}(\hat Y_4)$, and even to relations between equivalence classes of complex 2-cycles within $\mathrm{CH}^2(\hat Y_4)$. It had already been observed in \cite{Lin:2016vus} that the cancellation of gauge anomalies can be interpreted as a consequence of suitable homological relations between the matter surfaces up to terms orthogonal to all gauge invariant fluxes. The results of this section are a systematic generalisation of these observations to a considerably stronger set of relations which hold even up to rational equivalence. 

To make the argument it suffices to focus on  (\ref{G4dotanomaly1}). As recalled above, the expression in brackets is orthogonal to all gauge fluxes $G_4$ satisfying (\ref{verticality1}). Combined with the manifest orthogonality properties of the various classes appearing in brackets, this implies orthogonality to the space
\bea \label{span2}
V_1 = \text{span}_{\mathbb{C}} \left\{ [ E_{i_I} \cdot D^{\mathrm{b}}_\alpha ], [ U_A \cdot D^{\mathrm{b}}_\alpha ],\, [ S_0 \cdot D^{\mathrm{b}}_\alpha ], \, [ D^{\mathrm{b}}_\alpha \cdot D^{\mathrm{b}}_\beta ] \right\} \subset H^{2,2}(\hat Y_4)\,.
\eea

$V_1$ is a subspace of the vertical cohomology group $H^{2,2}_{\mathrm{vert}}(\hat Y_4) = H^{1,1}(\hat Y_4) \wedge H^{1,1}(\hat Y_4)$ appearing in (\ref{H22decompo}). 
By the Shioda-Tate-Wazir theorem (\ref{STW}) we can express this space as\footnote{Note that $S_0 \wedge E_{i_I}$ vanishes on $\hat Y_4$ since the zero-section does not intersect the resolution divisors.}
\bea\label{defV4}
H^{2,2}_{\mathrm{vert}} ( \hat Y_4 ) = V_1 \cup V_2, \qquad V_2 = \text{span}_{\mathbb{C}} \left\{ \left[ E_{i_I} \cdot E_{j_J} \right], \, \left[ S_A \cdot E_{i_I} \right], \, \left[ S_A \cdot S_B \right], \, \left[ S_0 \cdot S_A \right] \right\} \,.
\eea
The important point to notice is now that to each element in $V_2$ we can associate a gauge invariant 4-form flux element $G_4 \in H^{2,2}_{\mathrm{vert}}(\hat Y_4)$ by adding suitable correction terms such that the flux satisfies (\ref{verticality1}) as well as (\ref{gaugeinvariantflux}). As we prove generally in appendix \ref{app_fluxconstruction}, the necessary correction terms lie in $V_1$. Together with the fact that the terms in brackets in (\ref{G4dotanomaly1}) are orthogonal to the set of all $G_4$, this implies that they are orthogonal to all elements in $H^{2,2}_{\mathrm{vert}}(\hat Y_4)$. 

To proceed recall the orthogonal decomposition (\ref{H22decompo}) of $H^{2,2}(\hat Y_4)$. We need to distinguish two qualitatively distinct situations. First, suppose that the cohomology groups associated with all matter surfaces lie in the primary vertical subspace, \ie suppose that
\bea
\left[ S^a_{\mathbf{R}} \right] \in H^{2,2}_{\mathrm{vert}} ( \hat Y_4 ) \qquad \forall \, \, S^a_\bfR \,.
\eea 
This is in fact the situation in most explicit examples of elliptically fibred Calabi-Yau 4-folds studied in the literature as of this writing, with the exception of the construction in \cite{Braun:2014pva}. Since the  three subspaces in the decomposition (\ref{H22decompo}) are mutually orthogonal, the expression in brackets in (\ref{G4dotanomaly1}) is orthogonal to $H^{2,2}_{\mathrm{hor}}(\hat Y_4)$ and $H^{2,2}_{\mathrm{rem}}(\hat Y_4)$. Furthermore, as just shown, it is orthogonal on $H^{2,2}_{\mathrm{vert}}(\hat Y_4)$. All of this together implies the following relations in cohomology
\bea
\sum_{\mathbf{R} \neq \mathbf{adj}} \sum_{a} n^a_{\Lambda \Sigma \Gamma} \left( \mathbf{R} \right) \, \left[ S^a_{\mathbf{R}} \right] \, + \frac{1}{2} \sum_{\rho_I} n^{\rho_I}_{\Lambda \Sigma \Gamma} \, \left[ S^{\rho_I} \right] - 3 \, \cdot \left[ F_{(\Gamma} \right] \cdot \left[ \hat{\pi}_* \left( F_{\Lambda} \cdot F_{\Sigma)} \right) \right] &=& 0 \label{Anomaly4dCohom1} \\
\sum_{\mathbf{R} \neq \mathbf{adj}} \sum_{a} \, \beta^a_\Lambda(\mathbf{R}) \, \left[ S^a_{\mathbf{R}} \right] + \frac{1}{2} \sum_{\rho_I} \, \beta^{\rho_I}_\Lambda \, \left[ S^{\rho_I} \right] + 6 \, \left[ \overline{K}_{B_3} \right] \cdot \left[ F_{\Lambda} \right] &=& 0 \label{Anomaly4dCohom2} \,,
\eea
where (\ref{Anomaly4dCohom2}) follows by applying similar reasoning to (\ref{G4dotanomaly2}).

The second situation corresponds to configurations where some of the matter surface classes have a part in the remainder $H^{2,2}_{\mathrm{rem}}(\hat Y_4)$. 
In this case, we can split the classes of the matter surfaces into orthogonal components
\[ \left[S^a(\mathbf{R})\right] = \left[S^a(\mathbf{R})\right]_{\mathrm{vert}} + \left[S^a(\mathbf{R})\right]_{\mathrm{rem}} \,. \]
We will convince ourselves in \autoref{app_NonVertMatter} that in this case 
\begin{subequations} \label{summary4dCoho1}
\begin{empheq}[box=\widefbox]{align}
\sum_{\mathbf{R} \neq \mathbf{adj}} \sum_{a} n^a_{\Lambda \Sigma \Gamma} \left( \bfR \right) \, \left[ S^a_{\bfR} \right]_\mathrm{vert} \, + \frac{1}{2} \sum_{\rho_I } n^{\rho_I}_{\Lambda \Sigma \Gamma} \, \left[ S^{\rho_I} \right] &= 3 \, \left[ F_{(\Gamma} \right] \cdot \left[ \hat{\pi}_* \left( F_{\Lambda} \cdot F_{\Sigma )} \right) \right] \label{summary4dCoho1-a} \\
\sum_{\mathbf{R} \neq \mathbf{adj}} \sum_{a} \, \beta^a_\Lambda \left( \mathbf{R} \right) \, \left[ S^a_{\mathbf{R}} \right]_\mathrm{vert} + \frac{1}{2} \sum_{\rho_I} \, \beta^{\rho_I}_\Lambda \, \left[ S^{\rho_I} \right] &= - 6 \, \left[ \overline{K}_{B_3} \right] \cdot \left[ F_{\Lambda} \right] \label{summary4dCoho1-b} \,.
\end{empheq}
\end{subequations}

The cohomological relations (\ref{summary4dCoho1}) fully incorporate the cancellation of gauge and mixed gravitational anomalies in 4-dimensional F-theory compactifications. Interestingly, we can derive, in addition to  (\ref{summary4dCoho1}), another type of cohomological relations from (\ref{G4dotanomaly1}) and (\ref{G4dotanomaly2}) which are of some relevance by themselves. To arrive at these, note first that to each matter surface 2-cycle $S^a_\bfR$ one can associate a flux 2-cycle class
\bea \label{Aageneral1}
A^a \left( \mathbf{R} \right) = S^a_\bfR + \Delta^a \left( \mathbf{R} \right) \in \mathrm{CH}^2 ( \hat Y_4 ) \,.
\eea
The correction factor $\Delta^a({\mathbf{R}})$ is chosen such that the cohomology class $[A^a({\mathbf{R}})] \in H^{2,2}(\hat Y_4)$ defines a bona-fide 4-form flux $G_4$ satisfying the two constraints (\ref{verticality1}) and (\ref{gaugeinvariantflux}). The transversality condition (\ref{verticality1}) holds automatically by construction of the $S^a({\mathbf{R}})$.\footnote{In particular, $S^a_\bfR$ does not contain the components of the fibre intersected by the zero-section as the associated wrapped M2-brane states correspond to KK non-zero modes in the dual M-theory vacuum.} To implement (\ref{gaugeinvariantflux}), it suffices to add the correction term \cite{Borchmann:2013hta,Bies:2017fam}
\bea \label{Aageneral2}
\Delta^a \left( \mathbf{R} \right) = + \left( \beta^a \left( \mathbf{R} \right)^T_{i_I} {\mathfrak C}^{-1}_{i_I j_I} \right) \, \left. E_{j_I} \right|_{C_\bfR} \in \mathrm{CH}^2 ( \hat Y_4 ) \,.
\eea
Here we recall that ${\mathfrak C}_{i_I j_I}$ governs the intersection numbers of the resolution divisors as in (\ref{intersectionEiEj}), with summation over repeated indices understood. Slightly modifying our arguments that lead to (\ref{summary4dCoho1}) we show in \autoref{App_homrelproof} that the following relations hold in cohomology on $\hat Y_4$,\footnote{Since $A^a(\mathbf{adj}) = 0$ by construction we can sum over all representations.}
\begin{subequations} \label{summary4dCoho2}
\begin{empheq}[box=\widefbox]{align}
\sum_{\mathbf{R}} \sum_{a} n^a_{i_I j_K k_K} \left( \mathbf{R} \right) \, \left[ A^a(\mathbf{R}) \right]_\mathrm{vert} \, & = 0 \in H^{2,2} ( \hat Y_4 ) \label{naijkA-1} \\
\sum_{\mathbf{R}} \sum_{a} n^a_{A \Sigma \Gamma} \left( \mathbf{R} \right) \, \left[ A^a \left( \mathbf{R} \right) \right]_{\mathrm{vert}} \, - 3\, \left[ U_{(A} \right] \cdot \left[ \hat{\pi}_* \left( F_{\Sigma} \cdot F_{\Gamma)} \right) \right] &= 0 \in H^{2,2} ( \hat Y_4 ) \label{naijkA-2} \\
\sum_{\mathbf{R}} \sum_{a} q_A \, \left[ A^a \left( \mathbf{R} \right) \right]_{\mathrm{vert}} \, + 6 \,\left[ U_A \right] \cdot \left[ \overline{K}_{B_3} \right] &= 0 \in H^{2,2} ( \hat Y_4 ) \,. \label{naijkA-3}
\end{empheq}
\end{subequations}
Note that the cohomological relations (\ref{summary4dCoho2}) are in general independent of (\ref{summary4dCoho1}). For instance pick $\Lambda, \Sigma, \Gamma = i_I, j_J, k_K$
and consider the difference of (\ref{summary4dCoho1-a}) and (\ref{naijkA-1}). This gives
\[ \label{equ:DifferenceAnomalies}
\sum_{\mathbf{R} \neq \mathbf{adj}} \sum_a - n^a_{i_I j_J k_K} \left( \mathbf{R} \right) \, \left[ \Delta^a \left( \mathbf{R} \right) \right] \, + \frac{1}{2} \sum_{\rho_I} n^{\rho_I}_{i_I j_J k_K} \left[ S^{\rho_I} \right] \, - 3 \, \left[ F_{(i_I} \right] \cdot \left[ \hat{\pi}_* \left( F_{j_J} \cdot F_{k_K)} \right) \right] = 0  \,. \]
This relation is trivial when intersected with $[U_A] \wedge [D_\alpha^{\rm b}]$, but its intersection with $[E_{i_I}] \wedge [D_\alpha^{\rm b}]$ is equivalent to the intersection of $[E_{i_I}] \wedge [D_\alpha^{\rm b}]$ with (\ref{summary4dCoho1-a}).
In particular it is  (\ref{summary4dCoho1-a}) which encodes anomaly cancellation for the remnant gauge group after breaking $G_I \rightarrow H_I \times U(1)_{i_I}$ via $G_4 = [E_{i_I}] \wedge [D_\alpha^{\rm b}]$.

To summarise anomaly cancellation in F-theory compactified on a smooth, flat elliptically fibred Calabi-Yau 4-fold $\hat Y_4$ implies the relations (\ref{summary4dCoho1}) and (\ref{summary4dCoho2}) within the cohomology ring (over the rationals) of $\hat Y_4$. In fact, we conjecture that under suitable conditions these relations are valid not only in cohomology, but at the level of the Chow group, at least with rational coefficients. This means they hold as relations among rational equivalence classes of algebraic cycles of complex codimension two, which form the elements of  $\mathrm{CH}^2(\hat Y_4) \otimes \mathbb Q$. More precisely, our claim is that the relations
\begin{subequations} \label{summary4dChow1}
\begin{empheq}[box=\widefbox]{align}
\sum_{\mathbf{R} \neq \mathbf{adj}} \sum_{a} n^a_{\Lambda \Sigma \Gamma} \left( \mathbf{R} \right) \, \left. S^a_{\mathbf{R}} \right|_\mathrm{vert} \, + \frac{1}{2} \sum_{\rho_I} n^{\rho_I}_{\Lambda \Sigma \Gamma} \, S^{\rho_I} &= 3 F_{(\Gamma} \cdot \hat{\pi}_* \left( F_{\Lambda} \cdot F_{\Sigma)} \right) \label{summary4dChow1-a} \\
\sum_{\mathbf{R} \neq \mathbf{adj}} \sum_{a} \, \beta^a_\Lambda \left( \mathbf{R} \right) \, \left. S^a_{\mathbf{R}} \right|_\mathrm{vert} + \frac{1}{2} \sum_{\rho_I} \, \beta^{\rho_I}_\Lambda \, S^{\rho_I} &= - 6 \, \overline{K}_{B_3} \cdot F_{\Lambda} \label{summary4dChow1-b} \,
\end{empheq}
\end{subequations}
and 
\begin{subequations} \label{summary4dChow2}
\begin{empheq}[box=\widefbox]{align}
\sum_{\mathbf{R}} \sum_{a} n^a_{i_I j_K k_K} \left( \mathbf{R} \right) \, \left. A^a \left( \mathbf{R} \right) \right|_{\mathrm{vert}} \, &= 0 \label{finalChow1} \\
\sum_{\mathbf{R}} \sum_{a} n^a_{A \Sigma \Gamma} \left( \mathbf{R} \right) \, \left. A^a \left( \mathbf{R} \right) \right|_{\mathrm{vert}} \, - 3\, U_{(A} \cdot \hat{\pi}_* \left( F_{\Sigma} \cdot F_{\Gamma)} \right) &= 0 \, \label{finalChow2} \\
\sum_{\mathbf{R}} \sum_{a} q_A \left( \mathbf{R} \right) \, \left. A^a \left( \mathbf{R} \right) \right|_{\mathrm{vert}} \, + 6 \,U_{A} \cdot \overline{K}_{B_3} &= 0 \, \label{finalChow3}
\end{empheq}
\end{subequations}
hold as relations in $\mathrm{CH}^2(\hat Y_4) \otimes \mathbb Q$, at least in the following situation: The fibre of the smooth resolution $\hat Y_4$ of the elliptic fibration possesses a locally closed embedding into a toric fibre ambient space, possibly with orbifold singularities. This embedding must be such that the fibral component of every matter surface cohomology class $[S^a(\bfR)_{\mathrm{vert}}]$ can be expressed as the pullback of a sum of cohomology classes on the fibre ambient space given by the product of ambient space divisor classes, and for these classes rational equivalence and cohomological equivalence on the fibre ambient space agree. 
These conditions are satisfied for instance when the underlying singular model $Y_4$ is given as a  crepant resolvable Tate model, or generalisations, over an arbitrary base $B_3$ with non-Abelian divisors $W_I$ represented as arbitrary hypersurfaces on $B_3$. We stress that a priori we only conjecture the above relations to be valid in the Chow group over the rationals, \ie ignoring potential torsional effects, but they may well be true more generally. From now on, whenever we write $\mathrm{CH}^\bullet(\hat Y_4)$ we understand that this space is taken over the rationals.

The importance of (\ref{summary4dChow2}) lies in the fact that it provides us with relations between different gauge backgrounds represented by the cycle classes $A^a(\mathbf{R})$. For instance, in \cite{Bies:2017fam} these relations have been used in order to facilitate the computation of massless matter states in the presence of such gauge backgrounds. While we do not have a general proof that (\ref{summary4dChow1}) and (\ref{summary4dChow2}) hold at the level of the Chow ring, and not merely in cohomology, we will exemplify the validity of our conjecture in \autoref{sec:ChowRelationsExamples}  in concrete setups (including those cases used in \cite{Bies:2017fam}). What simplifies this analysis is that all $A^a(\mathbf{R})$ give rise to the same 2-cycle class, \ie $A^a(\mathbf{R}) \equiv A(\mathbf{R})$ for all $a$. This allows us to simplify the expressions further. 

To avoid too heavy notation, let us spell this out for a gauge group of the form $G \times U(1)_A$. Consider for instance (\ref{finalChow2}): Let us take $\Sigma, \Gamma$ to be non-Abelian indices, \ie $\Sigma = i$ and $\Gamma = j$. Then (\ref{finalChow2}) takes the form
\[ \sum_{\mathbf{R}}\sum_{a}{n^a_{Aij} \left( \mathbf{R} \right) }   { \left. A \left( \mathbf{R} \right) \right|_{\text{vert}}}   - U_A \cdot \hat{\pi}_\ast \left( E_i \cdot E_j \right) = 0 \in {\rm{CH}}^2 ( \hat Y_4 ) \,. \label{equ:3.16bForSU(5)xU(1)}\]
From the definition (\ref{naRdef}), the trace relation (\ref{trace-coroot}) and the intersection numbers (\ref{intersectionEiEj}) it follows that
\bea
\sum_a n^a_{A ij} \left( \mathbf{R} \right) = q_{\mathbf{R}} \, \mathrm{tr}_{\mathbf{R}} \mathcal{T}_i \mathcal{T}_j = q_{\mathbf{R}} c_{\mathbf{R}}^{(2)} \, \lambda \, \mathfrak{C}_{ij} \, , \qquad \hat{\pi}_\ast \left( E_i \cdot E_j \right) = - \mathfrak{C}_{ij} \cdot W \, . 
\eea
Therefore (\ref{equ:3.16bForSU(5)xU(1)}) is equivalent to
\[ \sum_{\bf{R}} q_{\bf{R}} \, c^{(2)}_{\mathbf{R}} \, \left. A \left( \bf{R} \right) \right|_{\mathrm{vert}} - \frac{U_A}{\lambda} \cdot (-W) = 0 \in {\rm{CH}}^2 ( \hat Y_4 ) \, . \]
By evaluating (\ref{finalChow2}) for $\Sigma = A, \Gamma = A$ and for $\Sigma = i$, $\Gamma = A$ and applying similar reasoning to the other two equations in (\ref{summary4dChow2}), it is found that these three equations give rise to the following four anomaly conditions for gauge group $G \times U(1)_A$:
\begin{subequations} \label{anomGU1all}
\begin{align}
\sum_{\bf{R}} c^{(3)}_{\bf{R}} \, \left. A \left( \bf{R} \right) \right|_{\mathrm{vert}} &= 0 \in {\rm{CH}}^2 ( \hat Y_4 ) \, , \label{anom-cub-c} \\
\sum_{\rm{\bf{R}}} q_{\rm{\bf{R}}} \, c^{(2)}_{\mathbf{R}} \, \left. A \left( \rm{\bf{R}} \right) \right|_{\mathrm{vert}} + \frac{1}{\lambda} U_A \cdot W &= 0 \in {\rm{CH}}^2 ( \hat Y_4 ) \, , \label{anom-mixed2a} \\
\sum_{\rm{\bf{R}}} q^3_{\rm{\bf{R}}} \, {\rm{dim}} \left( \rm{\bf{R}} \right) \left. A \left( \rm{\bf{R}} \right) \right|_{\mathrm{vert}} - U_A \cdot 3 \hat{\pi}_* \left( U_A \cdot U_A \right) &= 0 \in {\rm{CH}}^2 ( \hat Y_4 ) \, , \label{anom-mixed2b} \\
\sum_{\rm{\bf{R}}} q_{\rm{\bf{R}}} \, {\rm{dim}} \left( \rm{\bf{R}} \right) \left. A \left( \rm{\bf{R}} \right) \right|_{\mathrm{vert}} + U_A \cdot (6 \overline{K}_{B_3} ) &= 0 \in {\rm{CH}}^2 ( \hat Y_4 ) \label{anom-mixed2c} \,.
\end{align}
\end{subequations}
In \autoref{sec:ChowRelationsExamples} we will prove these relations in non-trivial examples. 

Finally, let us point out that the relation (\ref{anom-cub-c}) derived from cancellation of the cubic non-Abelian anomalies can be only as powerful as the underlying constraints from absence of anomalies themselves. A representation $\bfR$ can only contribute to the cubic non-Abelian gauge anomalies if it is complex and if the anomaly coefficient $c^{(3)}_{\bfR}$ is non-vanishing. As is well-known, the only non-Abelian gauge groups with representations satisfying both of these conditions are $SU(N)$ with $N \geq 3$ and $SO(6)$. This does not mean that there may not exist similar interesting relations between complex 2-cycles in fibrations featuring different gauge groups, but their relation to constraints in the 4-dimensional effective field theory would necessarily have to be a different one.

\section{Implications for Calabi-Yau 3-Folds and 6-Dimensional Anomaly Cancellation} \label{sec_Implications6d}

The derivation of (\ref{summary4dCoho1}) and (\ref{summary4dCoho2}) for elliptically fibred Calabi-Yau 4-folds $\hat Y_4$ as presented here rests on the absence of anomalies in the 4-dimensional low-energy effective action obtained by compactifying F-theory on $\hat Y_4$. While we do not have a purely mathematical proof of these relations in general, we will verify the stronger set of relations (\ref{summary4dChow1}) and (\ref{summary4dChow2})  for explicit non-trivial examples in \autoref{sec:ChowRelationsExamples}. This analysis does not rely on specific properties of the base $B_3$, but only on the structure of the fibration over it. In particular the complex dimension of the base of the fibration does not enter explicitly. Therefore, (\ref{summary4dChow1}) and (\ref{summary4dChow2}) continue to hold as geometric relations in $\mathrm{CH}^2(\hat Y_{n+1})$ over a base $B_{n}$. We conjecture that this is the case not only for the explicit examples of \autoref{sec:ChowRelationsExamples}, but  more generally as long as the fibration satisfies the conditions stated after (\ref{summary4dChow2}).    

This raises the interesting question how to interpret these relations for general complex dimension $n$ of the base $B_n$. As we will now show, for $n=2$ the weaker version (\ref{summary4dCoho1}) is equivalent to the consistent cancellation of all gauge and mixed 
gauge-gravitational anomalies in the 6-dimensional $N=(1,0)$ theory obtained by compactification of F-theory on $\hat{\pi} \colon \hat Y_3 \twoheadrightarrow B_2$. This is an intriguing result because a priori the structure of loop-induced anomalies and their Green-Schwarz counterterms in six and four dimensions is very different (compare \autoref{figure:4DAnomaly} and \autoref{figure:6DAnomaly}). 
The geometric manifestation of anomaly cancellation in 6-dimensional F-theory models has been studied in great detail in the literature, most notably in \cite{Grassi:2000we,Grassi:2011hq} and \cite{Park:2011ji}, as reviewed already in the Introduction. The fact that the anomaly equations  in four and six dimensions are governed by a universal set of cohomological relations among algebraic codimension-2 cycles, however, is a new and stronger result.

\begin{figure}[tb]
\begin{center}
\begin{tikzpicture}[scale = 0.9]
  
  \def\shift{1};
  
  
  \draw[decorate, decoration=snake] (0,2) -- (1,1);
  \draw[decorate, decoration=snake] (4,2) -- (3,1);
  \draw[decorate, decoration=snake] (4,-2) -- (3,-1);
  \draw[decorate, decoration=snake] (0,-2) -- (1,-1);
  
  \draw (1,1) -- (3,1) -- (3,-1) -- (1,-1) -- (1,1);
    
  \draw [fill] (1,1) circle (2pt);
  \draw [fill] (3,1) circle (2pt);
  \draw [fill] (3,-1) circle (2pt);
  \draw [fill] (1,-1) circle (2pt); 

  \draw (\shift + 3,0) node {$+$};


  \draw[decorate, decoration=snake] (3 + 2 * \shift,0) -- (4 + 2 * \shift,0);
  \draw[dashed] (4 + 2 * \shift, 0) -- (4.9 + 2 * \shift, 0 );
  \draw[dashed] (5.1 + 2 * \shift, 0) -- (6 + 2 * \shift, 0 );
  \draw (5 + 2 * \shift,0) node {$\ast$};  
  \draw[decorate, decoration=snake] (6 + 2 * \shift,0) -- (6.5 + 2 * \shift,1);
  \draw[decorate, decoration=snake] (6 + 2 * \shift,0) -- (7 + 2 * \shift,0);
  \draw[decorate, decoration=snake] (6 + 2 * \shift,0) -- (6.5 + 2 * \shift,-1);
  \draw [fill] (4 + 2 * \shift,0) circle (2pt);
  \draw [fill] (6 + 2 * \shift,0) circle (2pt);
  
  \draw (7 + 3 * \shift,0) node {$+$};
  

  \draw[decorate, decoration=snake] (7 + 4 * \shift,1) -- (8 + 4 * \shift,0);
  \draw[decorate, decoration=snake] (7 + 4 * \shift,-1) -- (8 + 4 * \shift,0);
  \draw[dashed] (8 + 4 * \shift, 0) -- (8.9 + 4 * \shift, 0 );
  \draw[dashed] (9.1 + 4 * \shift, 0) -- (10 + 4 * \shift, 0 );
  \draw (9 + 4 * \shift,0) node {$\ast$}; 
  \draw[decorate, decoration=snake] (10 + 4 * \shift,0) -- (11 + 4 * \shift,1);
  \draw[decorate, decoration=snake] (10 + 4 * \shift,0) -- (11 + 4 * \shift,-1);
  \draw [fill] (8 + 4 * \shift,0) circle (2pt);
  \draw [fill] (10 + 4 * \shift,0) circle (2pt);
  
  \draw (11 + 5 * \shift,0) node {$= 0$};
  
\end{tikzpicture}
\end{center}
\caption{6d quartic anomalies and their cancellation via Green-Schwarz counterterms.}
\label{figure:6DAnomaly}
\end{figure}

For simplicity, assume a gauge group of the special form $G \times U(1)_A$. Generalisations to several gauge group factors will be immediate. The second Chow class $\mathrm{CH}^2(\hat Y_3)$ in which (\ref{summary4dChow1}) is valued now describes the rational equivalence class of curves on $\hat Y_3$. Specifically, the curve classes appearing (\ref{summary4dChow1}) are located in the fibre over isolated points on $B_2$. For $\bfR \neq \mathbf{adj}(G)$, $S^a_{\bfR}$ describes such fibral curves over isolated points, whose Chow class we collectively denote by $p_\bfR \in \mathrm{CH}_0(B_2)$. M2-branes wrapping the fibre of $S^a_{\bfR}$ (in both orientations) give rise to one hypermultiplet over each of these points with weight vector $\beta^a(\bfR)$ and $U(1)_A$ charge $q_A$. It will turn out useful to take the intersection product of (\ref{summary4dChow1}) with the divisors of $\hat Y_3$, beginning with the resolution divisors $E_i$ associated with the Cartan generators of $G$. The intersection product of $S^a_\bfR$ with $E_i$ within the Chow ring  on $\hat Y_3$ gives an element in $\mathrm{CH}_0(\hat Y_3)$, the Chow group of points on $\hat Y_3$. Its projection to the base describes the point class $p_\bfR \in \mathrm{CH}_0(B_2)$ with a multiplicity given by the intersection in the fibre. Since this fibral intersection reproduces the weight vector, we find 
\bea \label{Chowint1-6d}
\hat{\pi}_\ast \left( E_i \cdot S^a_{\mathbf{R}} \right) = \beta^a \left( \mathbf{R} \right) \, p_{\mathbf{R}} \,.
\eea

Similarly,  the intersection with the $U(1)_A$ divisor $U_A$ gives
\bea \label{Chowint2-6d}
\hat{\pi}_\ast \left( U_A \cdot S^a_{\mathbf{R}} \right) = q_{\mathbf{R}} \, p_{\mathbf{R}} \,.
\eea
The cohomology class $x_{\mathbf{R}}  = [p_\bfR] \in H_0(B_2,\mathbb Z)$ 
equals the number of points in Chow class $p_\bfR$. This coincides with the number of hypermultiplets in representation $\bfR$. Hence, at the level of cohomology,
\bea
\left[ E_{i} \right] \cdot \left[ S^a_{\mathbf{R}} \right] &= \beta^a_{i} \left( \mathbf{R} \right) \, x_{\mathbf{R}}, \qquad \left[ U_A \right] \cdot \left[ S^a_{\mathbf{R}} \right] &= q_{\mathbf{R}} \, x_{\mathbf{R}} \,.
\eea
By construction, all remaining divisor classes have trivial intersection with $S^a_{\mathbf{R}}$,
\bea
S_0 \cdot S^a_{\mathbf{R}} = 0 \,, \qquad D_\alpha^{\mathrm{b}} \cdot S^a_{\mathbf{R}} = 0 \quad \forall \quad D_\alpha^{\mathbf{b}} \in \mathrm{CH}^{1} \left( B_2 \right) \,. 
\eea

The cycle classes $S^\mathrm{\rho}$ associated with the adjoint representation are defined as in (\ref{def-SrhoI}), with $W$ now representing a curve on the base $B_2$. Hence $S^\mathrm{\rho}$ defines a class of fibral curves located over the canonical divisor $K_W$ of $W$. For a genus $g$ curve $W$, $K_W$ is the divisor class of degree
\bea
\mathrm{deg} \left( K_W \right) = - \Mint_{W}{c_1 \left( W \right) } = 2 g - 2 \,.
\eea
The intersections within the Chow ring are\footnote{$\iota \colon W \hookrightarrow B_2$ is the embedding of $W$ into the base $B_2$ of the elliptic fibration $\hat{Y}_3$.}
\bea \label{Chowint3-6d}
\hat{\pi}_\ast \left( E_i \cdot S^\rho \right) = \rho_i \, \iota_{\ast} \left( K_W \right) \,, \qquad \hat{\pi}_\ast \left( U_A \cdot S^\rho \right) = 0
\eea
and at the level of cohomology
\bea
\left[ E_{i} \right] \cdot \left[ S^\rho \right] &= \rho_{i} \, (2 g-2), \qquad \left[ U_A \right] \cdot \left[ S^\rho \right] &= 0 \,.
\eea
Here $\rho_i$ is the component of the root $\rho$ with respect to the coroot ${\cal T}_i$ defined in \autoref{sec_RevF}.

After this preparation, consider the intersection product of $E_l$ with (\ref{summary4dChow1-a}), for indices $\Lambda, \Sigma, \Gamma = i, j, k$. At the level of cohomology, this yields
\bea
\sum_{\mathbf{R} \neq \mathbf{adj}} \sum_a n^a_{lijk} \left( \mathbf{R} \right) \, x_{\mathbf{R}} + (g-1) \sum_{\rho} \, n^{\rho}_{lijk} - 3 \, \left[ \hat{\pi}_\ast \left( E_l \cdot E_{(i} \right) \right] \cdot_{B_2} \left[ \hat{\pi}_\ast \left( E_j \cdot E_{k)} \right) \right] = 0 \,,
\eea
where we have defined
\bea
n^a_{lijk}(\bfR) &= \beta^a_l \left( \bfR \right) \, \beta^a_i \left( \bfR \right) \, \beta^a_j \left( \bfR \right) \, \beta^a_k \left( \bfR \right), \qquad 
n^{\rho}_{lijk} = \rho_l \, \rho_i \rho_j \rho_k \,.
\eea
Summation over all weights and roots gives\footnote{The last identity is often written rather as $\mathrm{tr}_\bfR F^4 = c_\bfR^{(4)} \mathrm{tr}_\mathrm{fund} F^4 + d_\bfR^{(2)} \left(\mathrm{tr}_\mathrm{fund} F^2 \right)^2 \,$.}
\begin{align}
\begin{split}
\sum_{a=1}^{\mathrm{dim} \left( \bfR \right) } n^a_{lijk} \left( \bfR \right) &= \mathrm{tr}_{\bfR} {\mathcal T}_l \, {\mathcal T}_i \, {\mathcal T}_j \, {\mathcal T}_k = c_{\bfR}^{(4)} \mathrm{tr}_\mathrm{fund} {\mathcal T}_l \, {\mathcal T}_i \, {\mathcal T}_j \, {\mathcal T}_k + d_\bfR^{(2)} \mathrm{tr}_\mathrm{fund} {\mathcal T}_l \, {\mathcal T}_{(i} \, \mathrm{tr}_\mathrm{fund} {\mathcal T}_j \, {\mathcal T}_{k)} \, , \\
\sum_{\rho} n^{\rho}_{lijk} &= \mathrm{tr}_{\mathbf{adj}} {\mathcal T}_l \, {\mathcal T}_i \, {\mathcal T}_j \, {\mathcal T}_k = c_{\mathbf{adj}}^{(4)} \mathrm{tr}_\mathrm{fund} {\mathcal T}_l \, {\mathcal T}_i \, {\mathcal T}_j \, {\mathcal T}_k + d_{\mathbf{adj}}^{(2)} \mathrm{tr}_\mathrm{fund} {\mathcal T}_l \, {\mathcal T}_{(i} \, \mathrm{tr}_\mathrm{fund} {\mathcal T}_j \, {\mathcal T}_{k)} \,.
\end{split}
\end{align}
The quartic and the quadratic traces are, for general indices $l,i,j,k$, independent. Separating them results in the two equations
\begin{align}
\begin{split} \label{xRcRequ}
\sum_{\bfR \neq \mathbf{adj}} x_{\bfR} \, c_{\bfR}^{(4)} + (g-1) \, c_\mathbf{adj}^{(4)} &= 0 \, , \\
\sum_{\bfR \neq \mathbf{adj}} x_{\bfR} \, d_{\bfR}^{(2)} + (g-1)  \,  d_\mathbf{adj}^{(2)} &= 3 \, \frac{\left[ W \right]}{\lambda} \cdot_{B_2} \frac{\left[ W \right]}{\lambda} \,,
\end{split}
\end{align}
where we used (\ref{intersectionEiEj}) and (\ref{trace-coroot}) to bring the second equation into this form. (\ref{xRcRequ}) coincides with the conditions for cancellation of the non-factorisable and, respectively, factorisable quartic non-Abelian anomalies, as listed \eg in section 2 of \cite{Park:2011ji}. The RHS of the last equation is the Green-Schwarz counterterm for the factorisable part. If we intersect $[U_A]$ with (\ref{summary4dChow1-a}), for $\Lambda, \Sigma, \Gamma = i, j, k$, the same logic gives, at the level of cohomology,
\bea
\sum_{\bfR} q_{\bfR} \, c^{(3)}_{\bfR} \, x_{\bfR} = 0 \,.
\eea
This is nothing but the condition for cancellation of the mixed $U(1)_A - G^3 $ gauge anomaly in the 6-dimensional effective action.

This logic can be repeated for all conditions (\ref{summary4dChow1}) with the final result 
\begin{subequations} \label{Y3gaugeweak}
\bea
\sum_{\bfR \neq \mathbf{adj}} x_{\bfR} \, c_{\bfR}^{(4)} + (g-1) \, c_{\mathbf{adj}}^{(4)} &=& 0 \,, \\
\sum_{\bfR \neq \mathbf{adj}} x_{\bfR} \, d_{\bfR}^{(2)} + (g-1) \, c_{\mathbf{adj}}^{(2)} - 3 \, \frac{\left[ W \right]}{\lambda} \cdot_{B_2} \frac{ \left[ W \right] }{\lambda} &=& 0 \,, \\
\sum_{\bfR} q_{\bfR} \, c^{(3)}_{\bfR} \, x_{\bfR} &=& 0 \,, \\
\sum_{\bfR} q_{\bfR}^2 c_{\bfR}^{(2)}  \, x_{\bfR} + \frac{1}{\lambda} \hat{\pi}_\ast \left( \left[ U_A \right] \cdot \left[ U_A \right] \right) \cdot_{B_2} \left[ W \right] &=& 0 \,, \\
\sum_{\bfR} q_{\bfR}^4 \mathrm{dim} \left( \bfR \right) \, x_{\bfR} + 3 \, \hat{\pi}_\ast \left( \left[ U_A \right] \cdot \left[ U_A \right] \right) \cdot_{B_2} \hat{\pi}_\ast \left( \left[ U_A \right] \cdot \left[ U_A \right] \right) &=& 0
\eea
\end{subequations}
and
\begin{subequations} \label{Y3gaugebravweak}   
\bea
\sum_{\bfR \neq \mathbf{adj}} c_{\bfR}^{(2)} \, x_{\bfR} + (g-1) \, c_\mathbf{adj}^{(2)} + 6 \frac{\left[ W \right]}{\lambda} \cdot_{B_2} \left[ \overline{K}_{B_3} \right] &=& 0 \,, \\
\sum_{\bfR} q_{\bfR}^2 \mathrm{dim} \left( \bfR \right) \, x_{\bfR} + 6 \hat{\pi}_\ast \left( \left[ U_A \right] \cdot \left[ U_A \right] \right) \cdot_{B_2} \left[ \overline{K}_{B_3} \right] &=& 0 \,.
\eea
\end{subequations}
These are precisely the conditions for cancellation of the gauge and the mixed gauge-gravitational anomalies including the correct Green-Schwarz counterterms \cite{Park:2011ji}. Generalisations to several Abelian and non-Abelian gauge group factors are straightforward.

Let us summarize the logic so far: 4-dimensional anomaly cancellation implies (\ref{summary4dCoho1}) on any smooth elliptically fibred Calabi-Yau 4-fold $\hat Y_4$. Assuming that (\ref{summary4dCoho1}) holds more generally on any smooth elliptically fibred Calabi-Yau 4-fold $\hat Y_{n+1}$, as suggested by the considerations of \autoref{sec:ChowRelationsExamples}, we have derived the 6-dimensional anomaly cancellation conditions from (\ref{summary4dCoho1}) interpreted as relations in $H^{2,2}(\hat Y_3)$. We can now turn tables round and take anomaly cancellation in six dimensions as the \emph{starting point} to \emph{derive} (\ref{summary4dCoho1}) on $\hat Y_3$: Namely, the cohomology class of a curve in $H^{2,2}(\hat Y_3)$ is trivial if and only if its cohomological intersection with every element in $H^{1,1}(\hat Y_3)$ vanishes. By construction, (\ref{summary4dCoho1}) on $\hat Y_3$ is orthogonal to any base divisor class $[D_\alpha^{\mathrm b}]$ as well as to the zero-section $S_0$. Intersection with $[U_A]$ and $[E_{i_I}]$ gives rise to a set of equations which, as just shown, are nothing but the 6-dimensional anomaly equations. Hence anomaly cancellation implies (\ref{summary4dCoho1}) at the level of $H^{2,2}(\hat Y_3)$. This statement is of course in the spirit of the intersection theoretic identities derived from 6-dimensional anomaly cancellation in \cite{Park:2011ji}.

\begin{figure}[tb]
\begin{center}
\begin{tikzpicture}[scale = 0.9]

  \def\shift{3};
  
  
  \draw[decorate, decoration=snake] (-1.5,1) -- (0,1);
  \draw[decorate, decoration=snake] (-1.5,-1) -- (0,-1);
  \draw (2.25,0) node {$\mathbf{\times}$};
  
  \draw (0,1) -- (1.5,0) -- (0,-1) -- (0,1);
    
  \draw [fill] (0,1) circle (2pt);
  \draw [fill] (0,-1) circle (2pt);
  \draw [fill] (1.5,0) circle (2pt);
  \draw [fill, blue] (3,0) circle (2pt);
  
  \draw [blue] (3,0) node [above] {$F_\Lambda$};

  \draw[vecArrow] (4,0) -- (\shift + 3,0);   
   
  
  \draw[decorate, decoration=snake] (\shift + 4,2) -- (\shift + 5,1);
  \draw[decorate, decoration=snake] (\shift + 8,2) -- (\shift + 7,1);
  \draw[decorate, decoration=snake, blue] (\shift + 8,-2) -- (\shift + 7,-1);
  \draw[decorate, decoration=snake] (\shift + 4,-2) -- (\shift + 5,-1);
  
  \draw (\shift + 5,1) -- (\shift + 7,1) -- (\shift + 7,-1) -- (\shift + 5,-1) -- (\shift + 5,1);
    
  \draw [fill] (\shift + 5,1) circle (2pt);
  \draw [fill] (\shift + 7,1) circle (2pt);
  \draw [fill] (\shift + 5,-1) circle (2pt);
  \draw [fill] (\shift + 7,-1) circle (2pt);
  
  \draw [blue] (\shift + 8,-2) node [right] {$F_\Lambda$};

\end{tikzpicture}
\end{center}
\caption{The transition from the 4d to the 6d anomaly: Intersecting Chow classes encoding the structure of 4d cubic anomalies with the $U(1)_\Lambda$ divisor $F_\Lambda$  corresponds to adding an external leg to the loop diagram. A similar relation exists for the Green-Schwarz counterterms.}
\label{figure:4Dto6DAnomaly}
\end{figure}
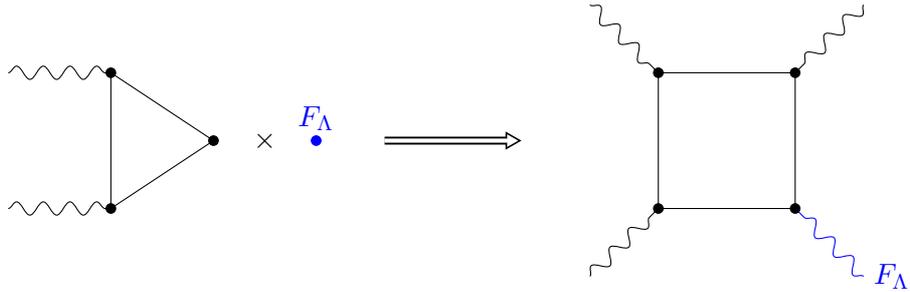

The fact that the same cohomological relations govern anomaly cancellation in four and six dimensions is intriguing, but in retrospect maybe not completely surprising: The gauge and mixed gauge gravitational anomalies in four dimensions are generated by cubic Feynman diagrams. Intuitively speaking, intersecting the cohomological relations on $\hat Y_4$ (which incorporate these diagrams geometrically) with the divisors $E_{i_I}$ and $U_A$ adds an extra external leg for the associated gauge field (see \autoref{figure:4Dto6DAnomaly}). The resulting relation hence encodes  the information of the quartic box diagrams underlying the structure of anomalies in six dimensions. 

Particularly interesting is furthermore the role of the cycles $S^{\rho}$ appearing in the relations (\ref{summary4dCoho1}): In four dimensions these terms are required for cancellation of all anomalies associated with the possible subgroups of a geometrically realized gauge group $G_I$ once it is broken by gauge flux. In six dimensions, no such breaking of the gauge group by fluxes can occur, but  at the same the states in the adjoint representation, which are accounted for by $S^\rho$, contribute to the $G_I$ anomalies due to their quartic nature. In this sense the more complicated structure of the anomalies in six dimensions is shadowed by the possibility of gauge group breaking flux in four dimensions.

We can, however, go even further: According to our conjecture, on any $\hat Y_3$ (\ref{summary4dChow1}) holds at the level of Chow classes (at least with rational coefficients and subject to the conditions stated after (\ref{summary4dChow2})). We can now consider the intersection of (\ref{summary4dChow1}) with $E_i$ and $U_A$ within the Chow ring and project the result to the base $B_2$, using (\ref{Chowint1-6d}), (\ref{Chowint2-6d}) and (\ref{Chowint3-6d}). This gives rise to a set of relations analogous to  (\ref{Y3gaugeweak}) and (\ref{Y3gaugebravweak}) (and their obvious generalisations) valued in $\mathrm{CH}_0(B_2)$, the Chow group of points on the base. These are obtained from  (\ref{Y3gaugeweak}) and (\ref{Y3gaugebravweak}) by replacing $x_\mathbf{R}$ by the point class $p_\mathbf{R}$ and $g-1$ by $\frac{1}{2} K_W$, interpreted via pushforward as a Chow class on $B_2$. Finally, the cohomological intersection product $\cdot_{B_2}$ in the Green-Schwarz terms is to be replaced by the intersection in the Chow ring of $B_2$. Though formulated slightly differently, a set of relations among Chow classes of points on $B_2$ has been discussed in \cite{Grassi:2011hq} and shown to imply (non-Abelian) anomaly cancellation.

\section{Chow Relations Exemplified} \label{sec:ChowRelationsExamples}

In this section we prove the Chow relations (\ref{summary4dChow1}) and (\ref{summary4dChow2}) for a prototypical class of elliptically fibred Calabi-Yau 4-folds $\hat Y_4$ over a generic base $B_3$. Apart from supporting our conjecture concerning the general validity of these equations within the Chow ring, the following analysis will illustrate the usefulness of  (\ref{summary4dChow1}) and (\ref{summary4dChow2}) for studying gauge backgrounds in F-theory. In order to exemplify the structure of all non-Abelian and mixed Abelian anomaly relations, the minimal requirement for $\hat Y_4$ is to contain a non-Abelian enhancement locus and at least one extra  rational section linearly independent of the zero-section, giving rise to an extra Abelian gauge group factor. The simplest geometry with this property is the $U(1)$ restricted Tate model \cite{Grimm:2010ez}, in which we furthermore engineer an additional non-Abelian gauge group factor. If we consider for instance Tate models with $I_n$ singularities, the existence of  vertical gauge fluxes in addition to the flux canonically associated with the extra rational section requires the gauge group to be at least $SU(5)$ \cite{Krause:2012yh}. For these two reasons we specialise, in \autoref{subsec:SpecialFTheoryGUTModel} - \autoref{subsec:AnomalyDifferencesForSU(5)xU(1)} to a $U(1)$ restricted Tate model of gauge group $G = {SU} ( 5 ) \times {U} ( 1 )_X$. This model, which has been introduced in \cite{Krause:2011xj}, suffices to exemplify the full structure of the relations (\ref{summary4dChow1}) and (\ref{summary4dChow2}). In \autoref{sec_SU4example} we will exemplify how the relations (\ref{summary4dChow2}) explain the aforementioned absence of vertical fluxes for $I_n$ Tate models with $n<4$, focussing for concreteness on the most interesting case $n=4$.

\subsection{\texorpdfstring{$\mathbf{\text{SU} ( 5 ) \times \text{U} ( 1 )_X}$}{SU(5)xU(1)} Fibration} \label{subsec:SpecialFTheoryGUTModel}

To fix notation, we very briefly recall the relevant properties of the $G = \text{SU} ( 5 ) \times \text{U} ( 1 )_X$ elliptic fibration. More details can be found in \cite{Krause:2011xj} and in the recent \cite{Bies:2017fam}, whose conventions we will be using throughout. For convenience of the reader the most crucial properties of this fibration are listed in \autoref{sec:FibreStructure}.

After resolving the singularities, the elliptic fibration is given by a smooth 4-fold $\hat{\pi} \colon \hat{Y}_4 \twoheadrightarrow B_3$ described as the vanishing locus of the polynomial
\begin{align}
\begin{split}
P_T^\prime &= y^2 s e_3 e_4 + a_1  x y z s + a_{3,2}  y z^3 e_0^2 e_1 e_4 - x^3 s^2 e_1 e_2^2 e_3 \\
           &\quad - a_{2,1}  x^2 z^2 s e_0 e_1 e_2 - a_{4,3}  x z^4 e_0^3 e_1^2 e_2 e_4 
\end{split}
\end{align}
inside an ambient space $\hat X_5$. Here $a_{i,j}$ is a section of $\overline{K}^i_{B_3} \otimes W^{-j}$ with $W$ the divisor class associated with the brane stack with gauge group $SU(5)$ and $\overline{K}_{B_3}$ the anti-canonical bundle of $B_3$. 
From the toric ambient space of the fibre we inherit the linear relations as encoded in the following table of toric weights:
\begin{center}
\begin{tabular}{|c|c|cccc|cccc|}
\toprule
 & $e_0$ & $e_1$ & $e_2$ & $e_3$ & $e_4$ & x & y & z & s \\
\hline \hline
$\overline{K}_{B_3}$ & $\cdot$ & $\cdot$ & $\cdot$ & $\cdot$ & $\cdot$ & 2 & 3 & $\cdot$ & $\cdot$ \\
W                    & 1 & $\cdot$ & $\cdot$ & $\cdot$ & $\cdot$ & $\cdot$ & $\cdot$ & $\cdot$ & $\cdot$ \\
\hline
$E_1$ & -1        & 1 & $\cdot$ & $\cdot$ & $\cdot$ & -1 & -1 & $\cdot$ & $\cdot$ \\
$E_1$ & -1        & $\cdot$ & 1 & $\cdot$ & $\cdot$ & -2 & -2 & $\cdot$ & $\cdot$ \\
$E_1$ & -1        & $\cdot$ & $\cdot$ & 1 & $\cdot$ & -2 & -3 & $\cdot$ & $\cdot$ \\
$E_1$ & -1        & $\cdot$ & $\cdot$ & $\cdot$ & 1 & -1 & -2 & $\cdot$ & $\cdot$ \\
\hline
Z & $\cdot$   & $\cdot$ & $\cdot$ & $\cdot$ & $\cdot$ & 2 &  3 & 1 & $\cdot$ \\
S & $\cdot$   & $\cdot$ & $\cdot$ & $\cdot$ & $\cdot$ & -1 & -1 & $\cdot$ & 1 \\
\bottomrule
\end{tabular}
\end{center}
Furthermore, the Stanley-Rei{s}ner ideal includes 
\begin{align}
\begin{split} \label{SRideal}
I_{\text{SR}} \left( \text{top} \right) = & \left\{ xy, x e_0 e_3, x e_1 e_3, x e_4, y e_0 e_3, y e_1, y e_2, z s, z e_1 e_4, z e_2 e_4, \right. \\
                & \qquad \left. z e_3, s e_0, s e_1, s e_4, e_0 e_2, z e_4, z e_1, z e_2, s e_2, e_0 e_3, e_1 e_3 \right\} \, .
\end{split}
\end{align}

The Cartan generators of $SU(5)$ are denoted by $E_i \in \mathrm{CH}^1(\hat Y_4)$, which at the level of analytic cycles corresponds to the vanishing locus $V(e_i)$.
Similarly, $Z\in \mathrm{CH}^1(\hat Y_4)$ and $S\in \mathrm{CH}^1(\hat Y_4)$ denote the zero-section and the extra independent rational section associated with $V(z)$ and $V(s)$, respectively. The Shioda map furthermore identifies the $U(1)_X$ generator as
\[ U_X := - \left( 5 \left( S - Z - \overline{K}_{B_3} \right) + 2 E_1 + 4 E_2 + 6 E_3 + 3 E_4 \right) \in \mathrm{CH}^1 ( \hat Y_4 ) \,.\]
The charged massless matter fields localise on the matter curves 
\begin{align}
  C_{\mathbf{10}_1} &= V \left( e_0, a_{1,0} \right) \, , \quad & C_{\mathbf{5}_3} &= V \left( e_0, a_{3,2} \right) \, , \cr
  C_{\mathbf{5}_{-2}} &= V \left( e_0, a_1 a_{4,3} - a_{2,1} a_{3,2} \right) \, , & C_{\mathbf{1}_{5}} &= V \left( a_{4,3}, a_{3,2} \right) \, .
\end{align}
The explicit form of the matter surfaces $S^a_\bfR$ associated with the various weights can be found in appendix B of \cite{Bies:2017fam}. Since for the applications of this paper we need the explicit form of the fibre of $\hat Y_4$ over the matter surfaces, we list the fibre components in \autoref{sec:FibreStructure} for completeness. In terms of these, the matter surface fluxes, whose general construction is given in (\ref{Aageneral1}), take the form
\begin{align}
\begin{split} \label{MSFexplicit}
A \left( \mathbf{10}_1 \right) &= A \left( \mathbf{10}_1 \right) \left( 0, 0, - \frac{2 \lambda}{5}, \frac{\lambda}{5}, -\frac{\lambda}{5}, \frac{2 \lambda}{5} \right) \in \mathrm{CH}^2 ( \hat Y_4 ) \, , \\
A \left( \mathbf{5}_3 \right) &= A \left( \mathbf{5}_3 \right) \left( 0, -\frac{\lambda}{5}, -\frac{2 \lambda}{5}, -\frac{3 \lambda}{5}, \frac{2 \lambda}{5}, \frac{\lambda}{5} \right)\in \mathrm{CH}^2 ( \hat Y_4 ) \, , \\
A \left( \mathbf{5}_{-2} \right) &= A \left( \mathbf{5}_{-2} \right) \left( 0, -\frac{\lambda}{5}, -\frac{2 \lambda}{5}, -\frac{3 \lambda}{5}, \frac{2 \lambda}{5}, \frac{\lambda}{5} \right) \in \mathrm{CH}^2 ( \hat Y_4 ) \, , \\
A \left( \mathbf{1}_{5} \right) &= \mathbb{P}^1_A \left( \mathbf{1}_{5} \right) = \lambda V \left( P^\prime_T, s, a_{3,2}, a_{4,3} \right)\in \mathrm{CH}^2 ( \hat Y_4 ) \, .
\end{split}
\end{align}
The notation is \eg
\[A \left( \mathbf{10}_1 \right) \left( a_0, a_1, \ldots, a_5 \right) = a_0 \mathbb{P}^1_{0A} \left( \mathbf{10}_1 \right) + a_1 \mathbb{P}^1_{14} \left( \mathbf{10}_1 \right) + \ldots + a_5 \mathbb{P}^1_{4D} \left( \mathbf{10}_1 \right) \]
in terms of the rational curves  over the respective matter curves, in the order as appearing in the list in \autoref{sec:FibreStructure}. The overall coefficients $\lambda \in \mathbb{Q}$ have to be chosen subject to the usual flux quantisation condition of \cite{oai:arXiv.org:hep-th/9609122}. For sake of simplicity, the following analysis makes the choice $\lambda = 1$.

As the final piece of information which we quote from \cite{Bies:2017fam} we note that these cycles can be expressed as elements $\mathcal{A} \in \text{CH}^2 ( \hat X_5)$ such that $\left. \mathcal{A} \right|_{\hat{Y}_4} = A$. The explicit form of $ \mathcal{A}$ -- for $\lambda = 1$ -- is
\begin{align}
\begin{split}
\label{equ:MatterSurfaceabcFluxes-X5abc}
\mathcal{A} \left( \mathbf{10}_1 \right) &= - \frac{1}{5} \, \left( 2 \cE_1 - \cE_2 + \cE_3 - 2 \cE_4 \right) \cdot\overline{\mathcal{K}}_{B_3} - \cE_2 \cdot \cE_4 \, , \\
\mathcal{A} \left( \mathbf{5}_3 \right) &= - \frac{1}{5} \, \left( \cE_1 + 2 \cE_2 - 2 \cE_3 - \cE_4 \right) \cdot\left( 3 \overline{\mathcal{K}}_{B_3} - 2 \mathcal{W} \right) - \cE_3 \cdot\mathcal{X} \, , \\
\mathcal{A} \left( \mathbf{5}_{-2} \right) &= - \frac{1}{5} \, \left( \cE_1 + 2 \cE_2 + 3 \cE_3 - \cE_4 \right) \cdot\left( 5 \overline{\mathcal{K}}_{B_3} - 3 \mathcal{W} \right) \\
& \qquad \qquad \qquad \qquad \qquad + \left( \cE_3 \, \overline{\mathcal{K}}_{B_3} + \cE_3 \cdot\mathcal{Y} - \cE_3 \cdot\cE_4 \right) \, , \\
\mathcal{A} \left( \mathbf{1}_{5} \right)  &=\mathcal{S} \, \left( 3 \overline{\mathcal{K}}_{B_3} - 2 \mathcal{W} \right) -  \mathcal{S} \cdot\mathcal{X} \, .
\end{split}
\end{align}
The $U(1)_X$ gauge background is described by
\[ \mathcal{A}_X \left( \mathcal{F} \right) = - \frac{1}{5} \, \mathcal{F} \cdot \left( 5 \mathcal{S} - 5 \mathcal{Z} - 5 \overline{\mathcal{K}}_{B_3} + 2 \cE_1 + 4 \cE_2 + 6 \cE_3 + 3 \cE_4 \right) \, . \label{equ:U1abcX-Fluxabc-X5} \]

\subsection{Relation (\ref{summary4dChow2}) for \texorpdfstring{$\mathbf{SU(5) \times U(1)_X}$}{SU(5)xU(1)}}

Let us now evaluate the relations (\ref{summary4dChow2}) for the $SU(5) \times U(1)_X$ model introduced in the previous section.
These equations have been specified to a gauge group of the form $G \times U(1)_A$ in (\ref{anomGU1all}), and it remains to evaluate these expressions in the case at hand. For $G= SU(N)$, the relevant values for group theoretic constants $c_{\mathbf{R}}^{(n)}$ and $\lambda$, defined in (\ref{cRn-def}) and (\ref{lambdadef}), are (see e.g. \cite{Erler:1993zy})
\bea
c^{(3)}_{\mathbf{\Lambda^2 N}} = N-4, \qquad c^{(2)}_{\mathbf{\Lambda^2 N}} = N-2, \qquad \lambda  = 1 \,.
\eea
For the $U(1)_X$ generator in the model under consideration
$\hat{\pi}_* \left( U_X \cdot U_X \right) = 30 \, W - 50 \, \overline{K}_{B_3} \label{projectionUX}$.\footnote{In particular, for every base divisor $D^{\mathbf{b}}_\alpha, D^{\mathbf{b}}_\beta$, $[U_X] \cdot_{\hat Y_4} [U_X] \cdot_{\hat Y_4} [D^{\mathbf{b}}_\alpha] \cdot_{\hat Y_4} [D^{\mathbf{b}}_\beta] = [ \hat{\pi}_*(U_X \cdot U_X )] \cdot_{B_3} [D^{\mathbf{b}}_\alpha] \cdot_{B_3} [D^{\mathbf{b}}_\beta]$. This integral has been evaluated for the model at hand in \cite{Krause:2011xj}, leading to $\hat{\pi}_* \left( U_X \cdot U_X \right) = 30 \, W - 50 \, \overline{K}_{B_3}$.}
With this information (\ref{anomGU1all}) becomes 
\bea
&&A \left( \mathbf{10}_1 \right) + A \left( \mathbf{5}_3 \right) + A \left( \mathbf{5}_{-2} \right) = 0  \label{equ:Chow1ForSU5xU1}  \\
&& 3  A \left( \mathbf{10}_1 \right) + 3 A \left( \mathbf{5}_3 \right) - 2  A \left( 5_{-2} \right) + U_A \cdot W = 0  \label{equ:Chow2ForSU5xU1} \\
&& 2 A \left( \mathbf{10}_1 \right) + 27  A \left( \mathbf{5}_3 \right) - 8 A \left( \mathbf{5}_{-2} \right) + 25  A \left( \mathbf{1}_5 \right) + U_A \cdot (30  \overline{K}_{B_3} - 18 W) = 0   \label{equ:Chow3ForSU5xU1} \\
&&10 A \left( \mathbf{10}_1 \right) + 15  A \left( \mathbf{5}_3 \right) - 10  A \left( 5_{-2} \right) + 5  A \left( \mathbf{1}_5 \right) + 6 \cdot U_A \cdot \overline{K}_{B_3}  = 0\, . \label{equ:Chow4ForSU5xU1} 
\eea
It is readily seen that (\ref{equ:Chow1ForSU5xU1}) - (\ref{equ:Chow4ForSU5xU1}) are not independent. Rather they are equivalent to the following three linearly independent relations within $\mathrm{CH}^2(\hat Y_4)$:
\bea
A \left( \mathbf{5}_3 \right) &=& - {A} \left( \mathbf{5}_{-2} \right) - {A} \left( \mathbf{10}_1 \right) \, , \label{relation1} \\
A \left( \mathbf{5}_{-2} \right) &=& {A}_X \left({W} \right) \, , \label{relation2} \\
A \left( \mathbf{1}_{5} \right) & =& - {A}_X \left( 6 \overline{K}_{B_3} - 5 {W} \right) + {A} \left( \mathbf{10}_1 \right) \,. \label{relation3}
\eea
In \autoref{sec:Proof} we prove that these relations indeed hold true as relations between equivalence classes of algebraic 2-cycles modulo rational equivalence, \ie between elements of $\mathrm{CH}^2(\hat Y_4)$. The proof rests on two important properties of the Chow groups.

First, on any algebraic variety $X$, two cycles $C_1, C_2$ are rationally equivalent if and only if one can find a rationally parametrized family of cycles which interpolates between $C_1$ and $C_2$. This means that we can find a cycle $\Gamma(t)$ on $\mathbb P^1 \times X$ such that 
\bea \label{Gammdefintion-general}
\Gamma \left( t=t_1 \right) = C_1, \qquad \quad \Gamma \left( t=t_2 \right) = C_2
\eea
with $t \in \mathbb P^1$ parametrising the interpolation between $C_1$ and $C_2$. In other words, $C_1$ and $C_2$ are related by a `rational homotopy'.

The second property we are using is specific to the fact that the elliptic fibre of $\hat Y_4$ is embedded into a toric fibre ambient space. Given a regular embedding $\iota \colon X \hookrightarrow Y$ between two algebraic varieties, the pullback map is a well-defined linear map \cite{FultonInt}
\bea \label{pullbackformula}
\iota^*: \mathrm{CH}^p \left( Y \right) \rightarrow \mathrm{CH}^p \left( X \right) \,.
\eea
This means that if two algebraic cycles $C_1, C_2 \subseteq Y$ are rationally equivalent on $Y$, then their pullbacks to $X$ are rationally equivalent on $X$.
The importance of this property of Chow groups is that for cycles arising as pullbacks from the ambient space we can use rational equivalence in the ambient space $Y$ to check for rational equivalence on $X$. This leads to drastic simplifications if the space $Y$ is a complete toric variety which in addition is simplicial or even smooth \cite{cox2011toric}: For a smooth and complete toric variety $Y$, rational equivalence coincides with homological equivalence, \ie $\mathrm{CH}^\bullet(Y)_{\mathbb{Z}}  \cong H^\bullet ( Y, \mathbb{Z} )$, and to  check rational equivalence for pullback cycles on $X$ we are hence allowed to perform operations on the pre-image of the cycle on $Y$ as long as these preserve its homology class on $Y$. Even if $Y$ is merely a complete and simplicial toric variety, we still have $\mathrm{CH}^\bullet(Y)_{\mathbb{Q}} \cong H^{\bullet}(Y, \mathbb{Q})$. These properties, which indeed hold in the example studied in this section, motivate the conditions stated after (\ref{summary4dChow2}) for our more general conjecture. 

Note that even though anomaly cancellation alone suffices to show that the relations (\ref{summary4dChow2}) and hence (\ref{anomGU1all}) hold true as relations in $H^\bullet(\hat Y_4)$, it is in general not the case that they hold already for the homology classes of the cycles $\mathcal{A}$ on the ambient space $\hat X_5$ from which the cycles on $\hat Y_4$ descend via pullback. If this were the case, then by the above reasoning it would be immediate that the relations hold in $\mathrm{CH}^\bullet(\hat Y_4)$. Instead, we have to have work harder to show this latter, stronger statement.

\subsection{Relation (\ref{summary4dChow1}) for \texorpdfstring{$\mathbf{SU(5) \times U(1)_X}$}{SU(5)xU(1)}} \label{subsec:AnomalyDifferencesForSU(5)xU(1)}

We now turn to the proof of (\ref{summary4dChow1}). Since (\ref{summary4dChow2}) has already been established (assuming the reader has dragged themselves through \autoref{sec:Proof}), it suffices to analyse the difference of both types of equations with compatible index structures. If all gauge indices are purely non-Abelian, the relevant expression is (\ref{equ:DifferenceAnomalies}). In the present example, the LHS of (\ref{equ:DifferenceAnomalies}) simplifies to 
\begin{align} 
\Xi_{ijk} = &\sum_{\mathbf{R} \neq \mathbf{adj}}{ \sum_{a}{- n^a_{ijk} \left( \mathbf{R} \right) \Delta^a \left( \mathbf{R} \right) } } + \frac{1}{2} \sum_{\rho}{\beta^\rho_i \beta^\rho_j \beta^\rho_k S^\rho}  - 3 E_{(i} \cdot \hat{\pi}_\ast \left( E_j \cdot E_{k)} \right). 
\end{align}
We will exemplify this computation for $i = j = k = 1$ and start by evaluating the first sum. With the help of the explicit form of the matter surfaces $S^a_\bfR$ and their associated weight vectors tabulated in \autoref{sec:FibreStructure} we arrive at 
\[ \sum_{\mathbf{R} \neq \mathbf{adj}}{ \sum_{a}{- \beta^a_1 \left( \mathbf{R} \right) \beta^a_1 \left( \mathbf{R} \right) \beta^a_1 \left( \mathbf{R} \right) \Delta^a \left( \mathbf{R} \right) } } = -3 \mathbb{P}^1_{14} \left( \mathbf{10}_1 \right) - \mathbb{P}^1_1 \left( \mathbf{5}_3 \right) - \mathbb{P}^1_1 \left( \mathbf{5}_{-2} \right) \, . \]
The second sum in (\ref{equ:DifferenceAnomalies}) refers to the adjoint representation of $SU(5) \times U(1)_X$. The matter surfaces associated with the negative roots of $SU(5)$ are given by
\begin{align}
\begin{split}
S^\rho_{-} \left( SU \left( 5 \right) \times U( 1 )_X \right) =&   \left\{ E_1, E_2, E_3, E_4, \right. \\
                                                               & \quad \left. E_1 + E_2, E_2 + E_3, E_3 + E_4, \right. \\
                                                               & \quad \left. E_1 + E_2 + E_3, E_2 + E_3 + E_4, \right. \\
                                                               & \quad  \left. \left. E_1 + E_2 + E_3 + E_4 \right\} \right|_{K_W} \, ,
\end{split}
\end{align}
and for the remaining 10 positive roots by $S_+^\rho = - S_-^\rho$. This leads to
\[ \frac{1}{2} \sum_{\rho}{\beta^\rho_1 \beta^\rho_1 \beta^\rho_1 \, S^\rho} = -11 \, E_1 \cdot \left( W - \overline{K}_{B_3} \right) \, . \]
As for the third term in (\ref{equ:DifferenceAnomalies}), the analogue of (\ref{intersectionEiEj}) in the Chow ring implies 
\[ - 3 E_{(1} \cdot \hat{\pi}_\ast \left( E_1 \cdot E_{1)} \right) = - 3 E_1 \cdot \hat{\pi}_\ast \left( E_1 \cdot E_1 \right) = - 3 E_1 \cdot \left( -2 \right) W = 6 E_1 \cdot W \]
and altogether we find
\[ \Xi_{111} = 11 \, E_1 \cdot  \overline{K}_{B_3} - 5 \, E_1 \cdot W -3 \mathbb{P}^1_{14} \left( \mathbf{10}_1 \right) - \mathbb{P}^1_1 \left( \mathbf{5}_3 \right) - \mathbb{P}^1_1 \left( \mathbf{5}_{-2} \right) \, . \]
Finally recall
\[ \mathbb{P}^1_{14} \left( \mathbf{10}_1 \right) = E_1 \cdot \overline{K}_{B_3}, \quad \mathbb{P}^1_{1} \left( \mathbf{5}_3 \right) = E_1 \cdot  \left( 3 \overline{K}_{B_3} - 2 W \right), \quad \mathbb{P}^1_{1} \left( \mathbf{5}_{-2} \right) = E_1 \cdot \left( 5 \overline{K}_{B_3} - 3 W \right) \, . \]
This leads to $\Xi_{111} = 0 \in \mathrm{CH}^\bullet(\hat Y_4)$. It is simple to repeat this analysis for all $( i,j,k ) \in \{ 1, 2, 3, 4 \}^3$ and to convince oneself that $\Xi_{ijk}$ is the trivial cycle in $\text{CH}^{\bullet} ( \hat{Y}_4 )$ for any such indices $i,j,k$.

The analogue of (\ref{equ:DifferenceAnomalies}) for the mixed Abelian-non-Abelian, cubic Abelian and mixed gravitational anomalies are the relations
\begin{align}
\begin{split}
\displaystyle \sum_{\mathbf{R} \neq \text{adj}}{ \sum_{a}{ n_{ijX}^a \left( \mathbf{R} \right) \Delta^a \left( \mathbf{R} \right) } } &= 0 \, , \qquad
\displaystyle \sum_{\mathbf{R} \neq \text{adj}}{ \sum_{a}{ n_{XXX}^a \left( \mathbf{R} \right) \Delta^a \left( \mathbf{R} \right) } } = 0 \, , \\
\displaystyle \sum_{\mathbf{R} \neq \text{adj}}{ \sum_{a}{ q_{\mathbf{R}} \, \Delta^a \left( \mathbf{R} \right) } } &= 0 \, .
\end{split}
\end{align}
In these equations the label `X' refers to the Abelian $U(1)_X$-group. Along the same strategy these sums can be shown to vanish identically in $\text{CH}^\bullet ( \hat Y_4 )$. This completes the proof of (\ref{summary4dChow1}) in the model under consideration, for any base $B_3$.

\subsection{Fluxless \texorpdfstring{$\mathbf{SU(4)}$}{SU(4)}} \label{sec_SU4example}

As a further amusing application we now exemplify that the absence of certain matter surface fluxes for F-theory models in $I_n$ Tate models with $n < 5$, observed already in \cite{Krause:2012yh}, can be traced back to the relation (\ref{anom-cub-c}).

 For brevity we only consider the model with $n=4$ in detail. A brief summary of the geometry of the $SU(4)$ Tate model is provided in  \autoref{fibre_structure_SU4}.
 The  $SU(4)$  divisor $W$ contains two matter curves
associated with massless matter in representations $\mathbf{4}$ and $\mathbf{6}$ of $SU(4)$. From the form of the $SU(N)$ index in the anti-symmetric representation, $c^{(3)}_{\mathbf{\Lambda^2 N}} = N-4$,  with $N=4$ it is clear that the matter surface flux $A(\mathbf{6})$ constructed from the matter surface $S^a_\mathbf{6}$ does not appear in (\ref{anom-cub-c}). Hence the relation in homology derived from absence of cubic anomalies is simply
\bea \label{A4=0}
\left[ A \left( \mathbf{4} \right) \right] = 0 \in H^{2,2} ( \hat Y_4 ) \,.
\eea
In fact, we show in \autoref{proofSU(4)} that this relation does hold also at the level of Chow groups, as stated by our more general conjecture. 

The result fits with the aforementioned observation of \cite{Krause:2012yh} that this model does not allow for any matter surface fluxes. Indeed we will show in \autoref{proofSU(4)} that, in addition to (\ref{A4=0}) the only other candidate for a  matter surface flux $A(\mathbf{6}) = 0 \in \mathrm{CH}^2(\hat Y_4)$. This result, or rather its a priori weaker version in cohomology, is also related to absence of cubic anomalies, but not quite in the way written in our relation (\ref{anom-cub-c}). Namely, since $\mathbf{6} = \mathbf{\overline{6}}$ the chiral index
\[ \chi(\mathbf{6}) = \left[ A \left( \mathbf{6} \right) \right] \cdot \left[ S \left( \mathbf{6} \right) \right] = 0 \, . \]
Then by anomaly cancellation clearly also 
\[ \chi(\mathbf{4}) =  \left[ A \left( \mathbf{6} \right) \right] \cdot \left[ S \left( \mathbf{4} \right) \right] = 0 \, . \]
We can now show that this implies
\[ \left[ A \left( \mathbf{6} \right) \right] = 0 \in H^{2,2} ( \hat Y_4 ) \label{A6is0} \]
using similar reasoning as in \autoref{sec:genrelChow}: By construction $[A(\mathbf{6})] $ is orthogonal on the subspace $V_1$ defined in (\ref{span2}) (recall that there are no extra sections). Furthermore, each of the generators of $V_2$ appearing in (\ref{defV4}) can be interpreted as the class of a matter surface plus a sum of terms in $V_1$. This is because each such element can be written as a vertical flux plus terms in $V_1$. In absence of a $U(1)$ the only vertical fluxes are the matter surface fluxes, and their explicit form $A({\bf R}) = S^a(\bfR) + \Delta^a(\bfR)$ with $\Delta^a(\bfR) \in V_1$ then implies the statement. Hence (\ref{A6is0}) even follows  without explicit computation.

\section{Conclusions and Outlook} \label{sec_Conclusions}

In this note we have presented a system of  cohomological relations on elliptic fibrations which are equivalent to the cancellation of all gauge and mixed gauge-gravitational anomalies in F-theory compactifications to four and six dimensions. These relations, given by  (\ref{summary4dCoho1}),  hold in $H^{2,2}(\hat Y_{n+1},\mathbb Q)$ with $n=2$ and $n=3$ referring to compactifications to six and four dimensions, respectively. To relate them to the structure of anomaly cancellation one must intersect (\ref{summary4dCoho1}) either with the set of possible gauge fluxes $G_4 \in H^{2,2}(\hat Y_4,\mathbb Q)$ for $n=3$ or with the divisors spanning the coroot/coweight lattice in compactifications to six dimensions ($n=2$). Interestingly, the consistent cancellation of local anomalies both in four and six dimensions is therefore governed by 
the same type of cohomological relations in F-theory. This illustrates the remarkably unifying amount of physics information encoded in the structure of the F-theory fibration as such, irrespective of details of the base. 

Our derivation of (\ref{summary4dCoho1}) relies on arguments from string theory, building on and extending the earlier work \cite{Cvetic:2012xn}  and reproducing the identities of \cite{Park:2011ji} as a corollary. The result generalises observations made in \cite{Lin:2016vus} in concrete example fibrations. Despite the string theoretic nature of their derivation, the relations (\ref{summary4dCoho1}) must hold as general mathematical identities valid on any elliptically fibered Calabi-Yau 3-fold and 4-fold which admits a smooth, flat, crepant resolution. This begs the question for a direct proof within `formal mathematics' which would complement the string theoretic arguments given within the realm of `physical mathematics' in this work.

Another immediate question for future research concerns the relation between the cohomological identities described in this work and the structure of anomalies in 2-dimensional $N=(0,2)$ supersymmetric F-theory compactifications on Calabi-Yau 5-folds \cite{Schafer-Nameki:2016cfr,Apruzzi:2016iac,Apruzzi:2016nfr,Lawrie:2016rqe}. A notable difference from compactifications to four and six dimensions is the appearance of \emph{chiral} charged zero modes at the intersection of 7-branes and D3-branes wrapping curves on the base of the fibration. These provide a crucial contribution to the gauge and gravitational anomalies \cite{Schafer-Nameki:2016cfr,Apruzzi:2016iac,Lawrie:2016rqe} which remains somewhat mysterious to date (cf. \cite{Lawrie:2016axq} and references therein).

Based on circumstantial evidence in concrete examples we have conjectured in this article that the cohomological relations (\ref{summary4dCoho1}) and (\ref{summary4dCoho2}) hold not only in $H^{2,2}(\hat Y_{n+1},\mathbb Q)$, but more generally as relations in the Chow ring of algebraic cycles modulo rational equivalence, at least with rational coefficients. We have described the specific assumptions underlying this conjecture after (\ref{summary4dChow2}). It would be very instructive to understand the difference between the relations in cohomology versus rational equivalence from a mathematical point of view: A priori the Chow ring contains considerably more information than cohomology. However, we cannot  completely exclude that under the assumptions stated after (\ref{summary4dChow2}) the two coincide, at least as far as the subspace of cycles is concerned for which our conjecture holds. If this is not the case, we face the question of the physical meaning of the extra amount of information encoded at the level of the Chow ring. For anomaly cancellation as such the relations (\ref{summary4dCoho1}) within  $H^{2,2}(\hat Y_{n+1},\mathbb Q)$  are both necessary and sufficient. On the other hand, the identities (\ref{summary4dChow2}) within the Chow ring have important consequences also in the context of F-theory, to the extent that they relate the possible gauge backgrounds on elliptic Calabi-Yau 4-folds. This has already been used in \cite{Bies:2017fam} and furthermore explains the absence of vertical gauge backgrounds in certain classes of elliptic fibrations \cite{Krause:2012yh}.

In this work we have focused on the interplay between the geometry of elliptic fibrations and the structure of \emph{local} gauge and mixed gauge-gravitational anomalies. It would be equally rewarding to investigate the structure of global anomalies from a similar perspective. For example, in perturbative D-brane models cancellation of global $SU(2)$ Witten anomalies is equivalent to cancellation of all K-theory charges carried by the D-branes \cite{Uranga:2000xp}. In \cite{GarciaEtxebarria:2005qc} the absence of Witten anomalies in F-theory has been attributed to suitable quantization of the dual M-theory flux $G_4$.
The latter in turn is determined geometrically by the divisibility properties of the second Chern class $c_2(\hat Y_4)$, which enters the Freed-Witten quantisation condition $G_4 + \frac{1}{2} c_2(\hat Y_4) \in H^{4}(\hat Y_4, \mathbb Z)$. Indeed, the absence of a Witten anomaly in F-theory models with gauge group $SU(2)$ (and further Abelian gauge group factors) has been explicitly traced back to such divisibility properties in \cite{Lin:2016vus}. It would be interesting to pursue a more general formulation of the connection between global anomalies and arithmetic properties of the cohomology ring of Calabi-Yau $n$-folds.

\paragraph{Ackowledgements}

We thank Eran Palti and  Wati Taylor for illuminating dicussions and Craig Lawrie, Ling Lin and Sakura Sch\"afer-Nameki  for important discussions and for collaboration on related topics. T.W. is grateful to MIT and Oxford University for hospitality, and C.M. and M.B. thank CERN for hospitality. The work of C.M. and M.B. are supported, respectively, by the Munich Excellence Cluster for Fundamental Physics `Origin and the Structure of the Universe' and by Studienstiftung des deutschen Volkes. This research has been funded in part by DFG under Transregio 33 `The Dark Universe' and GK `Particle Physics Beyond the Standard Model'.

\appendix

\section{Construction of Vertical Gauge Fluxes } \label{app_fluxconstruction}

In this appendix we prove the following assertion for a smooth elliptically fibred Calabi-Yau 4-fold $\hat Y_4$ arising as the resolution of a singular Weierstrass model: Given the decomposition of the primary vertical subspace 
\[ H^{2,2}_{\mathrm{vert}} ( \hat Y_4 ) = V_1 \cup V_2 \]
with 
\bea 
V_1 &=& \text{span} \left\{ [ E_{i_I} \cdot D^{\mathrm{b}}_\alpha ], \, [ D^{\mathrm{b}}_\alpha \cdot D^{\mathrm{b}}_\beta ], \, [ S_0 \cdot D^{\mathrm{b}}_\alpha ], \, [ S_A \cdot D^{\mathrm{b}}_\alpha ] \right\} \, , \\ \label{V3V4}
V_2 &=& \text{span} \left\{ \left[E_{i_I} \cdot E_{j_J} \right], \, \left[S_A \cdot E_{i_I} \right], \, \left[S_A \cdot S_B \right], \, \left[S_0 \cdot S_A \right] \right\} \,,
\eea
one can associate to each single generator of $V_2$ as listed in (\ref{V3V4}) a 4-form flux $G_4 \in H^{2,2}_{\mathrm{ vert}}$ satisfying the conditions (\ref{verticality1}) and (\ref{gaugeinvariantflux}) by adding elements of $V_1$.

The proof relies on general properties of the intersection numbers of elements in $H^{2,2}_{\mathrm{vert}}(\hat Y_4)$. Namely, for each element $[X \cdot Y] \in V_2$, the following intersection numbers hold: 
\begin{align} \label{XYrelations}
[ X ] \cdot [ Y ] \cdot [S_0 ] \cdot [D_\alpha^\mathrm{b} ] &= [S_0 ] \cdot [C_{XY} ] \cdot [D_\alpha^\mathrm{b} ] \,, \\
[ X ] \cdot [ Y ] \cdot [D_\alpha^\mathrm{b} ] \cdot [D_\beta^\mathrm{b} ] &= [S_0 ] \cdot [D_{XY} ] \cdot [D_\alpha^\mathrm{b} ] \cdot [D_\beta^\mathrm{b} ] \,, \\
[ X ] \cdot [ Y ] \cdot [E_{i_I} ] \cdot [D_\alpha^\mathrm{b} ] &= [S_0 ] \cdot [F_{XY,i_I} ] \cdot [W_I ] \cdot [D_\alpha^\mathrm{b} ] \,,
\end{align}
where $C_{XY}$ describes a curve class on the base $B_3$ (which may well be zero as \eg for $X= E_{i_I}$, $Y = E_{j_J}$) and $D_{XY}$ and $F_{XY,i_I}$ are divisor classes on $B_3$. The rationale behind (\ref{XYrelations}) is simply that the LHS can be expressed as an intersection product entirely on the base $B_3$, which is the hypersurface in $\hat Y_4$ defined by the zero-section $S_0$. For every given fibration over a general base $B_3$ the curve and divisor classes on the base appearing on the RHS can be explicitly computed, but their concrete form will not be needed for our argument. 

In terms of the classes $C_{XY}$, $D_{XY}$ and $F_{XY,i_I}$ we can define the algebraic 4-cycle class 
\[ A \left(X \cdot Y \right) = X \cdot Y - C_{XY} - \left(S_0 + D^{\rm b}_0 \right) \cdot D_{XY} + {\mathfrak C}_{l_L m_M}^{-1} E_{l_L} \cdot F_{XY,m_M} \in \mathrm{CH}^2 ( \hat Y_4 ) \,, \label{AXYclass} \]
where the matrix ${\mathfrak{C}}_{l_L m_M}$  
governs the intersections of the resolution divisors as in (\ref{intersectionEiEj}) 
and furthermore the divisor class $D_0^{\rm b}$ on $B_3$ is defined by the property that 
\[ [ S_0 ] \cdot [ S_0 + D^{\rm b}_0 ] \cdot [ D_\alpha^\mathrm{b} ] \cdot [ D_\beta^\mathrm{b} ] = 0 \,. \label{S0prop1} \]
The homology class associated with 4-cycle class $A(X \cdot Y)$ then gives rise to a transversal and gauge invariant flux
\[ G_4 \left( X \cdot Y \right) = \left[ A \left( X \cdot Y \right) \right] \in H^{2,2}_{\mathrm{ vert}} \,. \]
To show this we need to verify the transversality conditions (\ref{verticality1}) and gauge invariance (\ref{gaugeinvariantflux}). To this end, observe first that 
\[ G_4 \left( X \cdot Y \right) \cdot [ S_0 ] \cdot [ D_\alpha^\mathrm{b} ] = [ S_0 ] \cdot [ C_{XY} ] \cdot [ D_\alpha^\mathrm{b} ] - [ S_0 ] \cdot [ C_{XY} ] \cdot [ D_\alpha^\mathrm{b} ] = 0 \,, \label{transproof1} \]
where we used (\ref{S0prop1}) and $[S_0] \cdot [E_{l_L}] = 0$. Similarly
\[ G_4 \left( X \cdot Y \right) \cdot [ D_\alpha^\mathrm{b} ] \cdot [ D_\beta^\mathrm{b} ] = [ D_{XY} ] \cdot [ D_\alpha^\mathrm{b} ] \cdot [ D_\beta^\mathrm{b} ] \cdot [ S_0 ] - [ D_{XY} ] \cdot [ D_\alpha^\mathrm{b} ] \cdot [ D_\beta^\mathrm{b} ] \cdot [ S_0 ] = 0 \label{transproof2} \]
with the help of
\[ [ D_\alpha^\mathrm{b} ] \cdot [ D_\beta^\mathrm{b} ] \cdot [ D_\gamma^\mathrm{b} ] \cdot [ D_\delta^\mathrm{b} ] = 0 \,, \qquad [ E_{i_I} ] \cdot [ D^{\mathrm{b}}_\alpha ] \cdot [ D^{\mathrm{b}}_\beta ] \cdot [ D^{\mathrm{b}}_\gamma ] = 0 \,. \]
(\ref{transproof1}) and (\ref{transproof2}) establish transversality of the flux.

Finally, gauge invariance of the flux, (\ref{gaugeinvariantflux}), follows from (\ref{intersectionEiEj}), which can also be written as 
\bea
[ E_{i_I} ] \cdot [ E_{j_J} ] \cdot [ D_\alpha^{\rm b} ] \cdot [ D_\beta^{\rm b} ] &=&   - \delta_{IJ} \, {\mathfrak C}_{i_I j_I} \, [ S_0 ] \cdot [ W_I ] \cdot [ D_\alpha^{\rm b} ] \cdot [ D_\beta^{\rm b} ],
\eea
because
\begin{align}
\begin{split}
& G_4 \left( X \cdot Y \right) \cdot [ E_{i_I} ] \cdot [ D_\alpha^\mathrm{b} ] \\
& \qquad \qquad = [ S_0 ] \cdot [ W_I ] \cdot [ F_{XY,i_I} ]\cdot [ D_\alpha^\mathrm{b} ] - {\mathfrak C}^{-1}_{l_L m_M} {\mathfrak C}_{l_L i_I} [ W_I ] \cdot [ F_{XY,m_M} ] \cdot [ D_\alpha^\mathrm{b} ] \cdot [ S_0 ] = 0 \,.
\end{split}
\end{align}

Note that the fluxes constructed in this way from elements of $V_2$ are in general not linearly independent as elements of $H^{2,2}_\mathrm{vert}(\hat Y_4)$. In particular, it is a priori not guaranteed that the class $[A(X\cdot Y)]$ is non-trivial in $H^{2,2}_\mathrm{vert}(\hat Y_4)$. What is important for us, however, is that the homology classes associated with the correction terms in (\ref{AXYclass}) needed to render  $[A(X\cdot Y)]$ transversal and gauge invariant indeed lie in the subspace $V_1$.

\section{Details of the Proofs of  Section~\ref{sec:genrelChow}}
In this appendix we provide the missing proofs for some of the statements in \autoref{sec:genrelChow}.

\subsection{Cohomological Relations in Presence of Non-Vertical Matter Surfaces} \label{app_NonVertMatter}

We begin by considering a situation in which 
 some of the matter surface classes have a part in the remainder $H^{2,2}_{\mathrm{rem}}(\hat Y_4)$. 
An example would be a situation in which the matter curve $C_\bfR$ on $W_I$ splits into several components which are individually not complete intersections of $W_I$ with a divisor from the base $B_3$ \cite{Braun:2014pva}.
In this case, we can split the classes of the matter surfaces into orthogonal components
\[ \left[S^a(\mathbf{R})\right] = \left[S^a(\mathbf{R})\right]_{\mathrm{vert}} + \left[S^a(\mathbf{R})\right]_{\mathrm{rem}} \,. \]
and hence start from
\begin{align}
\begin{split} \label{G4dotanomaly1vert}
0 &= G_4 \cdot \left( \sum_{\mathbf{R} \neq \mathbf{adj}} \sum_{a} n^a_{\Lambda \Sigma \Gamma} \left( \mathbf{R} \right) \, \left[ S^a_{\mathbf{R}} \right]_\mathrm{vert} \, + \frac{1}{2} \sum_{\rho_I} n^{\rho_I}_{\Lambda \Sigma \Gamma} \, \left[ S^{\rho_I} \right] - 3 \, \cdot \left[ F_{(\Gamma} \right] \cdot \left[ \hat{\pi}_* \left( F_{\Lambda} \cdot F_{\Sigma )} \right) \right] \right) \\
& \qquad + G_4 \cdot \sum_{\mathbf{R} \neq \mathbf{adj}} \sum_{a} n^a_{\Lambda \Sigma \Gamma} \left( \mathbf{R} \right) \left[ S^a_{\mathbf{R}} \right]_\mathrm{rem} \,.
\end{split}
\end{align}
The terms in brackets of the first line all lie in $H^{2,2}_{\mathrm{vert}}(\hat Y_4)$ and are thus orthogonal to $H^{2,2}_{\mathrm{rem}}(\hat Y_4)$ and $H^{2,2}_{\mathrm{hor}}(\hat Y_4)$, while the terms next to $G_4$ in the second line are orthogonal to $H^{2,2}_{\mathrm{vert}}(\hat Y_4)$ and $H^{2,2}_{\mathrm{hor}}(\hat Y_4)$. Repeating the arguments leading to (\ref{Anomaly4dCohom1}) in \autoref{sec:genrelChow}, the expression in the first line is found to be orthogonal to the space $V_1$ and to all fluxes in $H^{2,2}_{\mathrm{vert}}(\hat Y_4)$, which is sufficient to conclude that the relations (\ref{summary4dCoho1}) hold in cohomology on $\hat Y_4$.
The second identity follows again from (\ref{G4dotanomaly2}). In addition, anomaly cancellation implies
\[ G_4 \cdot \sum_{\mathbf{R} \neq \mathbf{adj}} \sum_{a} n^a_{\Lambda \Sigma \Gamma} \left( \mathbf{R} \right) \left[ S^a_{\mathbf{R}} \right]_\mathrm{rem} = 0 \,, \qquad
G_4 \cdot \sum_{\mathbf{R} \neq \mathbf{adj}} \sum_{a} \beta^a_\Lambda \left( \mathbf{R} \right) \left[ S^a_{\mathbf{R}} \right]_\mathrm{rem} = 0 \]
for every choice of transversal flux $G_4$. In order to determine if a stronger set of relations can be deduced directly at the level of cohomology, more information is needed about the nature of $[S^a(\mathbf{R})]_{\mathrm{rem}} $ in non-trivial situations. As noted already, there do currently not exist any examples of fibrations with non-trivial $[S^a_\bfR]_{\mathrm{rem}}$ except those where a single $[S^a_\bfR]$ splits into various components, each with some contribution from $H^{2,2}_{\mathrm{rem}}(\hat Y_4)$, but such that the net $[S^a_\bfR]$ remains purely vertical. In particular, if the only contribution to $[S^a_\bfR]_\mathrm{rem}$ comes from a splitting of $C_\bfR$ into several curves which are not all obtained by intersection with a divisor, then for fixed $\bfR$ the sum over all distinct components of $[S^a_\bfR]$ is again purely vertical because the different components belonging to the same representation $\bfR$ come with the same prefactor. Hence in such examples the components along $H^{2,2}_{\mathrm{rem}}(\hat Y_4)$ trivially sum up to zero and there are no non-trivial relations among  $[S^a(\mathbf{R})]_{\mathrm{rem}}$ for different $\bfR$.

\subsection{Proof of Homological Relations for Flux Cycles} \label{App_homrelproof}

In this appendix we prove that the cohomological relations (\ref{summary4dCoho2}) are implied by anomaly cancellation in the 4-dimensional effective action obtained by F-theory compactified on a smooth elliptic fibration $\hat Y_4$. Our starting point  is the gauge anomaly relation (\ref{G4dotanomaly1}), rewritten in terms of (\ref{Aageneral1}) as\footnote{Note that for $\mathbf{R} = \mathbf{adj}$, $A^a(\mathbf{R}) = 0$ and therefore we can sum over all representations in the first line.}
\bea \label{gaugeanomrel2}
0 \, \, =&&G_4 \cdot \left( \sum_{\mathbf{R} } \sum_{a} n^a_{\Lambda \Sigma \Gamma} \left( \bfR \right) \, \left[ A^a \left( \mathbf{R} \right) \right] \, \right) \nonumber \\
+&& G_4 \cdot \left( \sum_{\mathbf{R} \neq \mathbf{adj}} \sum_a - n^a_{\Lambda \Sigma \Gamma} \left( \bfR \right) \, \left[ \Delta^a \left( \mathbf{R} \right) \right] \, + \frac{1}{2} \sum_{\rho_I} n^{\rho_I}_{\Lambda \Sigma \Gamma} \left[ S^{\rho_I} \right] \, \right) \\
+ && G_4 \cdot \left( - 3 \, \left[ F_{(\Gamma} \right] \cdot \left[ \hat{\pi}_* \left( F_{\Lambda} \cdot F_{\Sigma)} \right) \right] \right) \,. \nonumber
\eea
Similarly the gravitational anomaly relation can be formulated as 
\bea \label{gravanomrel2}
0 \, \, =&&G_4 \cdot \left( \sum_{\mathbf{R} } \sum_{a} \beta^a_{\Lambda} \left( \mathbf{R} \right) \, \left[ A^a \left( \mathbf{R} \right) \right] \, \right) \nonumber \\
+&& G_4 \cdot \left( \sum_{\mathbf{R} \neq \mathbf{adj}} \sum_a - \beta^a_{\Lambda} \left( \mathbf{R} \right)  \, \left[ \Delta^a \left( \mathbf{R} \right) \right] \, + \frac{1}{2} \sum_{\rho_I} \beta^{\rho_I}_{\Lambda} \, \left[ S^{\rho_I} \right] \, \, \right) \\
+ && G_4 \cdot \left( 6 \, \left[ F_{\Gamma} \right] \cdot \left[ \overline{K}_{B_3} \right] \right) \,. \nonumber
\eea
The expression in the first line of (\ref{gaugeanomrel2}) is by construction orthogonal to the subspace $V_3$ of $H^{2,2}(\hat Y_4)$ given by the span
\bea \label{span1}
V_3 = \text{span} \left\{ [ E_{i_I} \cdot D^{\mathrm{b}}_\alpha ], \, [ S_0 \cdot D^{\mathrm{b}}_\alpha ], \, [ D^{\mathrm{b}}_\alpha \cdot D^{\mathrm{b}}_\beta ] \right\} \subset H^{2,2}(\hat Y_4)
\eea
because the classes $[A^a(\mathbf{R})]$ represent gauge invariant fluxes which satisfy (\ref{verticality1}) as well as (\ref{gaugeinvariantflux}). The expression in  brackets in the second line is manifestly orthogonal to
\bea \label{span4}
V_4 = \text{span} \left\{ [ U_A \cdot D^{\mathrm{b}}_\alpha ], \, [ S_0 \cdot D^{\mathrm{b}}_\alpha ], \, [ D^{\mathrm{b}}_\alpha \cdot D^{\mathrm{b}}_\beta ] \right\} \subset H^{2,2}(\hat Y_4) \,,
\eea
and it is also manifestly orthogonal to the set of all gauge invariant fluxes $G_4$ satisfying (\ref{verticality1}) as well as (\ref{gaugeinvariantflux}).

The nature of the Green-Schwarz counterterms in the third line depends on the choice of indices $\Lambda, \Sigma, \Gamma$. Consider first the case where $\Lambda, \Sigma, \Gamma = i_I,j_J,k_K$ exclusively refer to the Cartan generators of the non-Abelian gauge group factors. 
In (\ref{span2}) we had defined the subspace $V_1$, which in fact is related to $V_3$ and $V_4$ via $V_1 = V_3 \cup V_4$.
As a result of  (\ref{DefcalC}) also the expression in brackets in the third line is orthogonal to $V_1$ and to the set of gauge invariant $G_4$. Since the sum of all three terms in brackets is orthogonal to the set of all gauge fluxes, which includes $ \text{span}_{\mathbb{C}} \{ [U_A \cdot D^{\mathrm{b}}_\alpha] \}$ and $ \text{span}_{\mathbb{C}} \{ [E_{i_I}] \cdot D^{\mathrm{b}}_\alpha] \}$, this implies that in this case the expression in brackets in the first line is by itself orthogonal to $V_1$. By the same arguments as those given after (\ref{span2}), this implies orthogonality to $H^{2,2}_\mathrm{vert}(\hat Y_4)$. With the same caveat as in the main text regarding the possibility that some of the matter surfaces might have a contribution in $H^{2,2}_\mathrm{rem}(\hat Y_4)$ we conclude
\bea\label{naijkA}
\sum_{\mathbf{R}} \sum_{a} n^a_{i_I j_K k_K} \left( \bfR \right) \, \left[ A^a \left( \mathbf{R} \right) \right]_{\mathrm{vert}} \, = 0 \quad \in H^{2,2} ( \hat Y_4 ) \,.
\eea
If one or several of the generators $F_\Sigma$ are associated with a non-Cartan $U(1)_A$, the expression in brackets in the third line has the same orthogonality properties of as those in the first line. The same arguments as before now imply that 
\[ \sum_{\mathbf{R} } \sum_{a} n^a_{A \Sigma \Gamma} \left( \bfR \right) \, \left[ A^a \left( \mathbf{R} \right) \right]_{\mathrm{vert}} \, - 3\, \left[ U_{(A} \right] \cdot \left[ \hat{\pi}_* \left( F_{\Sigma} \cdot F_{\Gamma)} \right) \right] = 0 \in H^{2,2} ( \hat Y_4 ) \label{mixedChow2} \]
and similarly
\[ \sum_{\mathbf{R} } \sum_{a} q_A \, \left[ A^a \left( \mathbf{R} \right) \right]_{\mathrm{vert}} \, + 6 \, \left[ U_A \right] \cdot \left[ \overline{K}_{B_3} \right] = 0 \in H^{2,2}( \hat Y_4 ) \,. \label{mixedChow3} \]
Note that the analogue of (\ref{mixedChow3}) with the index $A$ replaced by a Cartan index would be $\sum_{\mathbf{R} } \sum_{a} \beta^a_{i_I}(\bfR) \,  [A^a(\mathbf{R})]_{\mathrm{vert}} =0$ (without a Green-Schwarz term). This is a trivial relation since $A^a(\mathbf{R})$ is independent of $a$ and $\sum_{a} \beta^a_{i_I}(\bfR)=0$ for a semi-simple Lie algebra.

\section{Fibre Structure of the \texorpdfstring{$\mathbf{SU(5) \times U(1)_X}$}{SU(5) x U(1)} Model} \label{sec:FibreStructure}

In this section we present a brief summary of the $\mathbb{P}^1$-fibrations realised over the matter curves in the model discussed in \autoref{subsec:SpecialFTheoryGUTModel}. For further information we refer the interested reader to Appendix B of \cite{Bies:2017fam
} or the original literature \cite{Krause:2011xj}.

\subsection*{\texorpdfstring{$\mathbf{\mathbb{P}^1}$}{P1}-Fibrations over \texorpdfstring{$\mathbf{C_{\mathbf{10}_1}}$}{C10}}

Over $C_{\mathbf{10}_1}$ the resolution divisors $E_i$ split according to the pattern
\begin{align}
\begin{split}
\label{splitting101}
{E_0}|_{C_{\mathbf{10}_1}} &= \mathbb{P}_{0A}^1\left( \mathbf{10}_1 \right) \, , \qquad \qquad \qquad \quad \,
{E_1}|_{C_{\mathbf{10}_1}} = \mathbb{P}_{14}^1\left( \mathbf{10}_1 \right) \, , \\
{E_2}|_{C_{\mathbf{10}_1}} &= \mathbb{P}_{24}^1\left( \mathbf{10}_1 \right) + \mathbb{P}_{2B}^1\left( \mathbf{10}_1 \right) \, , \qquad
{E_3}|_{C_{\mathbf{10}_1}} = \mathbb{P}_{3C}^1\left( \mathbf{10}_1 \right) \, , \\
{E_4}|_{C_{\mathbf{10}_1}} &= \mathbb{P}_{4D}^1\left( \mathbf{10}_1 \right) + \mathbb{P}_{14}^1\left( \mathbf{10}_1 \right) \, .
\end{split}
\end{align}

The so-defined surfaces are explicitly given by
\begin{itemize}
 \item $\mathbb{P}_{0A}^1 \left( \mathbf{10}_1 \right) = V \left( a_{1,0} \left( z_i \right), e_0, y^2 e_4 - x^3 s e_1 e_2^2 \right)$,
 \item $\mathbb{P}_{14}^1 \left( \mathbf{10}_1 \right) = V \left( a_{1,0} \left( z_i \right), e_1, e_4 \right)$,
 \item $\mathbb{P}_{24}^1 \left( \mathbf{10}_1 \right) = V \left( a_{1,0} \left( z_i \right), e_2, e_4 \right)$,
 \item $\mathbb{P}_{2B}^1 \left( \mathbf{10}_1 \right) = V \left( a_{1,0} \left( z_i \right), e_2, y s e_3 + a_{3,2} \left( z_i \right) z^3 e_0^2 e_1 \right)$,
 \item $\mathbb{P}_{3C}^1 \left( \mathbf{10}_1 \right) = V \left( a_{1,0} \left( z_i \right), e_3, a_{3,2} \left( z_i \right) y z e_0 e_4 - a_{2,1} \left( 
      z_i \right) x^2 s e_2 - a_{4,3} \left( z_i \right) x z^2 e_0^2 e_1 e_2 e_4 \right)$,
 \item $\mathbb{P}^1_{4D} \left( \mathbf{10}_1 \right) = V \left( a_{1,0} \left( z_i \right), e_4, x s e_2 e_3 + a_{2,1} \left( z_i \right) z^2 e_0 
      \right)$.
\end{itemize}

The matter surfaces $S_{\mathbf{10}_1}^{(a)}$ over $C_{\mathbf{10}_1}$ are linear combinations of these fibrations. Our notation is that  $\vec{P}$ is a list of the multiplicites with which these $\mathbb{P}^1$-fibrations appear in the above order. Hence
\[ \vec{P} = \left( 0, 1, 0, 4, 0, 0 \right) \quad \leftrightarrow \quad 1 \cdot \mathbb{P}^1_{14} \left( \mathbf{10}_1 \right) + 4 \cdot \mathbb{P}^1_{2B} \left( \mathbf{10}_1 \right) \, . \]
$\vec{\beta}$ indicates the Cartan charges of such a linear combination, \ie lists the intersection numbers with the resolution divisors $E_i$, $1 \leq i \leq 4$. The matter surfaces over $C_{\mathbf{10}_1}$ are as follows:
{\small
\begin{align}
\begin{tabular}{|c||c|c||c||c|c|}
\toprule
Label & $\vec{P}$ & $\vect{\beta}$ & Label & $\vec{P}$ & $\vect{\beta}$ \\
\hline \hline
$S_{\mathbf{10}_1}^{(1)}$ & $\left( 0, -1, -2, -1, -1, 0 \right)$ & $\left( 0, 1, 0, 0 \right)$ & $S_{\mathbf{10}_1}^{(6)}$ & $\left( 0, 0, 0, 0, 0, 1 \right)$ & $\left( 1, 0, 0, -1 \right)$ \\
$S_{\mathbf{10}_1}^{(2)}$ & $\left( 0, -1, -1, 0, -1, 0 \right)$ & $\left( 1, -1, 1, 0 \right)$ & $S_{\mathbf{10}_1}^{(7)}$ & $\left( 0, 0, 0, 1, 0, 0 \right)$ & $\left( 0, -1, 0, 1 \right)$ \\
$S_{\mathbf{10}_1}^{(3)}$ & $\left( 0, 0, -1, 0, -1, 0 \right)$ & $\left( -1, 0, 1, 0 \right)$ & $S_{\mathbf{10}_1}^{(8)}$ & $\left( 0, 1, 0, 0, 0, 1 \right)$ & $\left( -1, 1, 0, -1 \right)$ \\
$S_{\mathbf{10}_1}^{(4)}$ & $\left( 0, -1, -1, 0, 0, 0 \right)$ & $\left( 1, 0, -1, 1 \right)$ & $S_{\mathbf{10}_1}^{(9)}$ & $\left( 0, 1, 1, 1, 0, 1 \right)$ & $\left( 0, -1, 1, -1 \right)$ \\
$S_{\mathbf{10}_1}^{(5)}$ & $\left( 0, 0, -1, 0, 0, 0 \right)$ & $\left( -1, 1, -1, 1 \right)$ & $S_{\mathbf{10}_1}^{(10)}$ & $\left( 0, 1, 1, 1, 1, 1 \right)$ & $\left( 0, 0, -1, 0 \right)$ \\
\bottomrule
\end{tabular}
\end{align}
}

\subsection*{\texorpdfstring{$\mathbf{\mathbb{P}^1}$}{P1}-Fibrations over \texorpdfstring{$\mathbf{C_{\mathbf{5}_3}}$}{C53}}

Over $C_{\mathbf{5}_3}$ the splitting of the resolution divisors takes the form
\begin{align}
\begin{split}
\label{splitting53}
{E_0}|_{C_{\mathbf{5}_3}} = \mathbb{P}_{0}^1\left( \mathbf{5}_3 \right) \, , \qquad \qquad \qquad \quad \, \,
{E_1}|_{C_{\mathbf{5}_3}} &= \mathbb{P}_{1}^1\left( \mathbf{5}_3 \right) \, , \qquad
{E_2}|_{C_{\mathbf{5}_3}} = \mathbb{P}_{2E}^1\left( \mathbf{5}_3 \right) \, , \\
{E_3}|_{C_{\mathbf{5}_3}} = \mathbb{P}_{3x}^1\left( \mathbf{5}_3 \right) + \mathbb{P}_{3F}^1\left( \mathbf{5}_3 \right) \, , \qquad
{E_4}|_{C_{\mathbf{5}_3}} &= \mathbb{P}_{4}^1\left( \mathbf{5}_3 \right)
\end{split}
\end{align}
with
\begin{itemize}
 \item $\mathbb{P}_{0}^1 \left( \mathbf{5}_3 \right) = V \left( a_{3,2}, e_0, a_{1,0} x y z - e_1 e_2^2 e_3 s x^3 + e_3 e_4 y^2 \right)$,
 \item $\mathbb{P}_{1}^1 \left( \mathbf{5}_3 \right) = V \left( a_{3,2}, e_1, e_3 e_4 y + a_{1,0} x z \right)$,
 \item $\mathbb{P}_{2E}^1 \left( \mathbf{5}_3 \right) = V \left( a_{3,2}, e_2, e_3 e_4 y + a_{1,0} x z \right)$,
 \item $\mathbb{P}_{3x}^1 \left( \mathbf{5}_3 \right) = V \left( a_{3,2}, e_3, x \right)$,
 \item $\mathbb{P}_{3F}^1 \left( \mathbf{5}_3 \right) = V \left( a_{3,2}, e_3, a_{1,0} s y - a_{2,1} e_0 e_1 e_2 s x z - a_{4,3} e_0^3 e_1^2 e_2 e_4 z^3 \right)$, 
 \item $\mathbb{P}_{4}^1 \left( \mathbf{5}_3 \right) = V \left( a_{3,2}, e_4, a_{1,0} y z - e_1 e_2^2 e_3 s x^2 - a_{2,1} e_0 e_1 e_2 x z^2 \right)$ \,.
\end{itemize}
The matter surfaces over $C_{\mathbf{5}_3}$ are as follows:
{\small
\begin{align}
\begin{tabular}{|c||c|c||c||c|c|}
\toprule
Label & $\vec{P}^1$ & $\vect{\beta}$ & Label & $\vec{P}^1$ & $\vect{\beta}$ \\
\hline \hline
$S_{\mathbf{5}_3}^{(1)}$ & $\left( 0, -1, -1, -1, 0, 0 \right)$ & $\left( 1, 0, 0, 0 \right)$ & $S_{\mathbf{5}_3}^{(4)}$ & $\left( 0, 0, 0, 0, 1, 0 \right)$ & $\left( 0, 0, -1, 1 \right)$ \\
$S_{\mathbf{5}_3}^{(2)}$ & $\left( 0, 0, -1, -1, 0, 0 \right)$ & $\left( -1, 1, 0, 0 \right)$ & $S_{\mathbf{5}_3}^{(5)}$ & $\left( 0, 0, 0, 0, 1, 1 \right)$ & $\left( 0, 0, 0, -1 \right)$ \\
$S_{\mathbf{5}_3}^{(3)}$ & $\left( 0, 0, 0, -1, 0, 0 \right)$ & $\left( 0, -1, 1, 0 \right)$ & & & \\
\bottomrule
\end{tabular}
\end{align}
}

\subsection*{\texorpdfstring{$\mathbf{\mathbb{P}^1}$}{P1}-Fibrations over \texorpdfstring{$\mathbf{C_{\mathbf{5}_{-2}}}$}{C5-2}}

Over $C_{\mathbf{5}_{-2}}$ the splitting follows the pattern
\begin{align}
\begin{split} \label{splitting5-2}
{E_0}|_{C_{\mathbf{5}_{-2}}} = \mathbb{P}_{0}^1\left( \mathbf{5}_{-2} \right), \qquad \qquad \qquad \qquad \,
{E_1}|_{C_{\mathbf{5}_{-2}}} &= \mathbb{P}_{1}^1\left( \mathbf{5}_{-2} \right), \qquad
{E_2}|_{C_{\mathbf{5}_{-2}}} = \mathbb{P}_{2}^1\left( \mathbf{5}_{-2} \right), \\
{E_3}|{C_{\mathbf{5}_{-2}}} = \mathbb{P}_{3G}^1\left( \mathbf{5}_{-2} \right) + \mathbb{P}_{3H}^1\left( \mathbf{5}_{-2} \right) , \qquad
{E_4}|{C_{\mathbf{5}_{-2}}} &= \mathbb{P}_{4}^1\left( \mathbf{5}_{-2} \right)
\end{split}
\end{align}
with\footnote{These results differ slightly from \cite{Krause:2011xj}, which did not make use of primary decompositions.}
\begin{align*}
\mathbb{P}_{0}^1 \left( \mathbf{5}_{-2} \right) &= V \left( a_{3,2} a_{2,1} - a_{4,3} a_{1,0}, e_0, e_3 e_4 y^2 + a_{1,0} x y z - e_1 e_2^2 e_3 s x^3 \right), \\
\mathbb{P}_{1}^1 \left( \mathbf{5}_{-2} \right) &= V \left( a_{3,2} a_{2,1} - a_{4,3} a_{1,0}, e_1, e_3 e_4 y + a_{1,0} x z \right), \\
\mathbb{P}_{2}^1 \left( \mathbf{5}_{-2} \right) &= V \left( a_{3,2} a_{2,1} - a_{4,3} a_{1,0}, e_2, e_0^2 z^3 e_1 e_4 a_{3,2} + y s e_3 e_4 + a_{1,0} x z s, \right. \\
      & \qquad \qquad \qquad \qquad \qquad \qquad \left. a_{1,0} a_{4,3} e_0^2 z^3 e_1 e_4 + a_{2,1} y s e_3 e_4 + a_{1,0} a_{2,1} x z s \right), \\
\mathbb{P}_{3G}^1 \left( \mathbf{5}_{-2} \right) &= V \left( a_{3,2} a_{2,1} - a_{4,3} a_{1,0}, e_3, a_{4,3} e_0^2 z^2 e_1 e_4 + a_{2,1} x s, a_{3,2} e_0^2 z^2 e_1 e_4 + a_{1,0} x s \right), \\
\mathbb{P}_{3H}^1 \left( \mathbf{5}_{-2} \right) &= V \left( a_{3,2} a_{2,1} - a_{4,3} a_{1,0}, e_3, a_{4,3} e_0 x z e_1 e_2 - a_{3,2} y, a_{2,1} e_0 x z e_1 e_2 - a_{1,0} y  \right), \\
\mathbb{P}_{4}^1 \left( \mathbf{5}_{-2} \right) &= V \left( a_{3,2} a_{2,1} - a_{4,3} a_{1,0}, e_4, a_{1,0} y z - a_{2,1} e_0 e_1 e_2 x z^2 - e_1 e_2^2 e_3 s x^2 \right).
\end{align*}

The matter surfaces over $C_{\mathbf{5}_{-2}}$ are as follows:
{\small
\begin{align}
\begin{tabular}{|c||c|c||c||c|c|}
\toprule
Label & $\vec{P}^1$ & $\vect{\beta}$ & Label & $\vec{P}^1$ & $\vect{\beta}$ \\
\hline \hline
$S_{\mathbf{5}_{-2}}^{(1)}$ & $\left( 0, -1, -1, -1, 0, 0 \right)$ & $\left( 1, 0, 0, 0 \right)$ & $S_{\mathbf{5}_{-2}}^{(4)}$ & $\left( 0, 0, 0, 0, 1, 0 \right)$ & $\left( 0, 0, -1, 1 \right)$ \\
$S_{\mathbf{5}_{-2}}^{(2)}$ & $\left( 0, 0, -1, -1, 0, 0 \right)$ & $\left( -1, 1, 0, 0 \right)$ & $S_{\mathbf{5}_{-2}}^{(5)}$ & $\left( 0, 0, 0, 0, 1, 1 \right)$ & $\left( 0, 0, 0, -1 \right)$ \\
$S_{\mathbf{5}_{-2}}^{(3)}$ & $\left( 0, 0, 0, -1, 0, 0 \right)$ & $\left( 0, -1, 1, 0 \right)$ & & & \\
\bottomrule
\end{tabular}
\end{align}
}

\subsection*{\texorpdfstring{$\mathbf{\mathbb{P}^1}$}{P1}-Fibrations over \texorpdfstring{$\mathbf{C_{\mathbf{1}_{5}}}$}{C15}}

{\small
\begin{align}
\begin{tabular}{|c|c|c|c|}
\toprule
Label & Vanishing Locus & $q$ & $\vect{\beta}$ \\
\hline \hline
$\mathbb{P}_{A}^1 \left( \mathbf{1}_{5} \right) = S_{\mathbf{1}_{5}}$ & $V \left( a_{3,2} \left( z_i \right), a_{4,3} \left( z_i \right), s \right)$ & $5$ & $\left( 0, 0, 0, 0 \right)$ \\
$\mathbb{P}_{B}^1 \left( \mathbf{1}_{5} \right)$ & \pbox{10cm}{$V \left( a_{3,2}, a_{4,3}, y^2 e_3 e_4 + a_{1,0} x y z \right.$ \\  \phantom{\hspace{5em}} $\left. - x^3 s e_1 e_2^2 e_3 - a_{2,1} x^2 z^2 e_0 e_1 e_2 \right)$} & $-5$ & $\left( 0,0,0,0 \right)$ \\
\bottomrule
\end{tabular}
\end{align}
}
Since $\mathbb{P}^1_B \left( \mathbf{1}_{5} \right)$ intersects with $z = 0$, it is discarded as matter surface.

\section{Proving the Chow Relations for \texorpdfstring{$\mathbf{SU(5) \times U(1)_X}$}{SU(5)xU(1)}} \label{sec:Proof}

In this appendix we prove the relations (\ref{summary4dChow2}) for the $SU(5) \times U(1)_X$ model considered in the main text.
In the concrete geometry at hand, (\ref{summary4dChow2}) reduces to the system of equations (\ref{relation1}), (\ref{relation2}),  (\ref{relation3}).

\subsection{Proof of Relation \texorpdfstring{\eqref{relation1}}{(5.14)}} \label{subsec:ProofFluxRelation1}

In order to prove the first relation (\ref{relation1}) we make use of the explicit representation of the involved fluxes as given in (\ref{MSFexplicit}) and recall that the $\mathbb{P}^1$-fibred surfaces appearing in these expressions are the (split components of  the) restriction of the resolution divisors as listed in (\ref{splitting101}), (\ref{splitting53}) and (\ref{splitting5-2}). The resolution divisors which do not split can be `summed up' and expressed as a pullback from the ambient space $\hat X_5$. We then see that (\ref{relation1}) is equivalent to
\[ \mathbb{P}^1_{3G} \left( \mathbf{5}_{-2} \right) - \mathbb{P}^1_{3F} \left( \mathbf{5}_3 \right) - \mathbb{P}^1_{2B} \left( \mathbf{10}_1 \right) = \left. \left( \mathcal{W} - 2 \overline{\mathcal{K}}_{B_3} \right) \cdot \left( \mathcal{E}_1 + 2 \mathcal{E}_2 - \mathcal{E}_4 \right) \right|_{\hat Y_4} \, . \label{relation1b} \]
The explicit form of the surfaces $\mathbb{P}^1_{3G} ( \mathbf{5}_{-2} )$, $\mathbb{P}^1_{3F} ( \mathbf{5}_3 )$ and $\mathbb{P}^1_{2B} ( \mathbf{10}_1 )$ as vanishing loci of certain ideals is given in \autoref{sec:FibreStructure}. To prove that (\ref{relation1b}) holds true in $\text{CH}^\bullet ( \hat{Y}_4)$ we proceed in various steps.
\begin{enumerate}
\item We first observe that up to rational equivalence in $\hat{Y}_4$, the involved surfaces satisfy
     \[ \mathbb{P}^1_{3G} \left( \mathbf{5}_{-2} \right) - \mathbb{P}^1_{3F} \left( \mathbf{5}_3 \right) + \mathbb{P}^1_{2B} \left( \mathbf{10}_1 \right) =  V \left(P_T^\prime, e_3, e_4 \right)  -  V \left(P_T^\prime, e_2, x \right) \, . \label{starting-rel1} \]
     We will justify this statement below. 
 \item Let us now subtract $2 \mathbb{P}^1_{2B} ( \mathbf{10}_1 )$ from both sides. By use of the explicit representation of $\mathbb{P}^1_{2B} ( \mathbf{10}_1 )$ given   
      in \autoref{sec:FibreStructure}, we  arrive at
      \begin{align}
      \begin{split}
      \label{relation1c}
      &\mathbb{P}^1_{3G} \left( \mathbf{5}_{-2} \right) - \mathbb{P}^1_{3F} \left( \mathbf{5}_3 \right) - \mathbb{P}^1_{2B} \left( \mathbf{10}_1 \right) 
      =  V \left( P_T^\prime, e_3, e_4 \right) \\ 
      &\qquad \qquad -  V \left(P_T^\prime, e_2, x \right)  - 2  V \left( P_T^\prime, \overline{K}_{B_3}, e_2 \right)  + 2  V \left( P_T^\prime, e_2, e_4 \right) \,.
      \end{split}
      \end{align}
 \item The final step consists in showing that the RHS can  be simplified further. As detailed in \autoref{subsec:AUsefulIdentity}, we have the following 
      identity in $\text{CH}^2 ( \hat X_5)$
      \[ \left( \mathcal{W} - 2 \overline{\mathcal{K}}_{B_3} \right) \cdot \left( \mathcal{E}_1 + 2 \mathcal{E}_2 - \mathcal{E}_4 \right) = \mathcal{E}_3 \cdot \mathcal{E}_4 - \mathcal{E}_2 \cdot \mathcal{X} - 2 \overline{\mathcal{K}}_{B_3} \cdot \mathcal{E}_2 + 2 \mathcal{E}_2 \cdot \mathcal{E}_4 \, . \label{usefulidentity} \]
      According to the discussion around (\ref{pullbackformula}), pulling back all terms to $\hat Y_4$ gives a corresponding relation in $\mathrm{CH}^2(\hat Y_4)$. In particular we show in \autoref{subsec:AUsefulIdentity} that after pullback to $\hat Y_4$ the RHS of (\ref{usefulidentity}) and of (\ref{relation1c}) coincide. Thereby we arrive at (\ref{relation1b}), as desired.
\end{enumerate}

To complete the proof, it remains to justify (\ref{starting-rel1}). By performing a primary ideal decomposition we obtain the following two identities in rational equivalence of $\hat{Y}_4$,
\begin{align}
\begin{split}
V \left(P_T^\prime, e_3, a_{3,2} e_0^2 z^2 e_1 e_4 + a_{1,0} x s \right) &= \mathbb{P}^1_{3x} \left( \mathbf{5}_3 \right) + \mathbb{P}^1_{3G} \left( \mathbf{5}_{-2} \right) \\
& \qquad + V \left( e_3, e_2, e_0^2 z^2 e_1 e_4 a_{3,2} + x s a_{1,0} \right)\,, \\
V \left(P_T^\prime, e_2, e_0^2 z^2 e_1 e_4 a_{3,2} + x s a_{1,0} \right) &= \mathbb{P}^1_{24} \left( \mathbf{10}_1 \right) + V \left(e_3, e_2, e_0^2 z^2 e_1 e_4 a_{3,2} + x s a_{1,0} \right).
\end{split}
\end{align}
As a consequence, we learn
\begin{align}
\begin{split}
\mathbb{P}^1_{3G} \left( \mathbf{5}_{-2} \right) + \mathbb{P}^1_{3x} \left( \mathbf{5}_3 \right) - \mathbb{P}^1_{24} \left( \mathbf{10}_1 \right) &= V \left(P_T^\prime, e_3, a_{3,2} e_0^2 z^2 e_1 e_4 + a_{1,0} x s \right) \\
& \qquad - V \left( P_T^\prime, e_2, a_{3,2} e_0^2 z^2 e_1 e_4 + a_{1,0} x s \right) \, . \label{equ:RelationAmongMatterSurfacesForInteractionY2}
\end{split}
\end{align}
At this stage we make use of the very definition of rational equivalence as recalled around equation (\ref{Gammdefintion-general}). To this end we define a cycle
$\Gamma_1(t)$ on $\mathbb P^1 \times \hat Y_4$ parametrized by $t \in \mathbb P^1$ such that $\Gamma_1(t) = V (P_T^\prime, e_3, a_{3,2} e_0^2 z^2 e_1 e_4 + t a_{1,0} x s )$. By definition, $\Gamma_1(t=0) = V ( P_T^\prime, e_3, a_{3,2} e_0^2 z^2 e_1 e_4 )$ and $\Gamma_1(t=1)$ are rationally equivalent cycles. Note also that
\[ V \left( P_T^\prime, e_3, a_{3,2} e_0^2 z^2 e_1 e_4\right) = \left. V \left( e_3, a_{3,2} e_0^2 z^2 e_1 e_4 \right) \right|_{\hat{Y}_4} = \left. V \left( e_3, a_{3,2} e_4 \right) \right|_{\hat{Y}_4} \]
where the second equality makes use of the Stanley-Rei{s}ner ideal (\ref{SRideal}) of the \emph{top}. Consequently we learn that in rational equivalence of $\hat{Y}_4$
\[ V \left( P_T^\prime, e_3, a_{3,2} e_0^2 z^2 e_1 e_4 + a_{1,0} x s \right) = V \left( P_T^\prime, e_3, a_{3,2} \right) + V \left( P_T^\prime, e_3, e_4 \right) \, . \label{zwischen-rel1a} \]
Similarly we can consider the cycle $\Gamma_2(t) = V ( P_T^\prime, e_2, t \, a_{3,2} e_0^2 z^2 e_1 e_4 + a_{1,0} x s )$ and exploit the rational equivalence of $\Gamma_2(t=1)$ and $\Gamma_2(t=0)$ to conclude
\[ V \left(P_T^\prime, e_2, a_{3,2} e_0^2 z^2 e_1 e_4 + a_{1,0} x s \right) = V \left( P_T^\prime, e_2, a_{1,0} \right) + V \left( P_T^\prime, e_2, x \right) \, . \label{zwischen-rel1b}  \]
The first terms on the RHS of (\ref{zwischen-rel1a}) and (\ref{zwischen-rel1b}) represent two roots restricted to the matter curves $C_{\mathbf{5}_3}$ and $C_{\mathbf{10}_1}$, respectively. As detailed in \autoref{sec:FibreStructure}, these are related to the relevant surfaces in (\ref{starting-rel1}) as follows 
\[ - \mathbb{P}_{24}^1\left( \mathbf{10}_1 \right) = \mathbb{P}_{2B}^1\left( \mathbf{10}_1 \right) -  V(P_T^\prime, e_2, a_{1,0}) \, , \qquad
\mathbb{P}_{3x}^1 \left( \mathbf{5}_3 \right) = - \mathbb{P}_{3F}^1 \left( \mathbf{5}_3 \right) + V(P_T^\prime, e_3, a_{3,2}) \, . \]
Plugging everything into (\ref{equ:RelationAmongMatterSurfacesForInteractionY2}), we recover (\ref{starting-rel1}) as claimed.

\subsection{Proof of Relation \texorpdfstring{\eqref{relation2}}{(5.15)}} \label{subsec:ProofFluxRelation2}

The second relation, (\ref{relation2}), is equivalent to the following statement in $\text{CH}^2( \hat{Y}_4)$:
\begin{align}
\begin{split}
\mathbb{P}^1_{3H} \left( \mathbf{5}_{-2} \right) &= \left. \overline{\mathcal{K}}_{B_3} \cdot \left( \mathcal{E}_1 + 2 \mathcal{E}_2 + 3 \mathcal{E}_3 - \mathcal{E}_4 \right) \right|_{\hat{Y}_4} - \left. \mathcal{W} \cdot \left( \mathcal{E}_1 + 2 \mathcal{E}_2 + 3 \mathcal{E}_3 \right) \right|_{\hat{Y}_4} \\
& \qquad + \left. \overline{\mathcal{K}}_{B_3} \mathcal{W} \right|_{\hat{Y}_4} - \left. \mathcal{S} \mathcal{W} \right|_{\hat{Y}_4} + \left. \mathcal{Z} \mathcal{W} \right|_{\hat{Y}_4} \, . \label{relation2-form2}
\end{split}
\end{align}
We proceed in several steps for the proof.
\begin{enumerate}
 \item First, a primary decomposition of $V \left( P_T^\prime, e_3, a_{2,1} e_0 x z e_1 e_2 - a_{1,0} y \right)$ yields
      \[ \mathbb{P}^1_{3H} \left( \mathbf{5}_{-2} \right) = V \left( P_T^\prime, e_3, a_{2,1} e_0 x z e_1 e_2 - a_{1,0} y \right) - V \left( P_T^\prime, e_3, e_4 \right) \, . \]
      By the method of  rational homotopy on $\mathbb P^1 \times \hat Y_4$ we find the relation
      \[ V \left( P_T^\prime, e_3, a_{2,1} e_0 x z e_1 e_2 - a_{1,0} y \right) = V(P_T^\prime,e_3,a_{1,0}) + V(P_T^\prime,e_3,y) \in \mathrm{CH}^2(\hat Y_4) \,. \]
      We combine these finding to conclude that in rational equivalence on $\hat Y_4$
      \[ \mathbb{P}^1_{3H} \left( \mathbf{5}_{-2} \right) = \left. \mathcal{E}_3 \cdot \overline{\mathcal{K}}_{B_3} \right|_{\hat Y_4} + \left. \mathcal{E}_3 \cdot \mathcal{Y} \right|_{\hat Y_4} - \left. \mathcal{E}_3 \cdot \mathcal{E}_4 \right|_{\hat Y_4} \,. \label{P13hrel2} \]
 \item To compare this with (\ref{relation2-form2}), first note that 
      \begin{align}
      \begin{split}
      \left. \mathcal{E}_3 \cdot \left( \mathcal{Y} - \mathcal{E}_4 \right) \right|_{\hat{Y}_4} &= \left. \overline{\mathcal{K}}_{B_3} \cdot \left( \mathcal{E}_1 + 2 \mathcal{E}_2 + 3 \mathcal{E}_3 - \mathcal{E}_4 \right) \right|_{\hat{Y}_4} - \left. \mathcal{W} \cdot \left( \mathcal{E}_1 + 2 \mathcal{E}_2 + 3 \mathcal{E}_3 \right) \right|_{\hat{Y}_4} \\
      &\qquad + \left. \overline{\mathcal{K}}_{B_3} \cdot \mathcal{W} \right|_{\hat{Y}_4} - \left. \mathcal{S} \cdot \mathcal{W} \right|_{\hat{Y}_4} + \left. \mathcal{Z} \cdot \mathcal{W} \right|_{\hat{Y}_4} \label{Ve3y}
      \end{split}
      \end{align}
      holds in rational equivalence on $\hat Y_4$ if and only if 
      \[ 0 = - \left. \mathcal{E}_0 \mathcal{E}_1 \right|_{\hat{Y}_4} + \left. \mathcal{E}_0 \overline{\mathcal{K}}_{B_3} \right|_{\hat{Y}_4} + 
      \left. \mathcal{E}_0 \mathcal{Z} \right|_{\hat{Y}_4} \,. \label{zwischen-rel2a} \]
      The equivalence of (\ref{Ve3y}) and (\ref{zwischen-rel2a})  follows readily from the linear relations induced by the $\text{SU} ( 5 ) \times U ( 1 )_X$-top and the trivial restrictions introduced in \autoref{subsec:TrivialRestrictions}. Since (\ref{Ve3y}) and (\ref{P13hrel2}) imply (\ref{relation2-form2}), it remains to prove (\ref{zwischen-rel2a}).
 \item Indeed, a primary decomposition shows
      \begin{align}
      \begin{split}
      V \left( P_T^\prime, e_0, x^3 s e_1 e_2^2 - y^2 e_4 \right) &= V \left( e_0, a_{1,0}, x^3 s e_1 e_2^2 - y^2 e_4 \right) \\
      & \qquad \qquad \qquad \qquad \qquad + V \left( e_0, z, x^3 s e_1 e_2^2 - y^2 e_4 \right) \\
      &= V \left( P_T^\prime, e_0, a_{1,0} \right) + V \left( P_T^\prime, e_0, z \right) \, . \label{equ:proof2last1}
      \end{split}
      \end{align}
      Using a suitable rational homotopy, we furthermore have the following identities in rational equivalence of $\hat{Y}_4$
      \[ V \left( P_T^\prime, e_0, x^3 s e_1 e_2^2 - y^2 e_4 \right) = V \left( P_T^\prime, e_0, x^3 s e_1 e_2^2 \right) = V \left( P_T^\prime, e_0, e_1 \right) \,, \label{equ:proof2last2} \]
      where the last step uses the Stanley-Reisner ideal on $\hat Y_4$. Combining (\ref{equ:proof2last1}) and (\ref{equ:proof2last2}) implies (\ref{zwischen-rel2a}), as desired.
\end{enumerate}

\subsection{Proof of Relation \texorpdfstring{\eqref{relation3}}{(5.16)}} \label{subsec:ProofFluxRelation3}

\begin{enumerate}
 \item First we make use of (\ref{equ:U1abcX-Fluxabc-X5}) and (\ref{equ:MatterSurfaceabcFluxes-X5abc}) to write the final relation (\ref{relation3}) as
      \begin{align}
      \begin{split}
             \left. \mathcal{S} \cdot \left( 3 \overline{\mathcal{K}}_{B_3} - 2 \mathcal{W} - \mathcal{X} \right) \right|_{\hat{Y}_4} &= \frac{1}{5} \left( 6 \overline{\mathcal{K}}_{B_3} - 5 \mathcal{W} \right) \\
             & \qquad \quad \left. \cdot \left( 2 \mathcal{E}_1 + 4 \mathcal{E}_2 + 6 \mathcal{E}_3 + 3 \mathcal{E}_4 + 5 \mathcal{S} - 5 \mathcal{Z} - 5 \overline{\mathcal{K}}_{B_3} \right) \right|_{\hat{Y}_4} \\
             &\qquad - \left. \frac{1}{5} \left( 2 \mathcal{E}_1 - \mathcal{E}_2 + \mathcal{E}_3 - 2 \mathcal{E}_4 \right) \overline{\mathcal{K}}_{B_3} \right|_{\hat{Y}_4} - \left. \mathcal{E}_2 \mathcal{E}_4 \right|_{\hat{Y}_4} \, ,
      \end{split}
      \end{align}
      which in turn is equivalent to      
      \begin{align}
      \begin{split}
      0 &= - \left. \mathcal{E}_2 \mathcal{E}_4 \right|_{\hat Y_4} + 2 \left. \mathcal{E}_1 \overline{\mathcal{K}}_{B_3} \right|_{\hat Y_4} + 5 \left. \mathcal{E}_2 \overline{\mathcal{K}}_{B_3} \right|_{\hat Y_4} + 7 \left. \mathcal{E}_3 \overline{\mathcal{K}}_{B_3} \right|_{\hat Y_4} + 4 \left. \mathcal{E}_4 \overline{\mathcal{K}}_{B_3} \right|_{\hat Y_4} \\
      & \qquad - 6 \left. \overline{\mathcal{K}}_{B_3}^2 \right|_{\hat Y_4} + 3 \left. \overline{\mathcal{K}}_{B_3} \mathcal{S} \right|_{\hat Y_4} - \left. 2 \mathcal{E}_1 \mathcal{W} \right|_{\hat Y_4} - \left. 4 \mathcal{E}_2 \mathcal{W} \right|_{\hat Y_4} - \left. 6 \mathcal{E}_3 \mathcal{W} \right|_{\hat Y_4} \\
      &\qquad - \left. 3 \mathcal{E}_4 \mathcal{W} \right|_{\hat Y_4} + \left. 5 \overline{\mathcal{K}}_{B_3} \mathcal{W} \right|_{\hat Y_4} - 3 \left. \mathcal{S} \mathcal{W} \right|_{\hat Y_4} + \left. \mathcal{S} \mathcal{X} \right|_{\hat Y_4} - \left. 6 \overline{\mathcal{K}}_{B_3} \mathcal{Z} \right|_{\hat Y_4} + \left. 5 \mathcal{W} \mathcal{Z} \right|_{\hat Y_4} \, . \label{equ:Proof3Equ1}
      \end{split}
      \end{align}
 \item By use of the linear relations induced from the $\text{SU} ( 5 ) \times \text{U} ( 1 )_X$-top and the trivial restrictions introduced in 
      \autoref{subsec:TrivialRestrictions}, it can be seen that (\ref{equ:Proof3Equ1}) is equivalent to
      \begin{align}
      \begin{split}
      0 &= 3 V \left( P_T^\prime, e_0, e_1 \right) -2 V \left( P_T^\prime, e_2, e_3 \right) - 3 V \left( P_T^\prime, e_0, e_4 \right) + V \left( P_T^\prime, e_3, e_4 \right) \\
      & \qquad \qquad - V \left( P_T^\prime, e_3, s \right) - 3 V \left( P_T^\prime, e_3, x \right) + 2 V \left( P_T^\prime, e_3, y \right) \, . 
      \label{equ:Proof3Equ2}
      \end{split}
      \end{align}
      As argued in \autoref{subsec:AUsefulIdentity}, $V ( P_T^\prime, e_0, e_1 ) = V ( P_T^\prime, e_0, e_4 )$ in rational equivalence of $\hat{Y}_4$. Therefore (\ref{equ:Proof3Equ2}) is equivalent to
      \begin{align}
      \begin{split}
      0 &= -2 V \left( P_T^\prime, e_2, e_3 \right) + V \left( P_T^\prime, e_3, e_4 \right) \\
      & \qquad \qquad \qquad - V \left( P_T^\prime, e_3, s \right) - 3 V \left( P_T^\prime, e_3, x \right) + 2 V \left( P_T^\prime, e_3, y \right) \, . 
      \label{equ:Proof3Equ3}
      \end{split}
      \end{align}
 \item We can rewrite (\ref{equ:Proof3Equ3}) as
      \[ 0 = - \left. \mathcal{E}_3 \left( 2 \mathcal{E}_2 - \mathcal{E}_4 + \mathcal{S} + 3 \mathcal{X} - 2 \mathcal{Y} \right) \right|_{\hat{Y}_4} \, . \]
      Upon use of the linear relations on $\hat{X}_\Sigma$ induced from the $\text{SU} ( 5 ) \times U ( 1 )_X$-top we see that this in turn is equivalent to $0 = \left. \mathcal{E}_1 \mathcal{E}_3 \right|_{\hat{Y}_4}$. And indeed this is true. Namely $V ( P_T, e_1, e_3 ) = \emptyset$ because $e_1 e_3 \in I_{\text{SR}} ( \text{top} )$. This completes the proof.
\end{enumerate}

\subsection{Auxiliary Identities}

\subsubsection{Trivial Restrictions} \label{subsec:TrivialRestrictions}

Consider the cycle $V ( e_1, x ) \in Z^2 ( \hat X_5 )$. This cycle is non-trivial in $\hat X_5$. Nonetheless we have $V ( P_T^\prime, e_1, x ) = V ( e_1, x, e_3 e_4 s y^2 ) = \emptyset$, so this cycle pulls back to give the trivial cycle in $\hat{Y}_4$. Similarly we have
\[ V \left( P_T^\prime, x, z \right) = V \left( P_T^\prime, y, z \right) = V \left( P_T^\prime, e_0, x \right) = V \left( P_T^\prime, e_0, y \right) = \emptyset \, . \]
Hence the restriction of all these non-trivial cycles in $\hat X_5$ gives trivial cycles on $\hat{Y}_4$.

\subsubsection{A Useful Identity} \label{subsec:AUsefulIdentity}

In this subsection we justify the following identity, which we made use of in (\ref{subsec:ProofFluxRelation3}).
\[ \left. \left( \mathcal{W} - 2 \overline{\mathcal{K}}_{B_3} \right) \left( \mathcal{E}_1 + 2 \mathcal{E}_2 - \mathcal{E}_4 \right) \right|_{\hat{Y}_4} = \left. \mathcal{E}_3 \mathcal{E}_4 \right|_{\hat{Y}_4} - \left. \mathcal{E}_2 \mathcal{X} \right|_{\hat{Y}_4} - 2 \left. \overline{\mathcal{K}}_{B_3} \mathcal{E}_2 \right|_{\hat{Y}_4} + 2 \left. \mathcal{E}_2 \mathcal{E}_4 \right|_{\hat{Y}_4} \, . \label{usefulrelation} \]
We proceed in a number of steps to justify this result.
\begin{enumerate}
 \item First of all we exploit the linear relations induced from the $\text{SU} ( 5 ) \times \text{U} ( 1 )_X$-top and its Stanley-Reisner ideal. Thereby one can 
      show that (\ref{usefulrelation}) is equivalent to
      \[ \left. \mathcal{E}_1 \mathcal{W} \right|_{\hat{Y}_4} + 4 \left. \mathcal{E}_2 \mathcal{W} \right|_{\hat{Y}_4} - 2 \left. \mathcal{E}_1 \overline{\mathcal{K}}_{B_3} \right|_{\hat{Y}_4} - 6 \left. \mathcal{E}_2 \overline{\mathcal{K}}_{B_3} \right|_{\hat{Y}_4} = \left. \mathcal{E}_1 \mathcal{E}_2 \right|_{\hat{Y}_4} - 2 \left. \mathcal{E}_2 \mathcal{E}_3 \right|_{\hat{Y}_4} + \left. \mathcal{E}_0 \mathcal{E}_4 \right|_{\hat{Y}_4} \, . \label{equ:ProofUsefulIdentityEqu1} \]
 \item By use of $\mathcal{W} = 2 \overline{\mathcal{K}}_{B_3} + ( \mathcal{W} - 2 \overline{\mathcal{K}}_{B_3} )$ and $\mathcal{W} = \frac{3}{2} 
      \overline{\mathcal{K}}_{B_3} + \frac{1}{2} ( 2 \mathcal{W} - 3 \overline{\mathcal{K}}_{B_3} )$ we find that (\ref{equ:ProofUsefulIdentityEqu1}) is equivalent to
      \begin{align}
      \begin{split}
      &- V \left( P_T^\prime, e_1, a_{2,1} \right) - 2 V \left( P_T^\prime, e_2, a_{3,2} \right) \\
      &\hspace{10em} = V \left( P_T^\prime, e_1, e_2 \right) - 2 V \left( P_T^\prime, e_2, e_3 \right) + V \left( P_T^\prime, e_0, e_4 \right) \, .
      \label{equ:ProofUsefulIdentityEqu2}
      \end{split}
      \end{align}
 \item The $\text{SU} ( 5 ) \times U ( 1 )_X$-\emph{top} is a complete and simplicial toric variety. Hence  
      $\text{CH}^\bullet ( \text{top} )_{\mathbb{Q}} \cong H^\bullet ( \text{top}, \mathbb{Q} )$  \cite{cox2011toric}. By exploiting this we find the following identities in $\text{CH}^\bullet ( \hat X_5 )_{\mathbb{Q}}$: 
      \begin{align}
      \begin{split}
      V \left( e_1, e_2 e_3 s x \right) &= V \left( e_1, e_0 a_{2,1} z^2 \right) \\
      \Leftrightarrow V \left( e_1, e_2 \right) + V \left( e_1, x \right) &= V \left( e_1, e_0 \right) + V \left( e_1, a_{2,1} \right) \\
      \Leftrightarrow V \left( e_1, e_2 \right) &= V \left( e_1, e_0 \right) + V \left( e_1, a_{2,1} \right) - V \left( e_1, x \right) \, .
      \end{split}
      \end{align}
      Similarly $V ( e_2, e_0^2 e_1 a_{3,2} z^3 ) = V ( e_2, e_3 s y )$ is equivalent to $V ( e_2, e_3 ) = V ( e_2, e_1 ) + V ( e_2, a_{3,2} )$. By use of these two identites it can be shown that (\ref{equ:ProofUsefulIdentityEqu2}) is equivalent to
      \[ 0 = V \left( P_T^\prime, e_0, e_4 \right) + V \left( P_T^\prime, e_1, x \right) - V \left( P_T^\prime, e_1, e_0 \right) \, . \label{equ:ProofUsefulIdentityEqu3} \]
 \item Next recall from \autoref{subsec:TrivialRestrictions} that $V ( P_T^\prime, e_1, x ) = \emptyset$. Hence all we are left to show is that
      $V ( P_T^\prime, e_0, e_1 ) = V ( P_T^\prime, e_0, e_4 )$ holds true in $\text{CH}^2( \hat{Y}_4 )$. To see this we start with a primary decomposition, which shows
      \[ V \left( P_T^\prime, e_0, e_1 \right) = V \left( e_0, e_1, y e_3 e_4 + x z a_{1,0} \right) = V \left( P_T^\prime, e_0, y s e_3 e_4 + x z a_{1,0} \right) \, . \]
      By use of a suitable interpolating cycle in $\mathbb{P}^1 \times \hat{Y}_4$ we thus arrive at $V ( P_T^\prime, e_0, e_1 ) = V ( P_T^\prime, e_0, x z a_{1,0} )$. A primary decomposition of $V ( P_T^\prime, e_0, x z a_{1,0} )$ now shows
      \[ V \left( P_T^\prime, e_0, e_1 \right) = V \left( a_{1,0}, e_0, x^3 s e_1 e_2^2 - y^2 e_4 \right) + V \left( z, e_0, x^3 s e_1 e_2^2 - y^2 e_4 \right) \, . \]
      For $V ( P_T^\prime, e_0, e_4 )$ we proceed along the very same lines. Namely first we obtain from primary decomposition
      \begin{align}
      \begin{split}
      V \left( P_T^\prime, e_0, e_4 \right) &= V \left( P_T^\prime, e_0, e_4, x^2 s e_1 e_2^2 e_3 - y z a_{1,0} \right) \\
                                            &= V \left( P_T^\prime, e_0, x^2 s e_1 e_2^2 e_3 - y z a_{1,0} \right) \, .
      \end{split}
      \end{align}
      By use of an interpolating cycle in $\mathbb{P}^1 \times \hat{Y}_4$ we see $V ( P_T^\prime, e_0, e_4 ) = V ( P_T^\prime, e_0, y z a_{1,0} )$. Finally a primary decomposition of $V ( P_T^\prime, e_0, y z a_{1,0} )$ yields
      \[ V \left( P_T^\prime, e_0, e_4 \right) = V \left( a_{1,0}, e_0, x^3 s e_1 e_2^2 - y^2 e_4 \right) + V \left( z, e_0, x^3 s e_1 e_2^2 - y^2 e_4 \right) \, . \]
      Indeed, these results imply that $V ( P_T^\prime, e_0, e_1 ) = V( P_T^\prime, e_0, e_4)$ holds true in $\text{CH}^2 ( \hat{Y}_4 )$, which completes our proof.
\end{enumerate}

\section{Details of the \texorpdfstring{$\mathbf{\mathbf{SU(4)}}$}{SU(4)} Model} \label{computations_SU4}

\subsection{Fibre Structure} \label{fibre_structure_SU4}

We consider the resolved $SU(4)$ Tate model defined by $\hat Y_4 = V(P_T) \subset \hat X_5$ with 
\[ P_T = y^2 e_3 + a_{1,0} x y z + a_{3,2} y z^3 e_0^2 e_1 e_3 - x^3 e_1 e_2^2 - a_{2,1} x^2 z^2 e_0 e_1 e_2 - a_{4,2} x z^4 e_0^2 e_1 - a_{6,4} z^6 e_0^4 e_1^2 e_3 \, . \]
The divisors $E_i= V (e_i)$, $i=1,2,3$, denote the $SU(4)$ Cartan divisors, and $E_0$ represents the fibration of the affine node over the $SU(4)$ divisor $W = V(w) \subset B_3$. The Stanley-Reisner ideal 
\[ I_{\text{SR}} \left( \text{top} \right) = \left\langle x y z, x y e_0, z e_1, z e_2, z e_3, y e_1, x e_3, e_0 e_2 \right\rangle \]
and the linear relations
\begin{align}\label{eq:linear_relations-su4}
\begin{split}
X - 2 Z - E_0 + E_2 - 2 \overline{K}_{B_3} + W &= 0 \, , \\
- X + Y - Z + E_3 - \overline{K}_{B_3} &= 0 \, , \\
- Y + 3Z + 2 E_0 + E_1 + 3 \overline{K}_{B_3} - 2 W &= 0
\end{split}
\end{align}
both follow from the toric data of the fibre ambient space collected in \autoref{toricambient-SU4}.
\begin{table}
\begin{center}
\begin{tabular}{|c|ccc|c|ccc|}
\toprule
& $x$ & $y$ & $z$ & $w \equiv e_0$ & $e_1$ & $e_2$ & $e_3$ \\
\hline \hline
$\overline{K}_{B_3}$ & 2 & 3 & 0 & 0 & 0 & 0 & 0 \\
$\mathcal{W}$ & 0 & 0 & 0 & 1 & 0 & 0 & 0 \\
\hline
&  2 &  3 & 1 &  0 & 0 & 0 & 0 \\
& -1 & -1 & 0 & -1 & 1 & 0 & 0 \\
& -1 & -2 & 0 & -1 & 0 & 0 & 1 \\
& -2 & -2 & 0 & -1 & 0 & 1 & 0 \\
\bottomrule
\end{tabular} \caption{Toric fibre ambient space for $SU(4)$ Tate model. } \label{toricambient-SU4}
\end{center}
\end{table}
From the discriminant of $P_T$ one can read off that the enhancement loci are given by the curves 
\[ C_{\mathbf{6}} = V \left(w, a_{1,0} \right), \quad  C_{\mathbf{4}} = V \left( w, a_{4,2} \left(a_{4,2} + a_{1,0} a_{3,2} \right) - a_{1,0}^2 a_{6,4} \right) \, . \label{C46SU4} \]
Over generic points of the matter curve $C_{\mathbf{6}} = V(w, a_{1,0})$ the resolution divisors behave as
\begin{align}
\begin{split}
E_0 |_{C_\mathbf{6}} &= \mathbb{P}_{0}^1 \left( \mathbf{6} \right) \, , \qquad \qquad
E_1 |_{C_\mathbf{6}} = \mathbb{P}_{13}^1 \left( \mathbf{6} \right) \, , \\
E_2 |_{C_\mathbf{6}} &= \mathbb{P}_{2}^1 \left( \mathbf{6} \right) \, , \qquad \qquad
E_3 |_{C_\mathbf{6}} = \mathbb{P}_{13}^1 \left( \mathbf{6} \right) + \mathbb{P}_{3A}^1 \left( \mathbf{6} \right) \, , \\
\end{split}
\end{align}
with
\begin{align}
\begin{split}
\mathbb{P}^1_0 \left( \mathbf{6} \right) &= V \left( a_{1,0}, e_0, x^3 e_1 e_2^2 - y^2 e_3 \right) \, , \\
\mathbb{P}^1_{13} \left( \mathbf{6} \right) &= V \left(a_{1,0}, e_1, e_3 \right) \, , \\
\mathbb{P}^1_2 \left( \mathbf{6} \right) &= V \left( a_{1,0}, e_2, e_0^4 z^6 e_1^2 e_3 a_{6,4} - e_0^2 y z^3 e_1 e_3 a_{3,2} + e_0^2 x z^4 e_1 a_{4,2} - y^2 e_3 \right) \, , \\
\mathbb{P}^1_{3A} \left( \mathbf{6} \right) &= V \left( a_{1,0}, e_3, e_0^2 z^4 a_{4,2} + e_0 x z^2 e_2 a_{2,1} + x^2 e_2^2 \right) \, .
\end{split}
\end{align}
The resulting intersection numbers in the fibre with the Cartan divisors are as follows:
\begin{align}
\begin{tabular}{|c||c|c|c|c|}
\toprule
& $\mathbb{P}^1_0 \left( \mathbf{6} \right)$ & $\mathbb{P}^1_{13} \left( \mathbf{6} \right)$ & $\mathbb{P}^1_2 \left( \mathbf{6} \right)$ & $\mathbb{P}^1_{3A} \left( \mathbf{6} \right)$ \\
\hline \hline
$E_0$ & -2 & 1  & 0  & 0 \\
$E_1$ & 1  & -2 & 1  & 2 \\
$E_2$ & 0  & 1  & -2 & 0 \\
$E_3$ & 1  & 0  & 1  & -2 \\
\bottomrule
\end{tabular}
\end{align}
Consequently the weight vector $\vect{\beta}$ associated with $\mathbb{P}^1_{3A} ( \mathbf{6} )$ satisfies $\vect{\beta} ( \mathbb{P}^1_{3A} ( \mathbf{6} ) ) = ( 2, 0, -2 )$.\footnote{Note that for this matter surface there are actually two $\mathbb P^1$'s over every point of the $\mathbf 6$-curve of the base. Since these $\mathbf P^1$'s are exchanged around the points $w=a_{1,0}=a_{2,1}^2-4\,a_{4,2}=0$ they are indistinguishable and  become identified. This is also the reason why we put a one-half in front of $\mathbb{P}^1_{3A} ( \mathbf{6} )$ to define the `right' matter surface. }
The matter surface associated with the $\mathbf{6}$ weight $( 1, 0, -1 )$ is therefore $\frac{1}{2} \mathbb{P}^1_{3A} ( \mathbf{6} )$, defined as an element in $\mathrm{CH}^2(\hat Y_4) \otimes \mathbb Q$.

Over $C_{\mathbf{4}} = V(Q, w)$ with $Q = a_{4,2} [ a_{4,2} + a_{1,0,} a_{3,2} ] - a_{1,0}^2 a_{6,4}$, the fibral structure is 
\begin{align}
\begin{split}
E_0 |_{C_\mathbf{4}} &= \mathbb{P}_{0}^1 \left( \mathbf{4} \right) \, , \qquad \qquad \qquad \qquad \quad
E_1 |_{C_\mathbf{4}} = \mathbb{P}_{1}^1 \left( \mathbf{4} \right) \, , \\
E_2 |_{C_\mathbf{4}} &= \mathbb{P}_{2B}^1 \left( \mathbf{4} \right) + \mathbb{P}_{2C}^1 \left( \mathbf{4} \right) \, , \qquad \qquad
E_3 |_{C_\mathbf{4}} = \mathbb{P}_{3}^1 \left( \mathbf{4} \right) \, , \\
\end{split}
\end{align}
with
\begin{align*}
\begin{split}
\mathbb{P}^1_0 \left( \mathbf{4} \right) &= V \left( e_0, Q, x^3 e_1 e_2^2 - x y z a_{1,0} - y^2 e_3 \right) \, , \\
\mathbb{P}^1_1 \left( \mathbf{4} \right) &= V \left(e_1, Q, x z a_{1,0} + y e_3, y e_3 a_{3,2} a_{4,2} - x z a_{4,2}^2 - y e_3 a_{1,0} a_{6,4} 
      \right) \, , \\
\mathbb{P}^1_{2B} \left( \mathbf{4} \right) &= V \left(e_2, Q, e_0^2 z^3 e_1 a_{4,2} - y a_{1,0}, e_0^2 z^3 e_1 a_{1,0} a_{6,4} - y a_{1,0} a_{3,2} - y a_{4,2}, \right. \\
  & \qquad \quad \left. e_0^4 z^6 e_1^2 a_{6,4} - e_0^2 y z^3 e_1 a_{3,2} - y^2 \right) \, , \\
\mathbb{P}^1_{2C} \left( \mathbf{4} \right) &= V \left(e_2, Q, e_0^2 z^3 e_1 e_3 a_{4,2} a_{6,4} - y e_3 a_{3,2} a_{4,2} + x z a_{4,2}^2 + y e_3 a_{1,0} a_{6,4}, \right. \\
& \qquad \quad e_0^2 z^3 e_1 e_3 a_{1,0} a_{6,4} + x z a_{1,0} a_{4,2} + y e_3 a_{4,2}, \\
& \qquad \quad e_0^2 z^3 e_1 e_3 a_{1,0} a_{3,2} + e_0^2 z^3 e_1 e_3 a_{4,2} + x z a_{1,0}^2 + y e_3 a_{1,0}, \\
& \qquad \quad \left. e_0^4 z^6 e_1^2 e_3 a_{6,4} - e_0^2 y z^3 e_1 e_3 a_{3,2} + e_0^2 x z^4 e_1 a_{4,2} - x y z a_{1,0} - y^2 e_3 \right) \, , \\
\mathbb{P}^1_3 \left( \mathbf{4} \right) &= V \left( e_3, Q, e_0^2 z^4 e_1 a_{4,2} + e_0 x z^2 e_1 e_2 a_{2,1} + x^2 e_1 e_2^2 - y z a_{1,0}, \right. \\
& \qquad \quad \left. e_0^2 z^4 e_1 a_{1,0}^2 a64 + e_0 x z^2 e_1 e_2 a_{1,0} a_{2,1} a_{3,2} +  e_0 x z^2 e_1 e_2 a_{2,1} a_{4,2} + x^2 e_1 e_2^2 a_{1,0} a_{3,2} \right. \\
& \qquad \quad \left. + x^2 e_1 e_2^2 a_{4,2} - y z a_{1,0}^2 a_{3,2} - y z a_{1,0} a_{4,2} \right) \, .
\end{split}
\end{align*} 
The fibral intersection numbers with the divisors $E_i$ are
\begin{align}
\begin{tabular}{|c||c|c|c|c|c|}
\toprule
& $\mathbb{P}^1_0 \left( \mathbf{4} \right)$ & $\mathbb{P}^1_{1} \left( \mathbf{4} \right)$ & $\mathbb{P}^1_{2B} \left( \mathbf{4} \right)$ & $\mathbb{P}^1_{2C} \left( \mathbf{4} \right)$ & $\mathbb{P}^1_{3} \left( \mathbf{4} \right)$ \\
\hline \hline
$E_0$ & -2 & 1 & 0 & 0 & 1 \\
$E_1$ & 1 & -2 & 0 & 1 & 0 \\
$E_2$ & 0 & 1 & -1 & -1 & 1 \\
$E_3$ & 1 & 0 & 1 & 0 & -2 \\
\bottomrule
\end{tabular} 
\end{align}
The resulting weight vectors associated with the split surfaces,
\[ \vect{\beta} \left( \mathbb{P}^1_{2B} \left( \mathbf{4} \right) \right) = \left( 0, -1, 1 \right), \qquad \vect{\beta} \left( \mathbb{P}^1_{2C} \left( \mathbf{4} \right) \right) = \left( 1, -1, 0 \right) \,, \]
identify the latter as matter surfaces for the $\mathbf{4}$ and $\mathbf{\overline{4}}$ representations, respectively.

\subsection{Proof of Fluxlessness} \label{proofSU(4)}

To verify explicitly that the matter surface fluxes $A(\mathbf{4})$ and $A(\mathbf{6})$ are trivial in the Chow ring, we use the fibral structure discussed in \autoref{fibre_structure_SU4}. Over $C_{\mathbf{6}}$, the fibre of $E_3$ splits into $\mathbb{P}^1_{13} ( \mathbf{6} ) + \mathbb{P}^1_{3A} ( \mathbf{6} )$, and the weight vector associated with $S^1(\mathbf{6}) = \frac{1}{2} \mathbb{P}^1_{3A} ( \mathbf{6} )$ is $ \mathbf{\beta} \left( S^1_{\mathbf{6}} \right) = \left( 1,0,-1 \right)$. The associated matter surface flux is therefore indeed trivial in $\mathbb{Q} \otimes \mathrm{CH}^2(\hat Y_4)$ because
\begin{align}
{A}\left( \mathbf{6} \right) &= \frac{1}{2} \mathbb{P}^1_{3A} \left( \mathbf{6} \right) + \mathfrak{\beta}^T({\mathbf{6}}) \, C^{-1} \cdot \left( \begin{array}{c} \mathbb{P}^1_{13} \left( \mathbf{6} \right) \\ \mathbb{P}^1_2 \left( \mathbf{6} \right) \\ \mathbb{P}^1_{13} \left( \mathbf{6} \right) + \mathbb{P}^1_{3A} \left( \mathbf{6} \right) \end{array} \right) \\
&= \frac{1}{2} \left( \mathbb{P}^1_{3A} \left( \mathbf{6} \right) - \left( - \mathbb{P}^1_{13} \left( \mathbf{6} \right) + \mathbb{P}^1_{13} \left( \mathbf{6} \right) + \mathbb{P}^1_{3A} \left( \mathbf{6} \right) \right)    \right)= 0 \, .
\end{align}
Here $\beta^T$ denotes the weights of the ${\bf 6}$ representation.

Over $C_\mathbf{4}$ it is $E_2$ which splits into two surfaces with weight-vectors
\[ \vect{\beta} \left( \mathbb{P}^1_{2B} \left( \mathbf{4} \right) \right) = \left( 0, -1, 1 \right), \qquad \vect{\beta} \left( \mathbb{P}^1_{2C} \left( \mathbf{4} \right) \right) = \left( 1, -1, 0 \right) \, . \]

We can for instance start with $S^1_{\mathbf{4}} = \mathbb{P}^1_{2B} ( \mathbf{4} )$ and deduced the gauge invariant flux
\[ {A} \left( \mathbf{4} \right)  = \frac{1}{4} \left( \mathbb{P}^1_{3} \left( \mathbf{4} \right) - \mathbb{P}^1_{1} \left( \mathbf{4} \right) \right) + \frac{1}{2} \left( \mathbb{P}^1_{2B} \left( \mathbf{4} \right) - \mathbb{P}^1_{2C} \left( \mathbf{4} \right) \right) \, . \]

With the help of
\begin{align}
\begin{split}
\mathbb{P}^1_1 \left( \mathbf{4} \right) &= V \left( P_T, Q, e_1 \right) \,, \\
\mathbb{P}^1_3 \left( \mathbf{4} \right) &= V \left( P_T, Q, e_3 \right) \,, \\
\mathbb{P}^1_{2B} \left( \mathbf{4} \right) &= V \left( P_T, e_2, e_0^2 z^3 e_1 a_{42} - y a_{1,0} \right) - V \left( P^\prime, P_T, e_2, e_3 \right) \, , \\
\mathbb{P}^1_{2C} \left( \mathbf{4} \right) &= V \left( P_T, Q, e_2, \right) - \mathbb{P}^1_{2B} \left( \mathbf{4} \right) \,,
\end{split}
\end{align}
with $Q = a_{4,2} [ a_{4,2} + a_{1,0,} a_{3,2} ] - a_{1,0}^2 a_{6,4}$ the polynomial cutting out $C_{\mathbf{4}}$ in (\ref{C46SU4}) we realise that this surface can be written as a restriction from $\hat X_5$ to $\hat Y_4$,
\begin{align}\label{eq:su4-flux-generator-A4}
\begin{split}
A \left( \mathbf{4} \right) &= \left( - \cE_2 \cdot \cE_3  - 2 \cE_1 \cdot \overline{\mathcal{K}}_{B_3} - 3  \cE_2 \cdot \overline{\mathcal{K}}_{B_3} + 2 \cE_3 \cdot  \overline{\mathcal{K}}_{B_3} \right. \\
& \qquad \qquad \left. + \cE_1 \cdot \mathcal{W} + 2 \cE_2 \cdot \mathcal{W} - \cE_3 \cdot \mathcal{W} + \cE_2 \cdot \mathcal{Y} \right) |_{\hat Y_4} \,.
\end{split}
\end{align}
We are now free to use the Chow relations on $\hat X_5$ to manipulate  this expression without changing the Chow class of $A(\mathbf{4})$ on $\hat Y_4$. Let us therefore 
employ the linear relations \eqref{eq:linear_relations-su4} and rewrite $A\left( \mathbf{4} \right)$ as
\begin{equation}
A \left( \mathbf{4} \right) =  \left(  \cE_0 \cdot (\cE_1 - \cE_3) - \mathcal{X} \cdot \cE_1  \right) |_{\hat Y_4} \,.
\end{equation}
Applying the linear relations \eqref{eq:linear_relations-su4} of the ambient space once more to rewrite $\cE_1 - \cE_3 = 2\,\mathcal{Y}-3\,\mathcal{X}-2\,\cE_2$, we find that $A ( \mathbf{4}) = 0 \in \mathrm{CH}^2(\hat Y_4)$. This is a consequence of the Stanley-Reisner ideal together with the fact that the vanishing sets 
\begin{equation}
 V \left( e_0,x \right)\,,\quad
 V \left( e_0,y \right)\,,\quad
 V \left( x,z \right)\,,\quad
 V \left( e_1,x \right)\,,\quad
 V \left( y,z \right)
\end{equation}
are empty once restricted to the hypersurface $\hat Y_4$.

\bibliography{papers}

\begin{thebibliography}{10}
\expandafter\ifx\csname url\endcsname\relax
  \def\url#1{{\tt #1}}\fi
\expandafter\ifx\csname urlprefix\endcsname\relax\def\urlprefix{URL }\fi
\providecommand{\eprint}[2][]{\url{#2}}

\bibitem{Green:1984sg}
M.~B. Green and J.~H. Schwarz, {\textit {Anomaly Cancellation in Supersymmetric
  D=10 Gauge Theory and Superstring Theory}\/}, {\textit Phys. Lett.\/}
  {\textbf B149} (1984) 117--122.

\bibitem{Green:1984ed}
M.~B. Green and J.~H. Schwarz, {\textit {Infinity Cancellations in SO(32)
  Superstring Theory}\/}, {\textit Phys. Lett.\/} {\textbf B151} (1985) 21--25.

\bibitem{Green:1984qs}
M.~B. Green and J.~H. Schwarz, {\textit {The Hexagon Gauge Anomaly in Type I
  Superstring Theory}\/}, {\textit Nucl. Phys.\/} {\textbf B255} (1985)
  93--114.

\bibitem{Polchinski:1987tu}
J.~Polchinski and Y.~Cai, {\textit {Consistency of Open Superstring
  Theories}\/}, {\textit Nucl. Phys.\/} {\textbf B296} (1988) 91--128.

\bibitem{Sagnotti:1995ga}
A.~Sagnotti, {\textit {Some properties of open string theories}\/}, in {\textit
  {Supersymmetry and unification of fundamental interactions. Proceedings,
  International Workshop, SUSY 95, Palaiseau, France, May 15-19, 1995}\/}, pp.
  473--484, 1995, \href{http://www.arxiv.org/abs/hep-th/9509080}{{\texttt
  [hep-th/9509080]}}.

\bibitem{Blumenhagen:2006ci}
R.~Blumenhagen, B.~Kors, D.~Lust and S.~Stieberger, {\textit {Four-dimensional
  String Compactifications with D-Branes, Orientifolds and Fluxes}\/}, {\textit
  Phys. Rept.\/} {\textbf 445} (2007) 1--193,
  \href{http://www.arxiv.org/abs/hep-th/0610327}{{\texttt [hep-th/0610327]}}.

\bibitem{Ibanez:2012zz}
L.~E. Ibanez and A.~M. Uranga, {\textit {String theory and particle physics: An
  introduction to string phenomenology}\/}, Cambridge University Press, 2012,
  ISBN 9780521517522, 9781139227421,
  \urlprefix\url{http://www.cambridge.org/de/knowledge/isbn/item6563092/?site_locale=de_DE}.

\bibitem{Kumar:2009us}
V.~Kumar and W.~Taylor, {\textit {String Universality in Six Dimensions}\/},
  {\textit Adv. Theor. Math. Phys.\/} {\textbf 15}, no.~2 (2011) 325--353,
  \href{http://www.arxiv.org/abs/0906.0987}{{\texttt [0906.0987]}}.

\bibitem{Kumar:2009ac}
V.~Kumar, D.~R. Morrison and W.~Taylor, {\textit {Mapping 6D N = 1
  supergravities to F-theory}\/}, {\textit JHEP\/} {\textbf 02} (2010) 099,
  \href{http://www.arxiv.org/abs/0911.3393}{{\texttt [0911.3393]}}.

\bibitem{Kumar:2010ru}
V.~Kumar, D.~R. Morrison and W.~Taylor, {\textit {Global aspects of the space
  of 6D N = 1 supergravities}\/}, {\textit JHEP\/} {\textbf 11} (2010) 118,
  \href{http://www.arxiv.org/abs/1008.1062}{{\texttt [1008.1062]}}.

\bibitem{Seiberg:2011dr}
N.~Seiberg and W.~Taylor, {\textit {Charge Lattices and Consistency of 6D
  Supergravity}\/}, {\textit JHEP\/} {\textbf 06} (2011) 001,
  \href{http://www.arxiv.org/abs/1103.0019}{{\texttt [1103.0019]}}.

\bibitem{Park:2011wv}
D.~S. Park and W.~Taylor, {\textit {Constraints on 6D Supergravity Theories
  with Abelian Gauge Symmetry}\/}, {\textit JHEP\/} {\textbf 01} (2012) 141,
  \href{http://www.arxiv.org/abs/1110.5916}{{\texttt [1110.5916]}}.

\bibitem{Grimm:2012yq}
T.~W. Grimm and W.~Taylor, {\textit {Structure in 6D and 4D N=1 supergravity
  theories from F-theory}\/}, {\textit JHEP\/} {\textbf 10} (2012) 105,
  \href{http://www.arxiv.org/abs/1204.3092}{{\texttt [1204.3092]}}.

\bibitem{Green:1984bx}
M.~B. Green, J.~H. Schwarz and P.~C. West, {\textit {Anomaly Free Chiral
  Theories in Six-Dimensions}\/}, {\textit Nucl. Phys.\/} {\textbf B254} (1985)
  327--348.

\bibitem{Sagnotti:1992qw}
A.~Sagnotti, {\textit {A Note on the Green-Schwarz mechanism in open string
  theories}\/}, {\textit Phys. Lett.\/} {\textbf B294} (1992) 196--203,
  \href{http://www.arxiv.org/abs/hep-th/9210127}{{\texttt [hep-th/9210127]}}.

\bibitem{Sadov:1996zm}
V.~Sadov, {\textit {Generalized Green-Schwarz mechanism in F theory}\/},
  {\textit Phys. Lett.\/} {\textbf B388} (1996) 45--50,
  \href{http://www.arxiv.org/abs/hep-th/9606008}{{\texttt [hep-th/9606008]}}.

\bibitem{Grassi:2000we}
A.~Grassi and D.~R. Morrison, {\textit {Group representations and the Euler
  characteristic of elliptically fibered Calabi-Yau threefolds}\/}
  \href{http://www.arxiv.org/abs/math/0005196}{{\texttt [math/0005196]}}.

\bibitem{Grassi:2011hq}
A.~Grassi and D.~R. Morrison, {\textit {Anomalies and the Euler characteristic
  of elliptic Calabi-Yau threefolds}\/}, {\textit Commun. Num. Theor. Phys.\/}
  {\textbf 6} (2012) 51--127,
  \href{http://www.arxiv.org/abs/1109.0042}{{\texttt [1109.0042]}}.

\bibitem{Park:2011ji}
D.~S. Park, {\textit {Anomaly Equations and Intersection Theory}\/}, {\textit
  JHEP\/} {\textbf 01} (2012) 093,
  \href{http://www.arxiv.org/abs/1111.2351}{{\texttt [1111.2351]}}.

\bibitem{Grimm:2011fx}
T.~W. Grimm and H.~Hayashi, {\textit {F-theory fluxes, Chirality and
  Chern-Simons theories}\/}, {\textit JHEP\/} {\textbf 03} (2012) 027,
  \href{http://www.arxiv.org/abs/1111.1232}{{\texttt [1111.1232]}}.

\bibitem{Bonetti:2011mw}
F.~Bonetti and T.~W. Grimm, {\textit {Six-dimensional (1,0) effective action of
  F-theory via M-theory on Calabi-Yau threefolds}\/}, {\textit JHEP\/} {\textbf
  05} (2012) 019, \href{http://www.arxiv.org/abs/1112.1082}{{\texttt
  [1112.1082]}}.

\bibitem{Cvetic:2012xn}
M.~Cvetic, T.~W. Grimm and D.~Klevers, {\textit {Anomaly Cancellation And
  Abelian Gauge Symmetries In F-theory}\/}, {\textit JHEP\/} {\textbf 02}
  (2013) 101, \href{http://www.arxiv.org/abs/1210.6034}{{\texttt [1210.6034]}}.

\bibitem{Grimm:2015zea}
T.~W. Grimm and A.~Kapfer, {\textit {Anomaly Cancelation in Field Theory and
  F-theory on a Circle}\/}, {\textit JHEP\/} {\textbf 05} (2016) 102,
  \href{http://www.arxiv.org/abs/1502.05398}{{\texttt [1502.05398]}}.

\bibitem{Esole:2015xfa}
M.~Esole and S.-H. Shao, {\textit {M-theory on Elliptic Calabi-Yau Threefolds
  and 6d Anomalies}\/} \href{http://www.arxiv.org/abs/1504.01387}{{\texttt
  [1504.01387]}}.

\bibitem{Lin:2016vus}
L.~Lin and T.~Weigand, {\textit {G 4 -flux and standard model vacua in
  F-theory}\/}, {\textit Nucl. Phys.\/} {\textbf B913} (2016) 209--247,
  \href{http://www.arxiv.org/abs/1604.04292}{{\texttt [1604.04292]}}.

\bibitem{Bies:2014sra}
M.~Bies, C.~Mayrhofer, C.~Pehle and T.~Weigand, {\textit {Chow groups, Deligne
  cohomology and massless matter in F-theory}\/}
  \href{http://www.arxiv.org/abs/1402.5144}{{\texttt [1402.5144]}}.

\bibitem{Bies:2017fam}
M.~Bies, C.~Mayrhofer and T.~Weigand, {\textit {Gauge Backgrounds and Zero-Mode
  Counting in F-Theory}\/} \href{http://www.arxiv.org/abs/1706.04616}{{\texttt
  [1706.04616]}}.

\bibitem{Krause:2012yh}
S.~Krause, C.~Mayrhofer and T.~Weigand, {\textit {Gauge Fluxes in F-theory and
  Type IIB Orientifolds}\/}, {\textit JHEP\/} {\textbf 1208} (2012) 119,
  \href{http://www.arxiv.org/abs/1202.3138}{{\texttt [1202.3138]}}.

\bibitem{Bershadsky:1996nh}
M.~Bershadsky, K.~A. Intriligator, S.~Kachru, D.~R. Morrison, V.~Sadov and
  C.~Vafa, {\textit {Geometric singularities and enhanced gauge symmetries}\/},
  {\textit Nucl. Phys.\/} {\textbf B481} (1996) 215--252,
  \href{http://www.arxiv.org/abs/hep-th/9605200}{{\texttt [hep-th/9605200]}}.

\bibitem{Shioda}
T.~Shioda, {\textit On elliptic modular surfaces\/}, {\textit J. Math. Soc.
  Japan\/} {\textbf 24} (1972) 20--59,
  \urlprefix\url{http://dx.doi.org/10.2969/jmsj/02410020}.

\bibitem{Tate1}
J.~T. Tate, {\textit Algebraic cycles and poles of zeta functions\/}, in
  {\textit Arithmetical {A}lgebraic {G}eometry ({P}roc. {C}onf. {P}urdue
  {U}niv., 1963)\/}, pp. 93--110, Harper \& Row, New York, 1965.

\bibitem{Tate2}
J.~Tate, {\textit On the conjectures of {B}irch and {S}winnerton-{D}yer and a
  geometric analog\/}, in {\textit S\'eminaire {B}ourbaki, {V}ol.\ 9\/}, pp.
  Exp.\ No.\ 306, 415--440, Soc. Math. France, Paris, 1995.

\bibitem{2001math.....12259W}
R.~{Wazir}, {\textit {Arithmetic on Elliptic Threefolds}\/}, {\textit ArXiv
  Mathematics e-prints\/} \href{http://www.arxiv.org/abs/math/0112259}{{\texttt
  [math/0112259]}}.

\bibitem{Intriligator:1997pq}
K.~A. Intriligator, D.~R. Morrison and N.~Seiberg, {\textit {Five-dimensional
  supersymmetric gauge theories and degenerations of Calabi-Yau spaces}\/},
  {\textit Nucl.Phys.\/} {\textbf B497} (1997) 56--100,
  \href{http://www.arxiv.org/abs/hep-th/9702198}{{\texttt [hep-th/9702198]}}.

\bibitem{Hayashi:2014kca}
H.~Hayashi, C.~Lawrie, D.~R. Morrison and S.~Schafer-Nameki, {\textit {Box
  Graphs and Singular Fibers}\/}, {\textit JHEP\/} {\textbf 05} (2014) 048,
  \href{http://www.arxiv.org/abs/1402.2653}{{\texttt [1402.2653]}}.

\bibitem{Esole:2014bka}
M.~Esole, S.-H. Shao and S.-T. Yau, {\textit {Singularities and Gauge Theory
  Phases}\/}, {\textit Adv. Theor. Math. Phys.\/} {\textbf 19} (2015)
  1183--1247, \href{http://www.arxiv.org/abs/1402.6331}{{\texttt [1402.6331]}}.

\bibitem{Greene:1993vm}
B.~R. Greene, D.~R. Morrison and M.~R. Plesser, {\textit {Mirror manifolds in
  higher dimension}\/}, {\textit Commun. Math. Phys.\/} {\textbf 173} (1995)
  559--598, [AMS/IP Stud. Adv. Math.1,745(1996)],
  \href{http://www.arxiv.org/abs/hep-th/9402119}{{\texttt [hep-th/9402119]}}.

\bibitem{Braun:2014xka}
A.~P. Braun and T.~Watari, {\textit {The Vertical, the Horizontal and the Rest:
  anatomy of the middle cohomology of Calabi-Yau fourfolds and F-theory
  applications}\/}, {\textit JHEP\/} {\textbf 01} (2015) 047,
  \href{http://www.arxiv.org/abs/1408.6167}{{\texttt [1408.6167]}}.

\bibitem{Beasley:2008dc}
C.~Beasley, J.~J. Heckman and C.~Vafa, {\textit {GUTs and Exceptional Branes in
  F-theory - I}\/}, {\textit JHEP\/} {\textbf 01} (2009) 058,
  \href{http://www.arxiv.org/abs/0802.3391}{{\texttt [0802.3391]}}.

\bibitem{Braun:2014pva}
A.~P. Braun, A.~Collinucci and R.~Valandro, {\textit {Hypercharge flux in
  F-theory and the stable Sen limit}\/}, {\textit JHEP\/} {\textbf 07} (2014)
  121, \href{http://www.arxiv.org/abs/1402.4096}{{\texttt [1402.4096]}}.

\bibitem{Borchmann:2013hta}
J.~Borchmann, C.~Mayrhofer, E.~Palti and T.~Weigand, {\textit {SU(5) Tops with
  Multiple U(1)s in F-theory}\/}, {\textit Nucl. Phys.\/} {\textbf B882} (2014)
  1--69, \href{http://www.arxiv.org/abs/1307.2902}{{\texttt [1307.2902]}}.

\bibitem{Grimm:2010ez}
T.~W. Grimm and T.~Weigand, {\textit {On Abelian Gauge Symmetries and Proton
  Decay in Global F-theory GUTs}\/}, {\textit Phys. Rev.\/} {\textbf D82}
  (2010) 086009, \href{http://www.arxiv.org/abs/1006.0226}{{\texttt
  [1006.0226]}}.

\bibitem{Krause:2011xj}
S.~Krause, C.~Mayrhofer and T.~Weigand, {\textit {$G_4$ flux, chiral matter and
  singularity resolution in F-theory compactifications}\/}, {\textit
  Nucl.Phys.\/} {\textbf B858} (2012) 1--47,
  \href{http://www.arxiv.org/abs/1109.3454}{{\texttt [1109.3454]}}.

\bibitem{oai:arXiv.org:hep-th/9609122}
E.~Witten, {\textit {On flux quantization in M theory and the effective
  action}\/}, {\textit J.Geom.Phys.\/} {\textbf 22} (1997) 1--13,
  \href{http://www.arxiv.org/abs/hep-th/9609122}{{\texttt [hep-th/9609122]}}.

\bibitem{Erler:1993zy}
J.~Erler, {\textit {Anomaly cancellation in six-dimensions}\/}, {\textit J.
  Math. Phys.\/} {\textbf 35} (1994) 1819--1833,
  \href{http://www.arxiv.org/abs/hep-th/9304104}{{\texttt [hep-th/9304104]}}.

\bibitem{FultonInt}
W.~Fulton, {\textit {Intersection Theory}\/}, {\textit Princeton University
  Press 1993\/} .

\bibitem{cox2011toric}
D.~Cox, J.~Little and H.~Schenck, {\textit Toric Varieties\/}, Graduate studies
  in mathematics, American Mathematical Soc., 2011, ISBN 9780821884263,
  \urlprefix\url{https://books.google.de/books?id=eXLGwYD4pmAC}.

\bibitem{Schafer-Nameki:2016cfr}
S.~Sch{\"a}fer-Nameki and T.~Weigand, {\textit {F-theory and 2d $(0, 2)$
  theories}\/}, {\textit JHEP\/} {\textbf 05} (2016) 059,
  \href{http://www.arxiv.org/abs/1601.02015}{{\texttt [1601.02015]}}.

\bibitem{Apruzzi:2016iac}
F.~Apruzzi, F.~Hassler, J.~J. Heckman and I.~V. Melnikov, {\textit {UV
  Completions for Non-Critical Strings}\/}, {\textit JHEP\/} {\textbf 07}
  (2016) 045, \href{http://www.arxiv.org/abs/1602.04221}{{\texttt
  [1602.04221]}}.

\bibitem{Apruzzi:2016nfr}
F.~Apruzzi, F.~Hassler, J.~J. Heckman and I.~V. Melnikov, {\textit {From 6D
  SCFTs to Dynamic GLSMs}\/}
  \href{http://www.arxiv.org/abs/1610.00718}{{\texttt [1610.00718]}}.

\bibitem{Lawrie:2016rqe}
C.~Lawrie, S.~Schafer-Nameki and T.~Weigand, {\textit {The gravitational sector
  of 2d (0, 2) F-theory vacua}\/}, {\textit JHEP\/} {\textbf 05} (2017) 103,
  \href{http://www.arxiv.org/abs/1612.06393}{{\texttt [1612.06393]}}.

\bibitem{Lawrie:2016axq}
C.~Lawrie, S.~Schafer-Nameki and T.~Weigand, {\textit {Chiral 2d Theories from
  N=4 SYM with Varying Coupling}\/}, {\textit JHEP\/} {\textbf 04} (2017) 111,
  \href{http://www.arxiv.org/abs/1612.05640}{{\texttt [1612.05640]}}.

\bibitem{Uranga:2000xp}
A.~M. Uranga, {\textit {D-brane probes, RR tadpole cancellation and K theory
  charge}\/}, {\textit Nucl. Phys.\/} {\textbf B598} (2001) 225--246,
  \href{http://www.arxiv.org/abs/hep-th/0011048}{{\texttt [hep-th/0011048]}}.

\bibitem{GarciaEtxebarria:2005qc}
I.~Garcia-Etxebarria and A.~M. Uranga, {\textit {From F/M-theory to K-theory
  and back}\/}, {\textit JHEP\/} {\textbf 02} (2006) 008,
  \href{http://www.arxiv.org/abs/hep-th/0510073}{{\texttt [hep-th/0510073]}}.

\end{thebibliography}
\bibliographystyle{custom1}

\end{document}